\documentclass[sigconf]{acmart}
\usepackage[ruled,vlined,linesnumbered]{algorithm2e}
\usepackage{subfigure}
\usepackage{balance}
\usepackage{comment}

\AtBeginDocument{%
  \providecommand\BibTeX{{%
    \normalfont B\kern-0.5em{\scshape i\kern-0.25em b}\kern-0.8em\TeX}}}

\setcopyright{acmcopyright}
\copyrightyear{2018}
\acmYear{2018}
\acmDOI{10.1145/1122445.1122456}

\acmConference[Woodstock '18]{Woodstock '18: ACM Symposium on Neural
  Gaze Detection}{June 03--05, 2018}{Woodstock, NY}
\acmBooktitle{Woodstock '18: ACM Symposium on Neural Gaze Detection,
  June 03--05, 2018, Woodstock, NY}
\acmPrice{15.00}
\acmISBN{978-1-4503-XXXX-X/18/06}

\begin{document}

\title{LiMITS: An Effective Approach for Trajectory Simplification}

\author{Yunheng Han}
\affiliation{
\institution{The University of Maryland}
  \city{College Park}
  \state{Maryland}
  \postcode{20742}
}
\email{yhhan@cs.umd.edu}

\author{Hanan Samet}
\affiliation{
\institution{The University of Maryland}
  \city{College Park}
  \state{Maryland}
  \postcode{20742}
}
\email{hjs@cs.umd.edu}


\begin{abstract}
Trajectories represent the mobility of moving objects and thus is of great value in  data mining applications. However, trajectory data is enormous in volume, so it is expensive to store and process the raw data directly. Trajectories are also redundant so data compression techniques can be applied. In this paper, we propose effective algorithms to simplify trajectories. We first extend existing algorithms by replacing the commonly used $L_2$ metric with the $L_\infty$ metric so that they can be generalized to high dimensional space (e.g., 3-space in practice). Next, we propose a novel approach, namely \underline{L}-\underline{i}nfinity \underline{M}ultidimensional \underline{I}nterpolation \underline{T}rajectory \underline{S}implification (LiMITS). LiMITS belongs to weak simplification and takes advantage of the $L_\infty$ metric. It generates simplified trajectories by multidimensional interpolation. It also allows a new format called compact representation to further improve the compression ratio. Finally, We demonstrate the performance of LiMITS through experiments on real-world datasets, which show that it is more effective than other existing methods.
\end{abstract}

\newcommand{\frechet}{Fr\'echet~}
\newcommand{\argmin}{\operatornamewithlimits{argmin}}
\setlength{\abovecaptionskip}{5pt}
\setlength{\belowcaptionskip}{-5pt}
\let\svthefootnote\thefootnote

\maketitle

\let\thefootnote\relax\footnote{This work was supported in part by the US National
     Science Foundation under grants IIS-13-20791 and
     IIS-1816889.}\addtocounter{footnote}{-1}\let\thefootnote\svthefootnote

\section{Introduction}
Trajectory data is ubiquitous and informative. Nowadays GPS navigation devices are constantly generating unprecedented amounts of trajectory data, from cell phones in hands to vehicles on the way. The enormous amounts of data make it possible to discover knowledge effectively as trajectories represent the mobility of the moving objects, e.g., many data driven approaches have been proposed to solve real-world problems in the past decades~\cite{Zheng2015}. However, such huge amounts also result in difficulty in data storage and retrieval, which influences the performance of all algorithms based on real trajectories. Trajectory data is also redundant because of inertia. Given the historical trajectory of a moving object, it is most likely that it keeps the current state (e.g., displacement, velocity and acceleration). The probability distribution of the trajectory, therefore, has relatively low information entropy, and data compression techniques can be adopted to reduce the storage cost.

Data compression could be either \textit{lossless} or \textit{lossy}. Lossless compression is widely used for text and images~\cite{ziv1978compression, witten1987arithmetic}, but it has been proved to be ineffective for trajectory data~\cite{cudre2010trajstore, nibali2015trajic, han2017compress, zhang2018trajectory}. In practice, most applications do not require as much accuracy as raw data has. As a result, many lossy approaches were proposed, namely trajectory simplification~\cite{douglas1973algorithms,IMAI198631, meratnia2004spatiotemporal, Muckell2014, Lin2019}. These approaches represent trajectories with fewer points while a certain data quality is still preserved. The output of trajectory simplification shares the same format as the input, so it is as convenient to process the simplified data. Furthermore, information retrieval is even more efficient on the simplified data: the higher the compression ratio is, the fewer points the simplified data contains, and thus the less time query processing costs. The simplification algorithms are classified into two categories, \textit{strong} and \textit{weak simplification}. Strong simplification produces a subset of the input data, while weak simplification may involve new points. Weak simplification is more challenging because the number of feasible solutions becomes uncountable (while it is finite in strong simplification). On the other hand, it could be more effective and hence a direction worth exploring.

The quality of trajectory data after simplification is guaranteed by the similarity between input and output trajectories: the more similar input and output are, the higher the quality~\cite{toohey2015similarity}. A trajectory is the path of a moving object in space as a function of time, and thus it contains both spatial and temporal information. In order to measure the similarity between trajectories, \frechet distance~\cite{alt1995frechet,bringmann2019walking} is usually used because it takes into account not only the spatial information---the locations of the points, but also the temporal information---the order they follow. In the scope of trajectory simplification, the error is often bounded by the \textit{synchronized distance}~\cite{meratnia2004spatiotemporal, cao2006spatio}, a specialized version of the general \frechet distance, which is defined as 
\begin{equation}
    d(\mathbf{S}, \mathbf{T}) = \max_{t} \| \mathbf{S}(t)- \mathbf{T}(t)\|
    \label{sd}
\end{equation}
where $\mathbf{S}$, $\mathbf{T}$ are two trajectories, $t$ is a time stamp and $\|\mathbf{x}-\mathbf{y}\|$ is the distance between $\mathbf{x}$ and $\mathbf{y}$. It measures distance between two objects at all time stamps and takes the maximum as the result. If $d(\mathbf{S}, \mathbf{T})$ is bounded by a given threshold, i.e., $d(\mathbf{S}, \mathbf{T})\leq\epsilon$, the derivation of output from input at any time stamp will be no greater than $\epsilon$, so the data quality is preserved. It is crucial to design a suitable and effective distance metric, as the problem of trajectory simplification can be perceived an instance of similarity learning~\cite{zhang2018trajectory}. Existing work considers the \textit{Euclidean distance} ($L_2$) in Equation~\ref{sd}, since it is the ordinary straight line distance. Notwithstanding, there are many other distance metrics, among which the Minkowski distance is an extension of the $L_2$ distance. For any value of $p$, the Minkowski distance is defined as
\begin{equation}
    \|\mathbf{x}-\mathbf{y}\| = \left(\sum^{m}_{i=1}{|x_i-y_i|^p}\right)^\frac{1}{p}.
    \label{mk}
\end{equation}
Note that only when $p \geq 1$ does the triangle inequality hold. Moreover, it is known as the \textit{Manhattan distance} or the city block distance ($L_1$) when $p=1$, and the \textit{Chebyshev distance} or the chessboard distance ($L_\infty$) when $p\to\infty$. Different from existing work, we use the \emph{synchronized $L_\infty$ distance} to bound the similarity between two trajectories.

$L_\infty$ is widely used in many applications including quadtree image representations~\cite{samet1983quadtree,samet1985computing}, corridor computation~\cite{ang1990new,amir1999efficient}, polygonal curve approximation~\cite{ihm1991piecewise, hiroshi1988polygonal}, warehouse systems~\cite{langevin2005logistics} and computer aided manufacturing~\cite{seitz1989advanced}. However, it has not been explored in the problem of trajectory simplification. $L_\infty$ is advantageous for the following reasons. First, it has the same properties as other Minkowski distances, such as the triangle inequality and translation invariance. Second, $L_\infty$ is a good approximation to other metrics. Third, it is the limiting case in Equation~\ref{mk} and thus smaller than any other distances, so it leads to the loosest error bounds and the highest compression ratio. Finally, $L_\infty$ makes it possible to extend existing simplification algorithms~\cite{IMAI198631,meratnia2004spatiotemporal,song2014press,han2017compress, Lin2019} originally designed for trajectories in 1-space and 2-space, including the optimal strong simplification and state-of-the-art weak simplification, to high dimensions (e.g., 3-space in practice), while it is difficult under other metrics. Hence $L_\infty$ is suitable for real applications on trajectory data.

In this paper, we propose an effective approach for trajectory simplification, namely \textit{\textbf{L}-\textbf{i}nfinity \textbf{M}ultidimensional \textbf{I}nterpolation \textbf{T}rajectory \textbf{S}implification} (LiMITS). It is weak simplification algorithm under the $L_\infty$ metric. Here we only present an overview of LiMITS. Given a trajectory $\mathbf{T}$ in $m$-space, LiMITS first projects it onto the planes spanned by the time axis and every spatial axis, forming $m$ trajectories $T_1, T_2, \cdots, T_m$ in 1-space. Since trajectory simplification in 1-space is feasible, we simplify them individually and obtain simplified trajectories $S_i$ for each $T_i$. Eventually, we combine $S_1, S_2, \cdots, S_m$ by interpolation to construct an output trajectory $\mathbf{S}$ in $m$-space. Note that the straightforward implementation is simple but not effective, and we leave the elaboration to the later sections. To sum up, the main contributions of our work include: 
\begin{enumerate}
\setlength\itemsep{0em}
  \item We explore the $L_\infty$ distance for trajectory simplification and extend existing algorithms to higher dimensions. This is crucial to real applications as most trajectories are in 3-space.
  \item LiMITS falls into the category of weak simplification, in which the constraints on the output with respect to the input are fewer, and thus is more effective than existing methods in terms of compression ratio.
  \item Besides the normal representation where trajectories are sequences of points, LiMITS allows a compact representation leading to an even higher compression ratio.
  \item LiMITS has a linear time complexity and hence is efficient for arbitrarily long trajectories. 
  \item We conducted extensive experiments on real-world trajectory datasets generated by humans and animals in 2-space and 3-space. The results show that the compression ratio of our approach is higher than that of existing algorithms.
\end{enumerate}

The rest of the paper is organized as follows. Section~\ref{sec:pre} introduces preliminaries and formalizes the problem of trajectory simplification. Section~\ref{approach2} proposes our approach based on multidimensional interpolation. Section~\ref{experiment} reports our experiments with real data to demonstrate the performance of the algorithms. Section~\ref{related} reviews related work and compares existing techniques with our approach. Section~\ref{conclusion} summarizes the paper and discusses directions for future research.

\section{Preliminaries}
\label{sec:pre}
\begin{figure}
\centering
\captionsetup{font=bf}
\begin{minipage}[t]{0.23\textwidth}
\includegraphics[trim=100 40 110 70,clip,width=\textwidth]{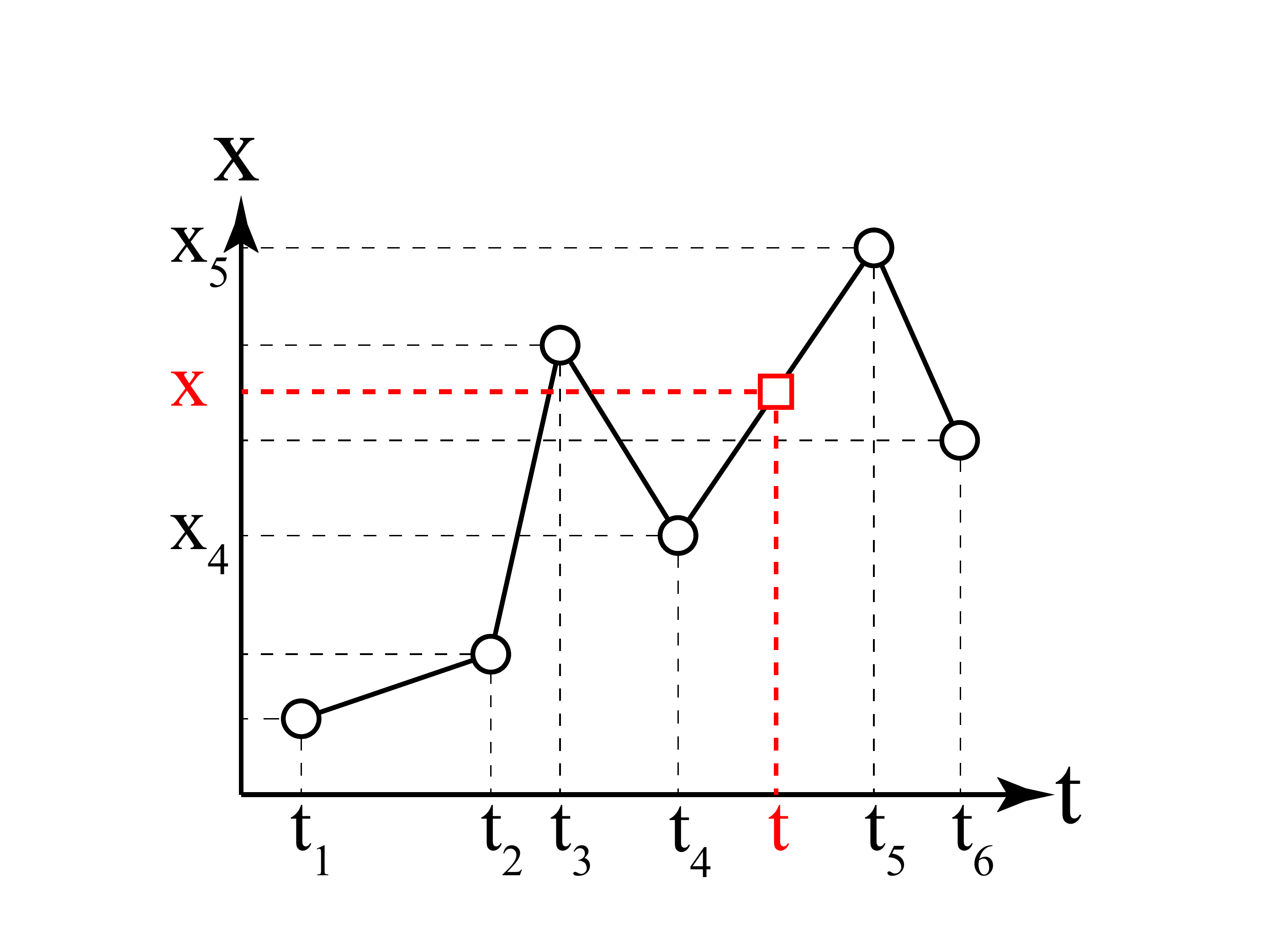}
\caption{Interpolation}
\label{fig:li}
\end{minipage}
\begin{minipage}[t]{0.23\textwidth}
\includegraphics[trim=100 40 110 70,clip,width=\textwidth]{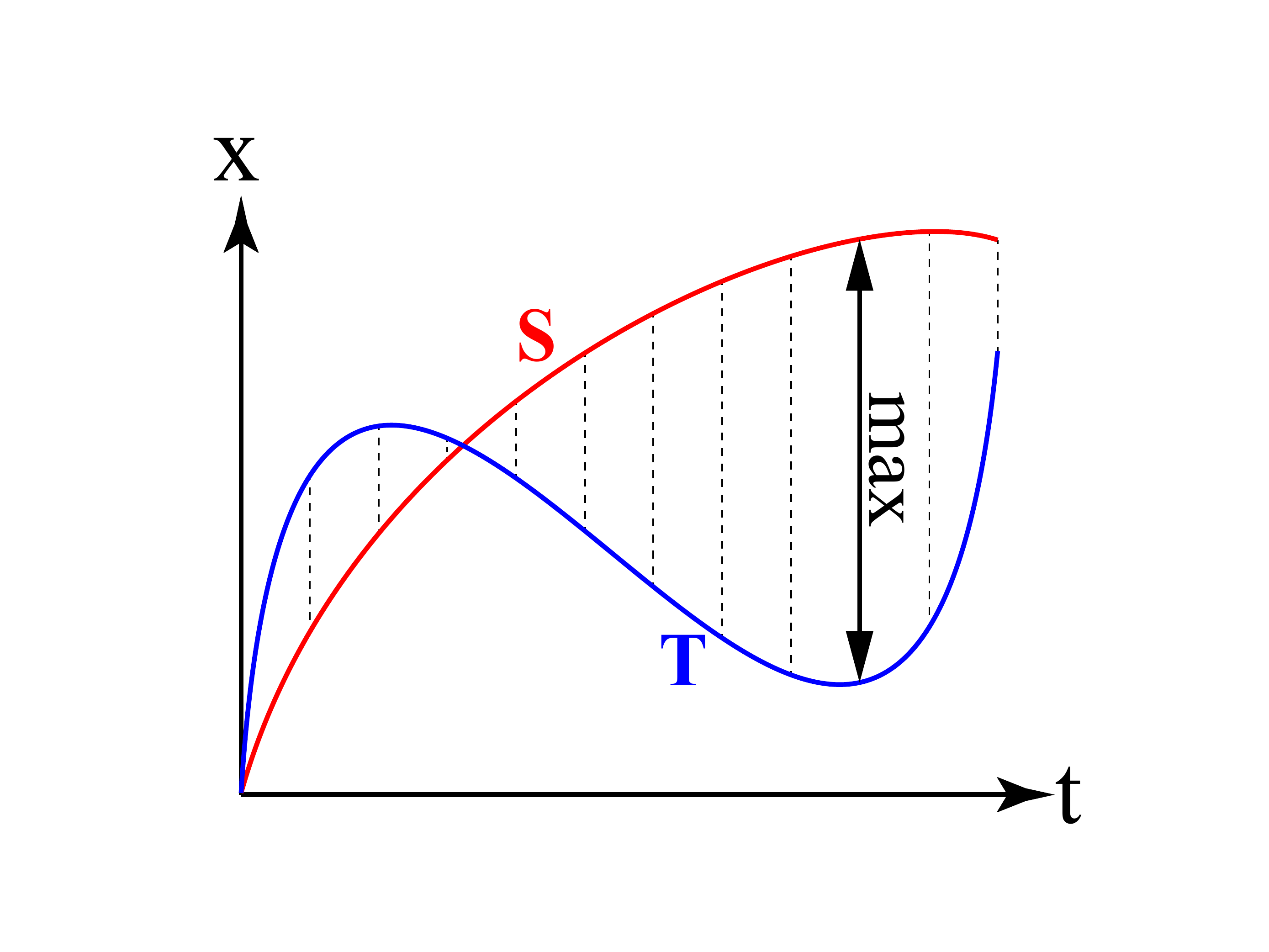}
\caption{Sync-distance}
\label{fig:sd}
\end{minipage}
\end{figure}

In this section, we present the formal definition of the problem. We first start with some preliminaries. According to the definition, a trajectory is the path of a moving object in space as a function of time. The path contains an infinite number of locations, but only a few of the locations are sampled in practice. Hence trajectory data is usually represented by a series of sample points: 
\begin{equation*}
    \mathbf{T} = \langle \mathbf{p}_1,t_1\rangle, \langle\mathbf{p}_2,t_2\rangle, \cdots, \langle\mathbf{p}_n,t_n\rangle,
\end{equation*}
where $\mathbf{T}$ is the trajectory and $n$ is the length. Moreover, the time stamps should increase strictly, i.e., $t_i < t_{i+1}$. This representation does not show the information of the locations between two consecutive time stamps since they are not sampled. Nonetheless, we are able to recover them through linear interpolation and the interpolation tends to be accurate when the sampling rate increases. A point of the trajectory in $m$-space is given by $m$ coordinate values: 
\begin{equation*}
    \mathbf{T}(t) = \langle T_1(t), T_2(t),\cdots, T_{m}(t)\rangle.
\end{equation*}
The $k$-th coordinate value of the interpolation point at any time $t$ is calculated as 
 \begin{equation*}
    T_k(t) =T_k(t_i)+ \frac{T_k(t_{i+1}) - T_k(t_i)}{t_{i+1} - t_i}\cdot (t - t_i),
\end{equation*}
where $1 \leq k \leq m$, $1 \leq i < n$ and $t_i \leq t \leq t_{i+1}$ (see Figure~\ref{fig:li}, where $i=4$).
This can be done in all dimensions and the interpolation point in $m$-space is 
\begin{equation}
\mathbf{T}(t) =\mathbf{T}(t_i)+ \frac{\mathbf{T}(t_{i+1}) - \mathbf{T}(t_{i})}{t_{i+1} - t_i}\cdot(t - t_i),
\end{equation}
where $\mathbf{T}(t_{i}) = \mathbf{p}_{i}$ and $\mathbf{T}(t_{i+1})=\mathbf{p}_{i+1}$ are sample points. For example, a location in 2-space can be represented by $(x_i,y_i)$, so a trajectory will be in the form of $\langle x_1,y_1,t_1\rangle, \langle x_2,y_2,t_2\rangle, \cdots, \langle x_n,y_n,t_n\rangle$. Furthermore, the location $\mathbf{T}(t) =\langle x(t), y(t)\rangle$ at an arbitrary time stamp $t$ is given by 
\begin{align*}
    x(t) = x_i + \frac{x_{i+1} - x_i}{t_{i+1} - t_i}\cdot(t - t_i),~
    y(t) = y_i + \frac{y_{i+1} - y_i}{t_{i+1} - t_i}\cdot(t - t_i),
\end{align*}
where $t_i \leq t \leq t_{i+1}$ and $1 \leq i < n$.
In a geometric view, a trajectory in $m$-space after interpolation is equivalent to an $m+1$-dimensional polygonal curve, because the interpolation forms straight line segments between sample points.

After interpolation, we can define distance at any $t$ in $[t_1,t_n]$ instead of merely time stamps of sample points. This is necessary for weak simplification, since it could involve new points at any $t$. In order to bound the errors, the synchronized distance is used:
\begin{equation}
        d(\mathbf{S}, \mathbf{T}) = \max_{t\in[t_1,t_n]}\|\mathbf{S}(t) - \mathbf{T}(t)\|.
        \label{sd2}
\end{equation}
In other words, the distance between two trajectories $\mathbf{S}$ and $\mathbf{T}$ is determined by the maximum distance between points corresponding to same time (see Figure~\ref{fig:sd}). Given an error tolerance $\epsilon$, the distance between any two points at any time will be no greater than $\epsilon$ if $d(\mathbf{S}, \mathbf{T}) \leq \epsilon$.

\begin{figure}
\centering
\subfigure[$L_p$ in 2-space]{
\includegraphics[width=0.3\linewidth]{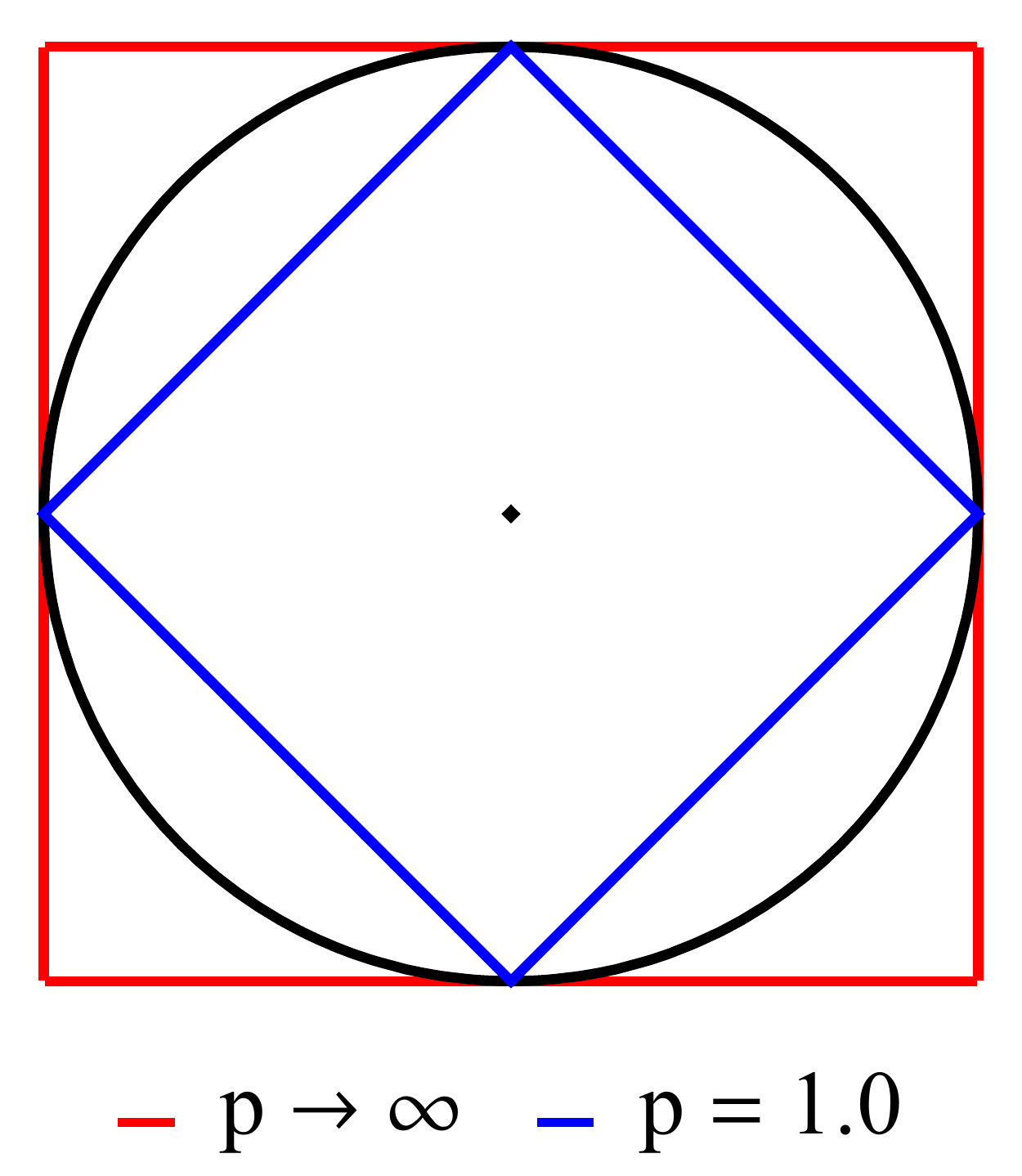}}
\subfigure[$L_1$ in 3-space]{
\includegraphics[width=0.3\linewidth]{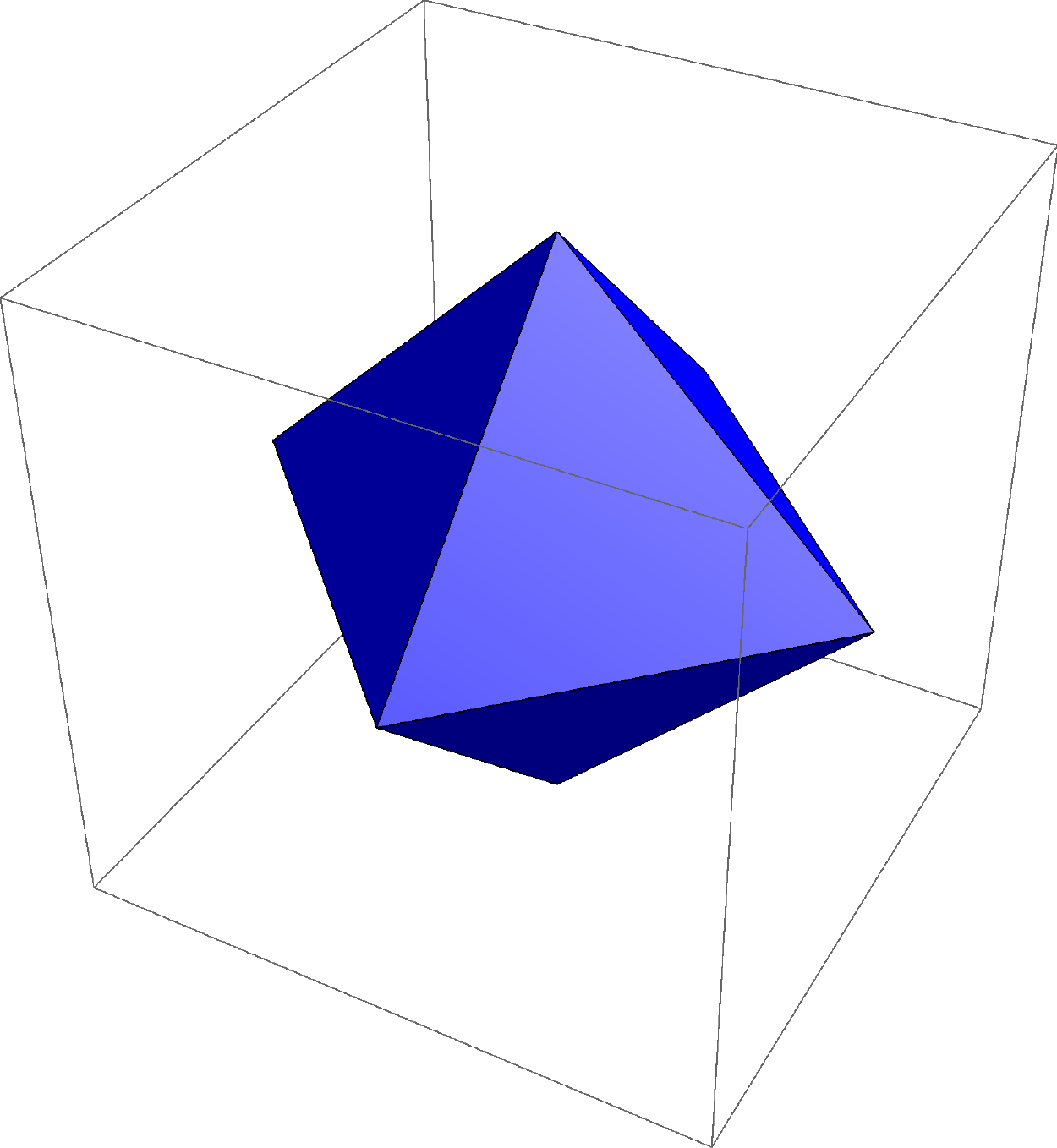}
\label{fig:metrics:l1}}
\subfigure[$L_\infty$ in 3-space]{
\includegraphics[width=0.3\linewidth]{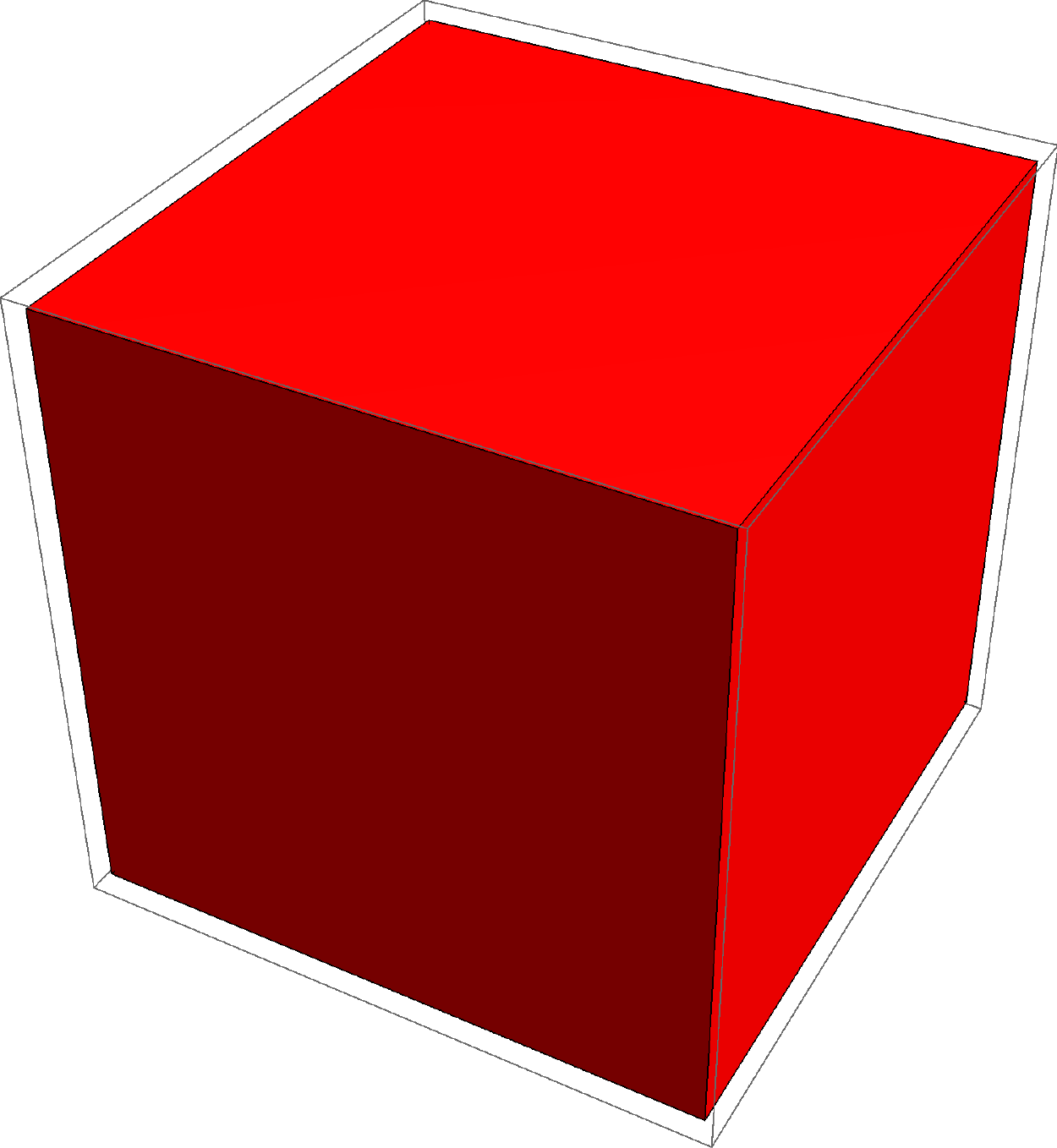}
\label{fig:metrics:loo}}
\captionsetup{font=bf}
\caption{Distance metrics}
\label{fig:metrics}
\end{figure}

All distance metrics can be used in Equation~\ref{sd2}, though existing work on trajectory simplification is limited to the $L_2$ distance. Here we consider all metrics. The synchronized distance with respect to any $L_p$ is 
\begin{align}
    d(\mathbf{S}, \mathbf{T})&=\max_{t\in[t_1,t_n]}\|\mathbf{S}(t) - \mathbf{T}(t)\|_p \nonumber\\
    &=\max_{t\in[t_1,t_n]}\left(\sum^{m}_{i=1}{|S_i(t) - T_i(t)|^p}\right)^\frac{1}{p}.
\end{align}
For any two points in space, the larger the value of $p$, the smaller the corresponding distance. Since all distances over time are smaller than the tolerance $\epsilon$, the larger values of $p$ lead to looser bounds. As a result, the synchronized distance using $L_\infty$ admits more simplification points, as shown in Figure~\ref{fig:metrics}. Furthermore, $L_\infty$ is a good approximation of other metrics, because $\lVert \mathbf{x} \rVert_\infty\leq\lVert \mathbf{x} \rVert_p \leq c \cdot\lVert \mathbf{x} \rVert_\infty$,
where $c = \sqrt[\uproot{3}p]{m}$ is a constant\footnote{\url{https://math.stackexchange.com/questions/218046}}. For example, the synchronized $L_2$ distance to the original trajectory will be no greater than $\sqrt{2}\cdot\epsilon$ if the synchronized $L_\infty$ distance is bounded by $\epsilon$ in $2$-space, and $\sqrt{3}\cdot\epsilon$ in $3$-space. Another interesting property is that all metrics are equivalent in 1-space. Furthermore, $L_1$ is equivalent to $L_\infty$ in 2-space after a  linear transformation, so simplification under $L_\infty$ is applicable to $L_1$ in 2-space. Note this does not hold true in higher dimensions, e.g., the unit sphere of $L_1$, an octahedron (Figure~\ref{fig:metrics:l1}), is dual but not equivalent to that of $L_\infty$, a cube (Figure~\ref{fig:metrics:loo}), in 3-space.

Finally, we present the formal definition of the problem. Trajectory simplification is to reduce the number of points of the trajectory, after which the distance between input and output is bounded, so that data is compressed while the quality is still guaranteed.
\newtheorem*{problem}{Problem Definition}
\begin{problem}
 Given an error bound $\epsilon$ and an input trajectory in $m$-space $\mathbf{T}$, we seek a trajectory $\mathbf{S}$ which consists of fewer points, and that the error between $\mathbf{S}$ and $\mathbf{T}$ is bounded by $\epsilon$, i.e., $d(\mathbf{S}, \mathbf{T}) = \max_{t} \|\mathbf{S}(t) - \mathbf{T}(t)\|_\infty \leq \epsilon$.
\end{problem}
The performance of simplification algorithms is evaluated by the compression ratio---the ratio of the input size to the output size. Under the normal representation, the size of the data depends on the length of the trajectory, so the compression ratio is $|\mathbf{T}|/|\mathbf{S}|$. Given the same error tolerance and input data, an algorithm with a higher compression ratio is more effective.

\begin{figure*}
    \centering
    \subfigure[X: step 1]{
    \includegraphics[trim=130 0 110 70,clip,width=0.15\textwidth]{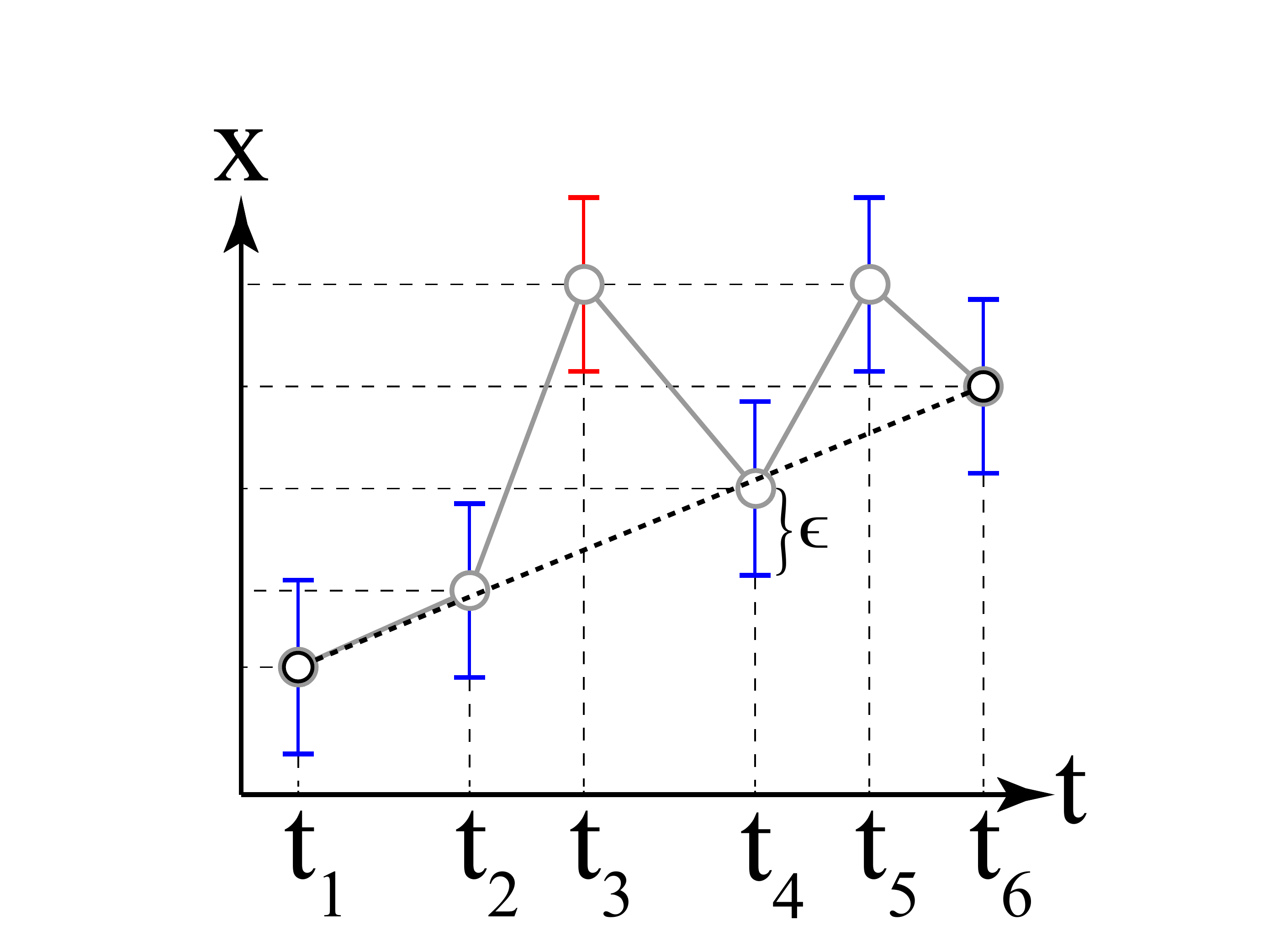}
    \label{fig:rdp1}}
    \subfigure[X: step 2]{
    \includegraphics[trim=130 0 110 70,clip,width=0.15\textwidth]{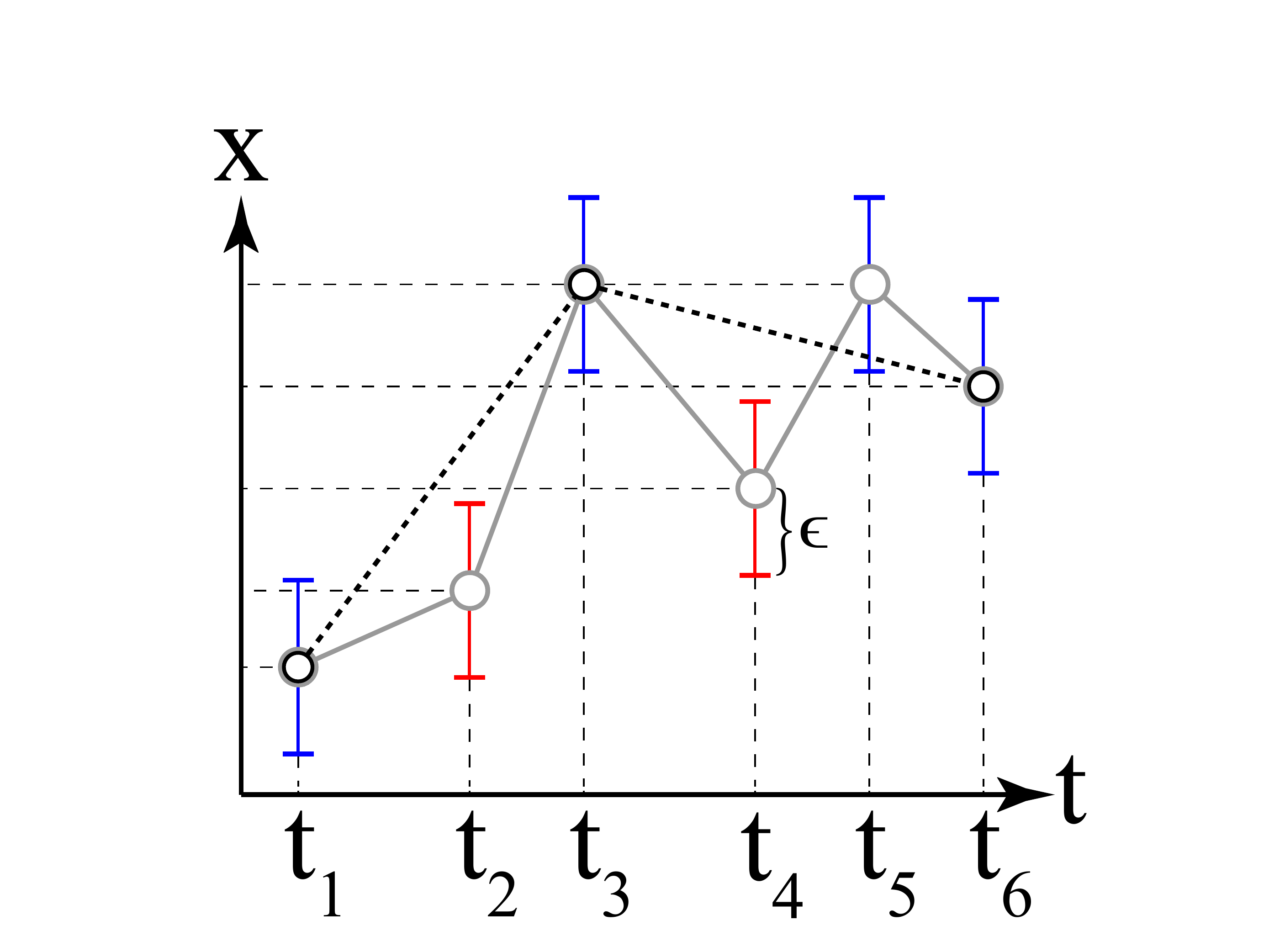}}
    \subfigure[X: step 3]{
    \includegraphics[trim=130 0 110 70,clip,width=0.15\textwidth]{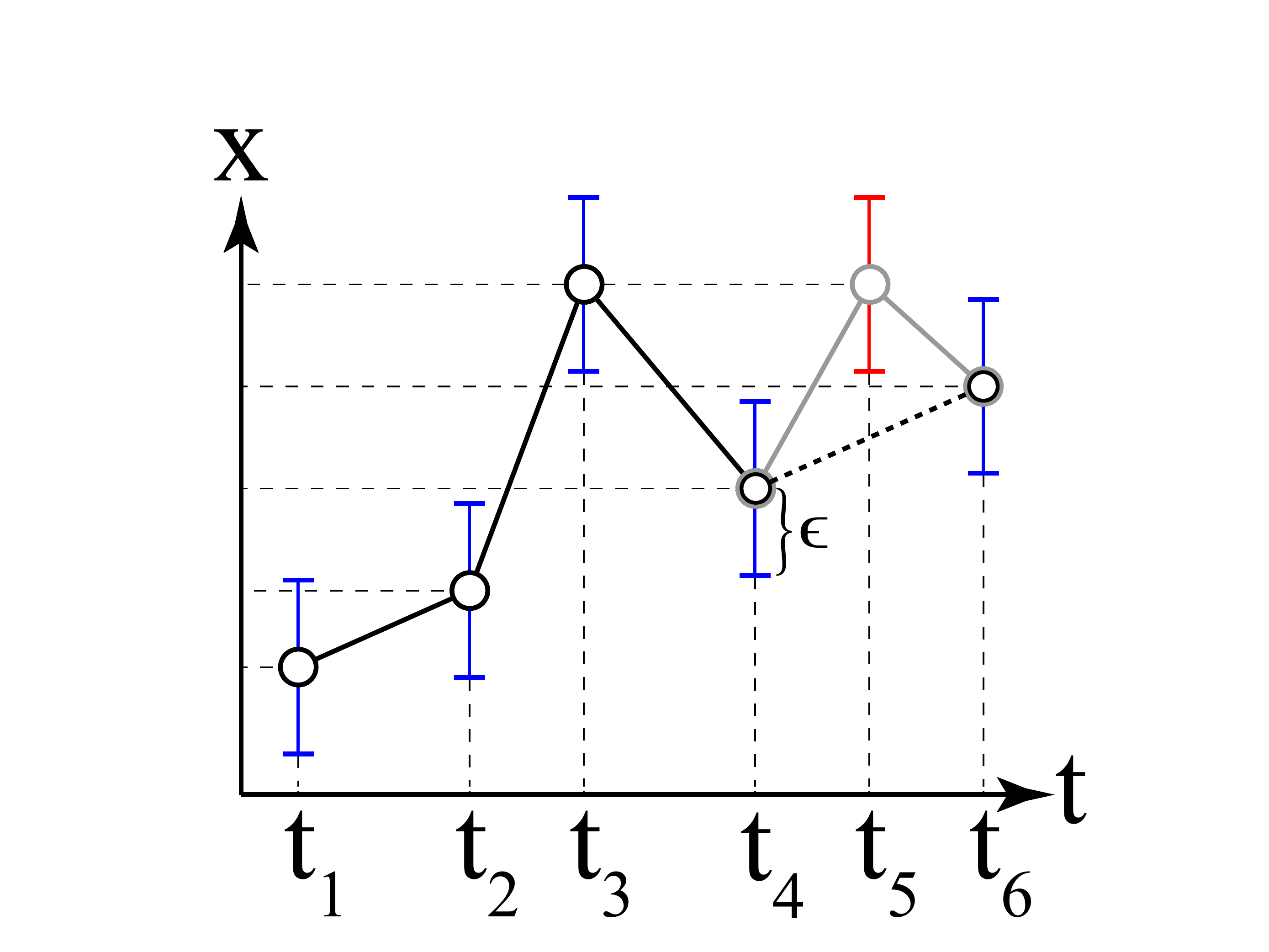}}
    \subfigure[X: step 4]{
    \includegraphics[trim=130 0 110 70,clip,width=0.15\textwidth]{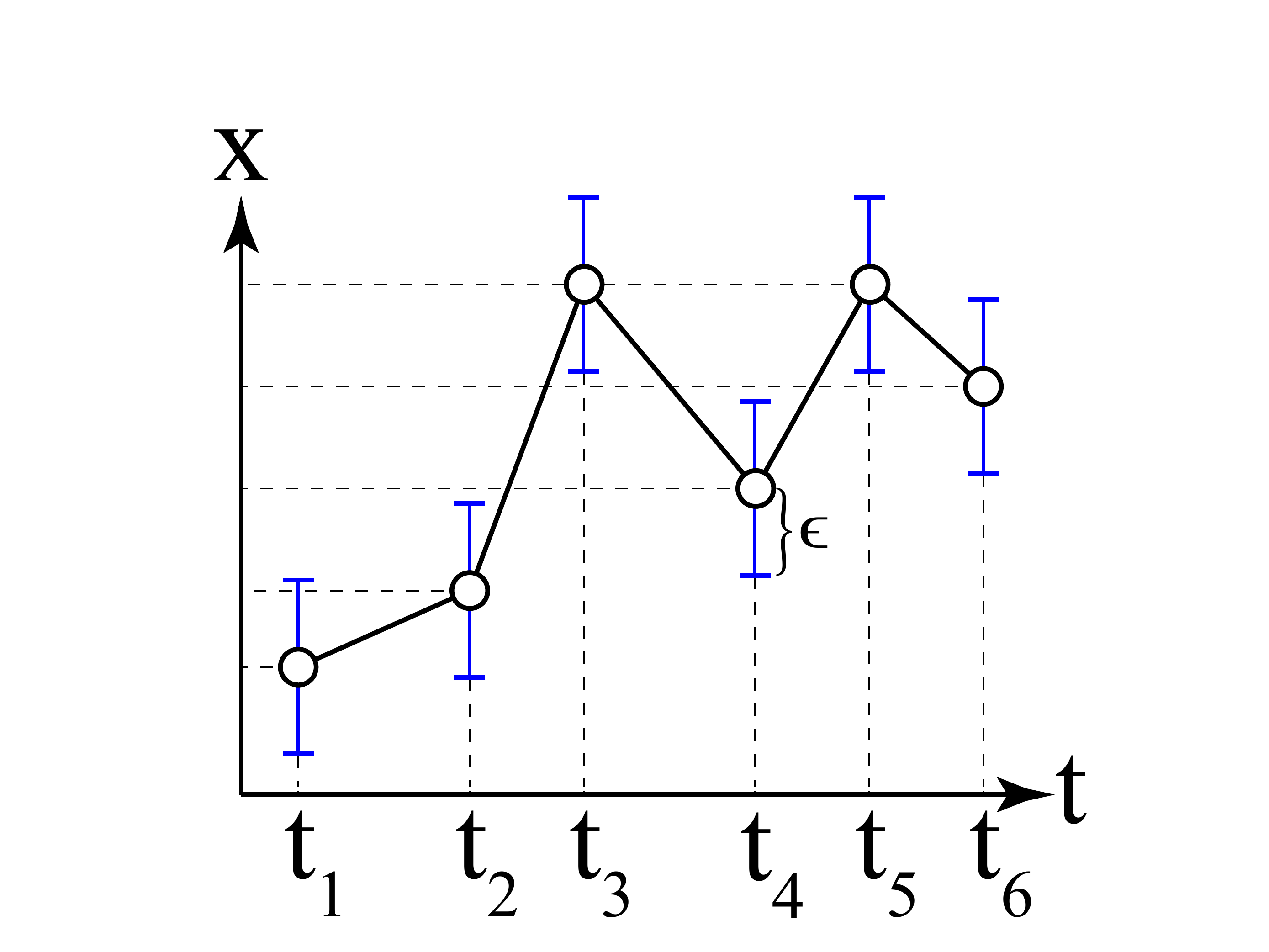}
    \label{fig:rdp4}}
    \subfigure[Y: step 1]{
    \includegraphics[trim=130 0 110 70,clip,width=0.15\textwidth]{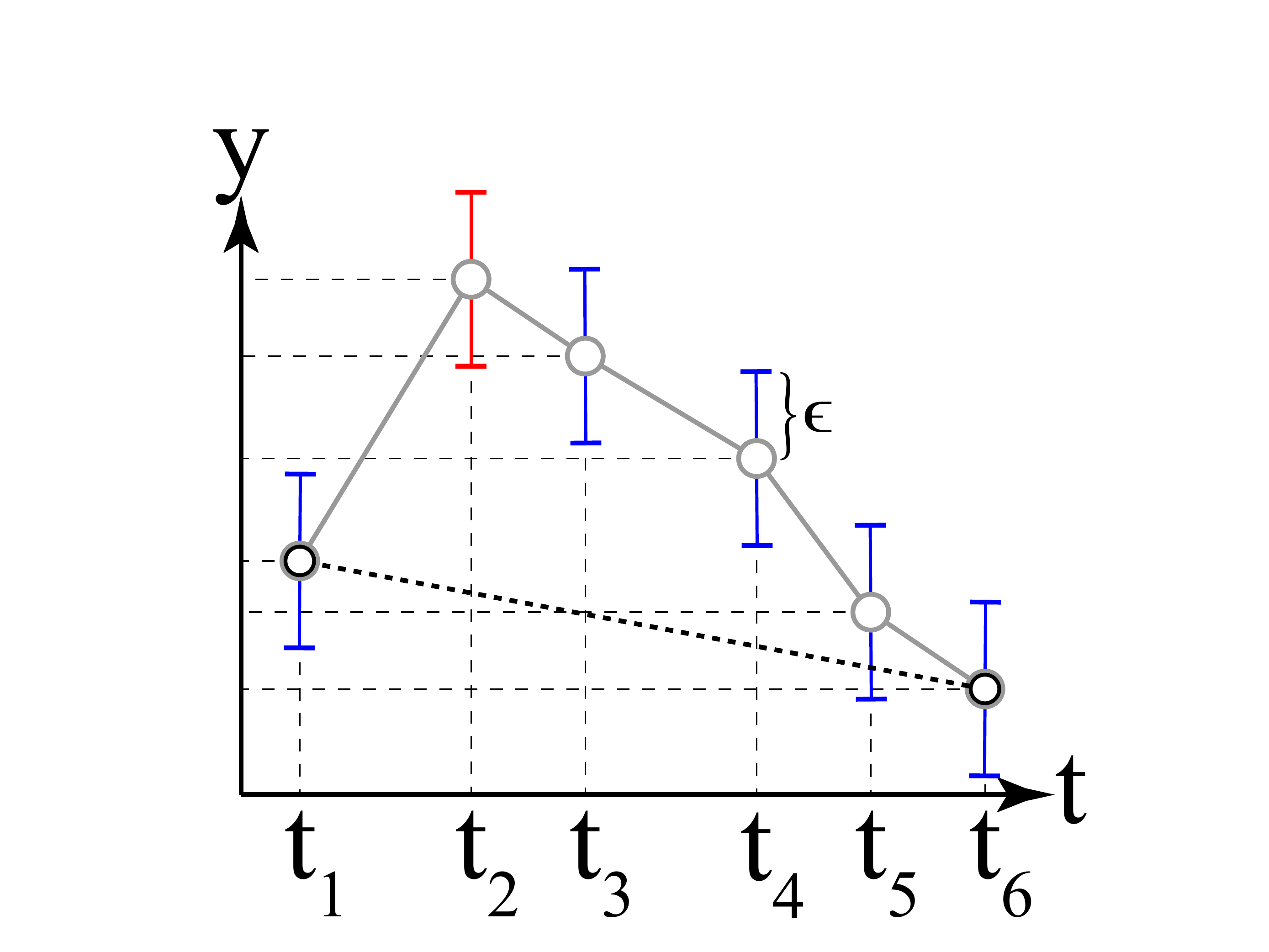}
    \label{fig:rdp5}}
    \subfigure[Y: step 2]{
    \includegraphics[trim=130 0 110 70,clip,width=0.15\textwidth]{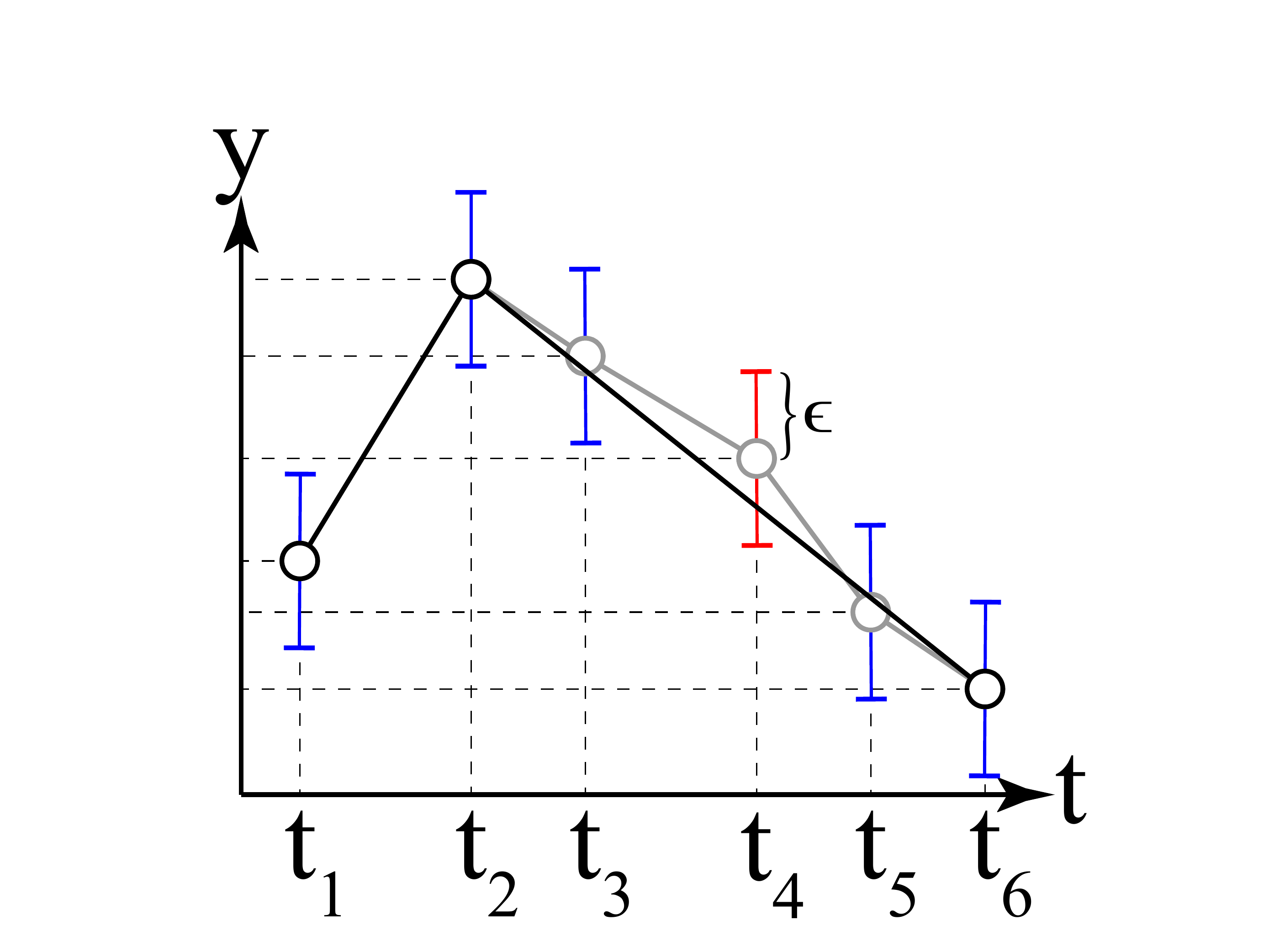}
    \label{fig:rdp6}}
    \captionsetup{font=bf}
    \caption{RDP algorithm}
    \label{fig:rdp}
\end{figure*}

\begin{figure*}
\centering
\subfigure[Input X]{
\includegraphics[trim=130 0 110 70,clip,width=0.15\textwidth]{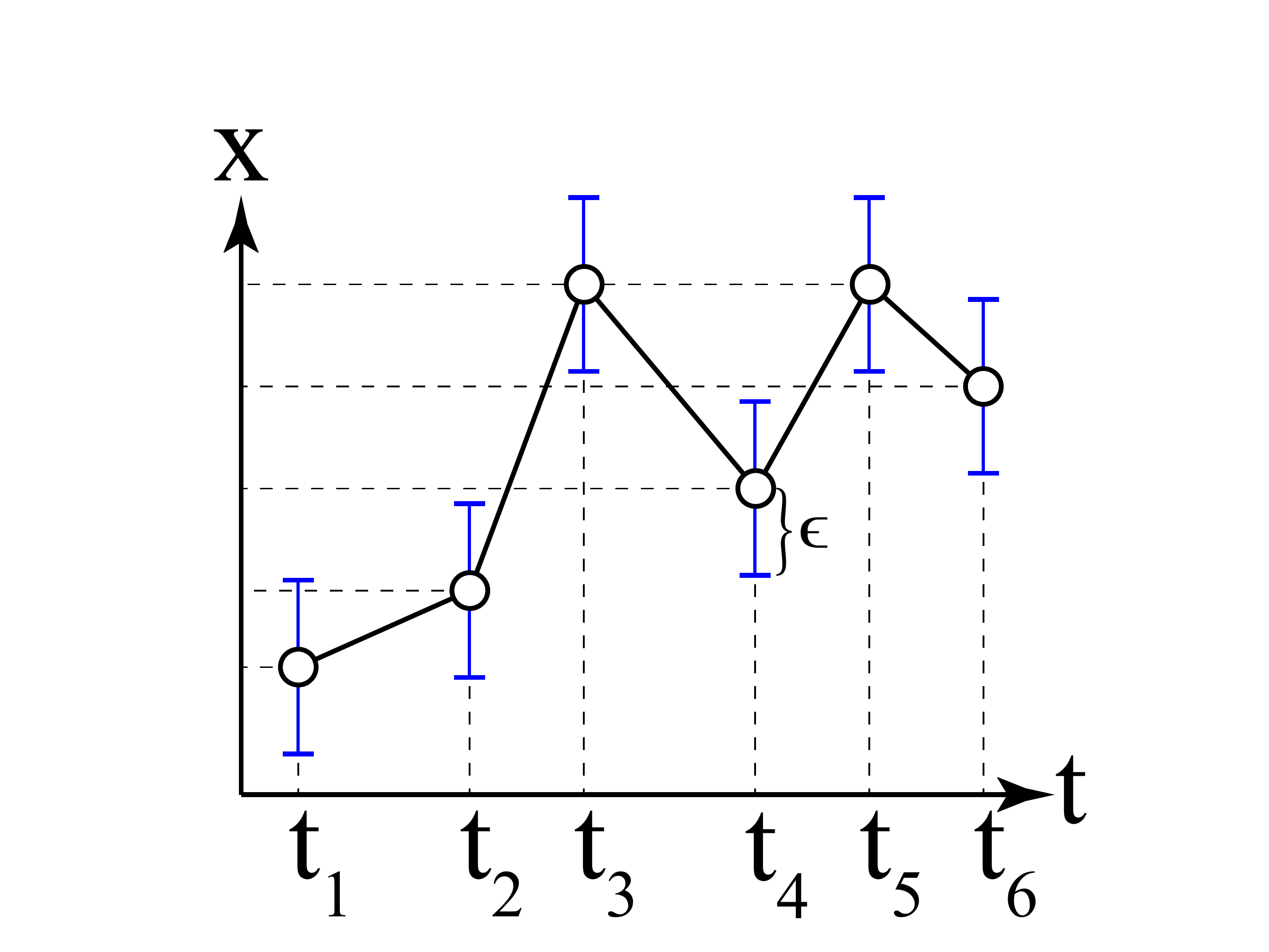}
\label{fig:ss11}}
\subfigure[X: step $t_1$]{
\includegraphics[trim=130 0 110 70,clip,width=0.15\textwidth]{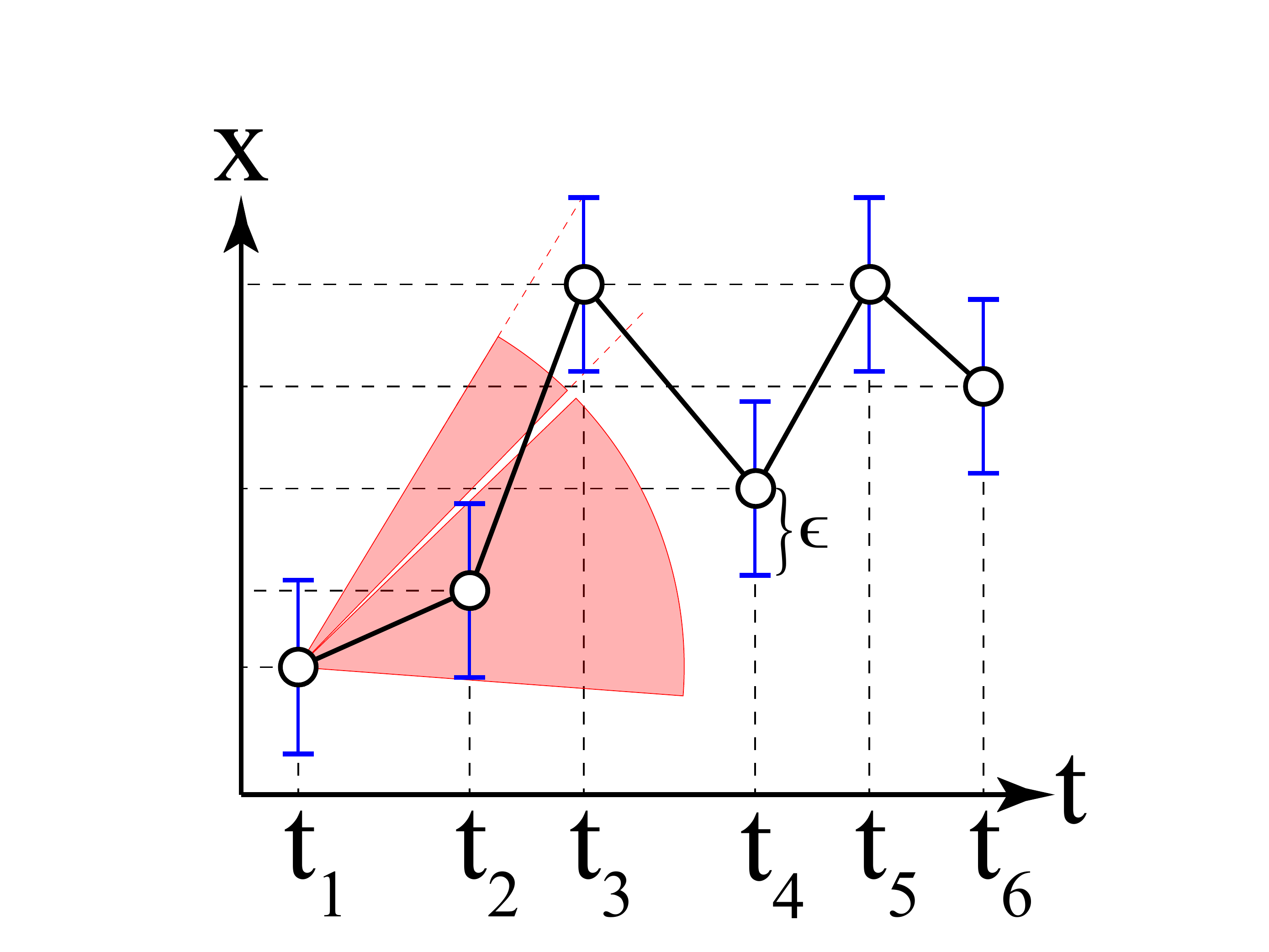}}
\subfigure[X: step $t_2$]{
\includegraphics[trim=130 0 110 70,clip,width=0.15\textwidth]{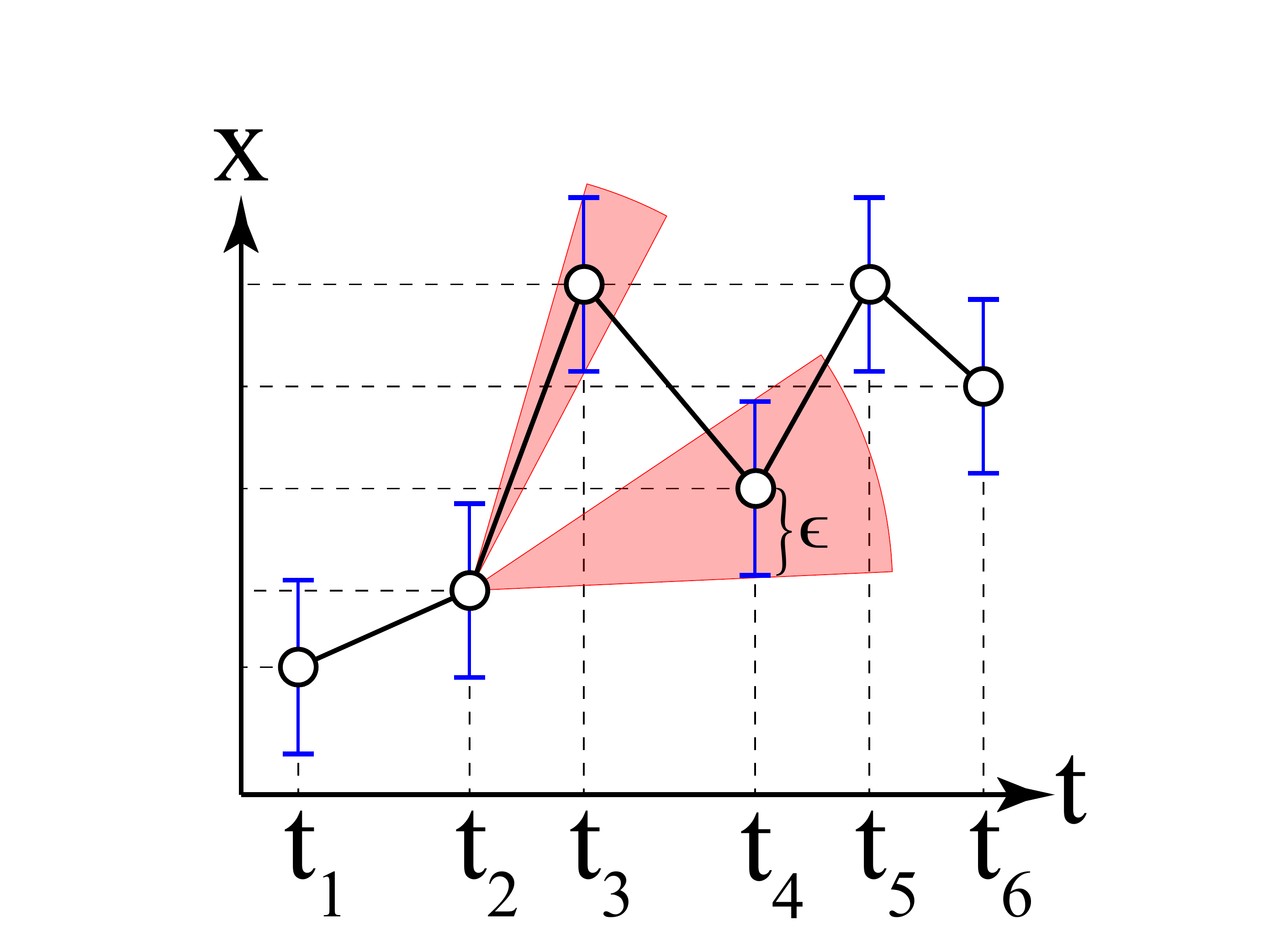}}
\subfigure[X: step $t_3$]{
\includegraphics[trim=130 0 110 70,clip,width=0.15\textwidth]{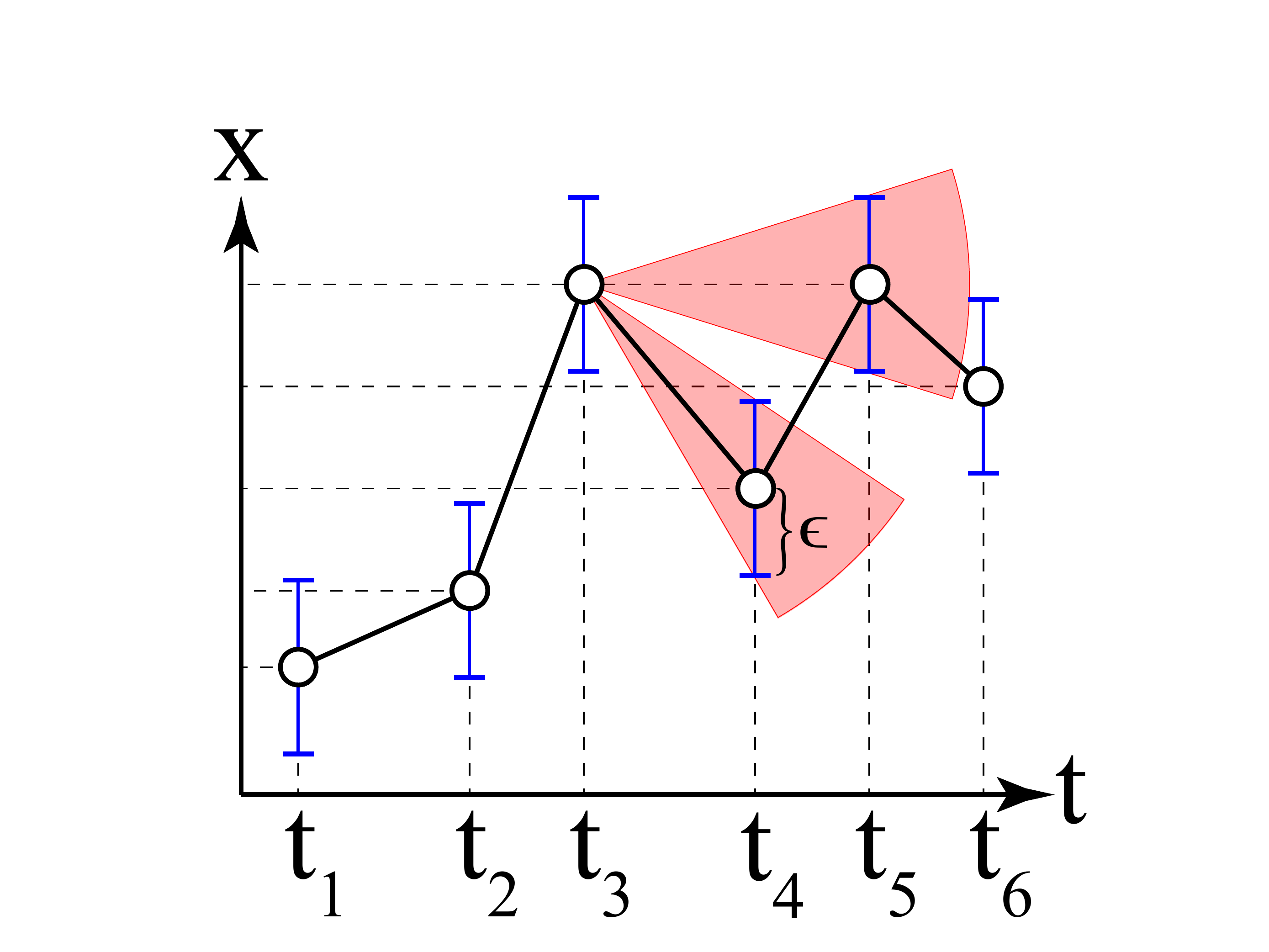}}
\subfigure[X: step $t_4$]{
\includegraphics[trim=130 0 110 70,clip,width=0.15\textwidth]{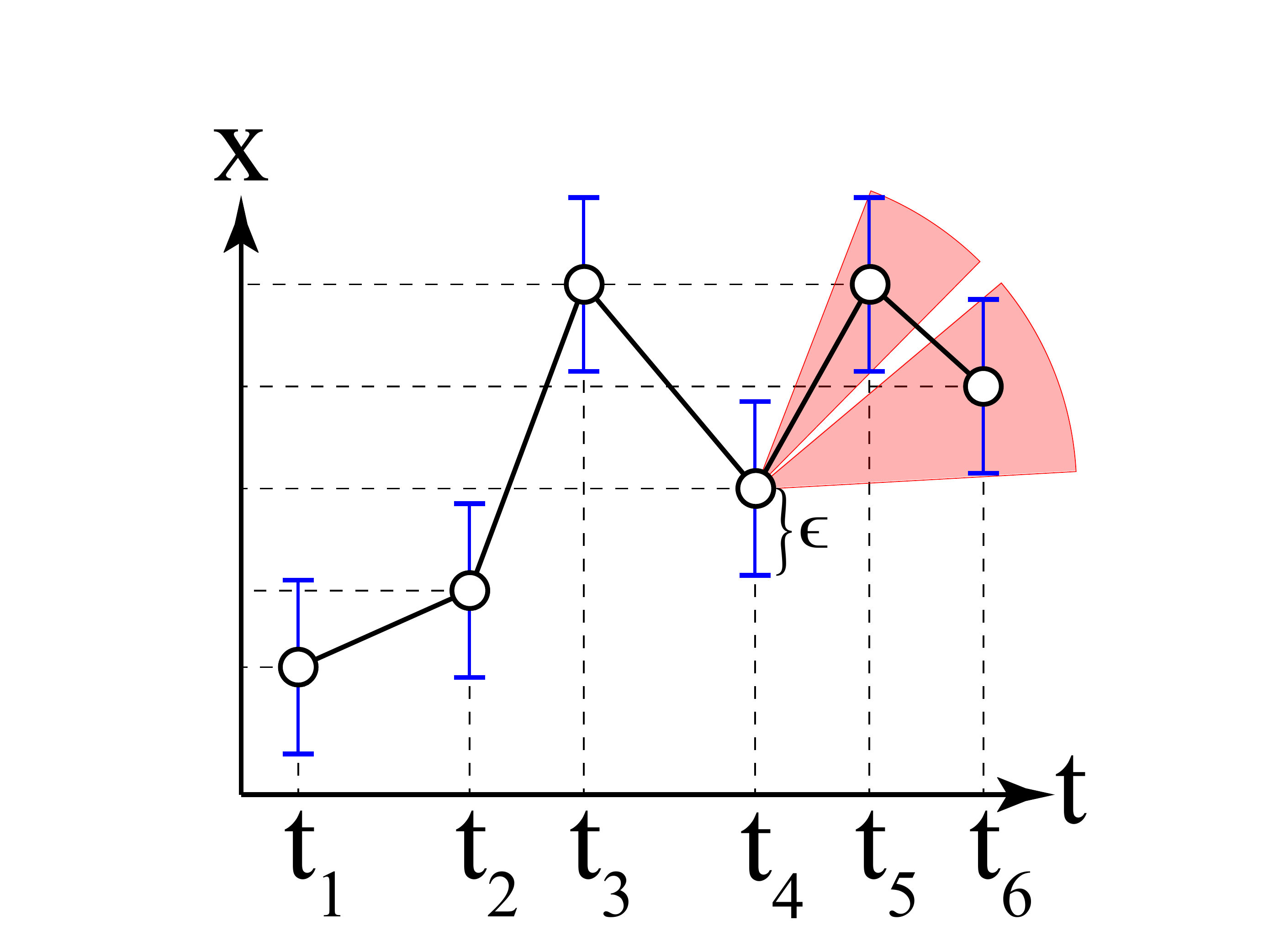}}
\subfigure[Output X]{
\includegraphics[trim=130 0 110 70,clip,width=0.15\textwidth]{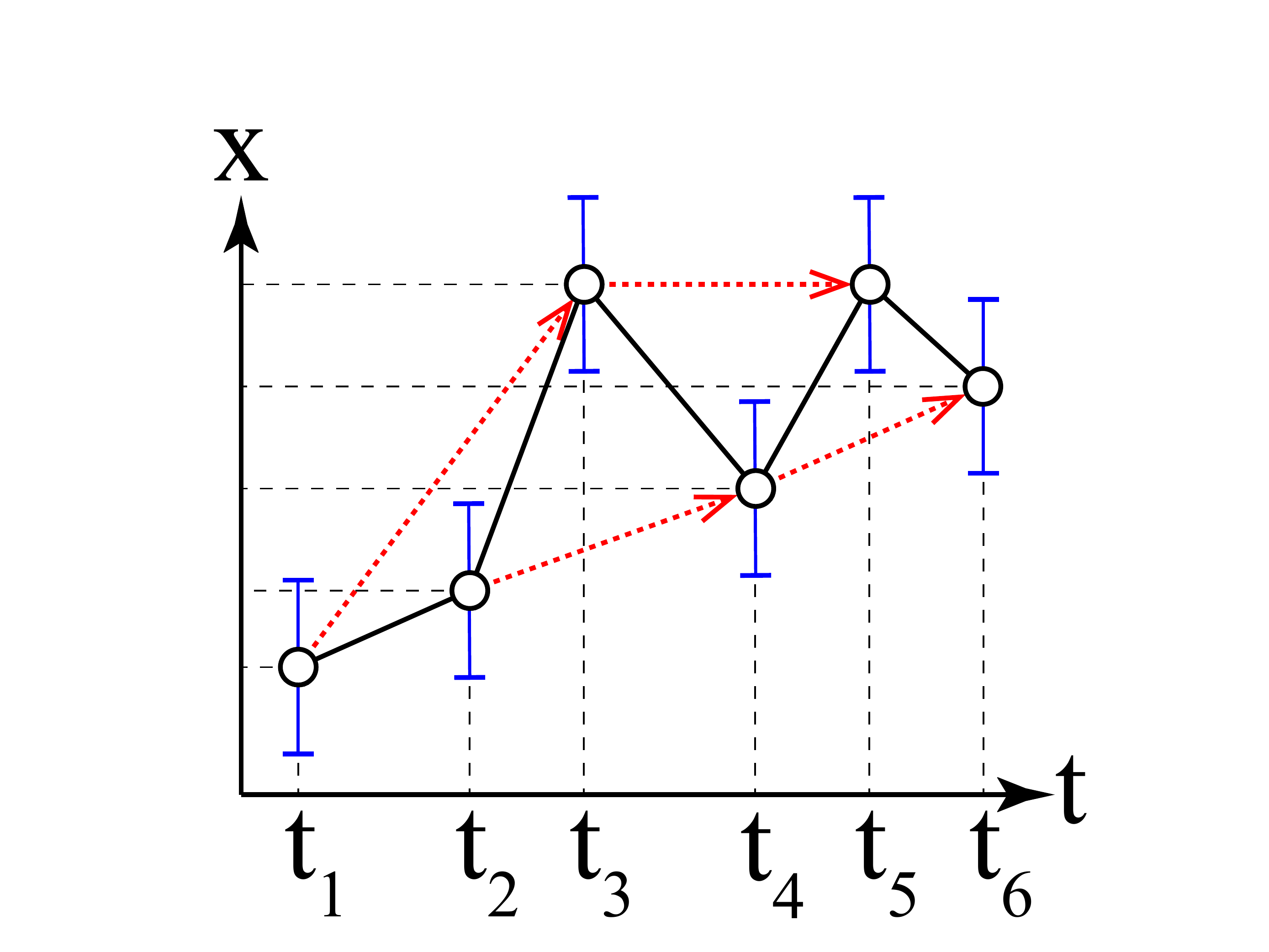}
\label{fig:ss16}}\\
\subfigure[Input Y]{
\includegraphics[trim=130 0 110 70,clip,width=0.15\textwidth]{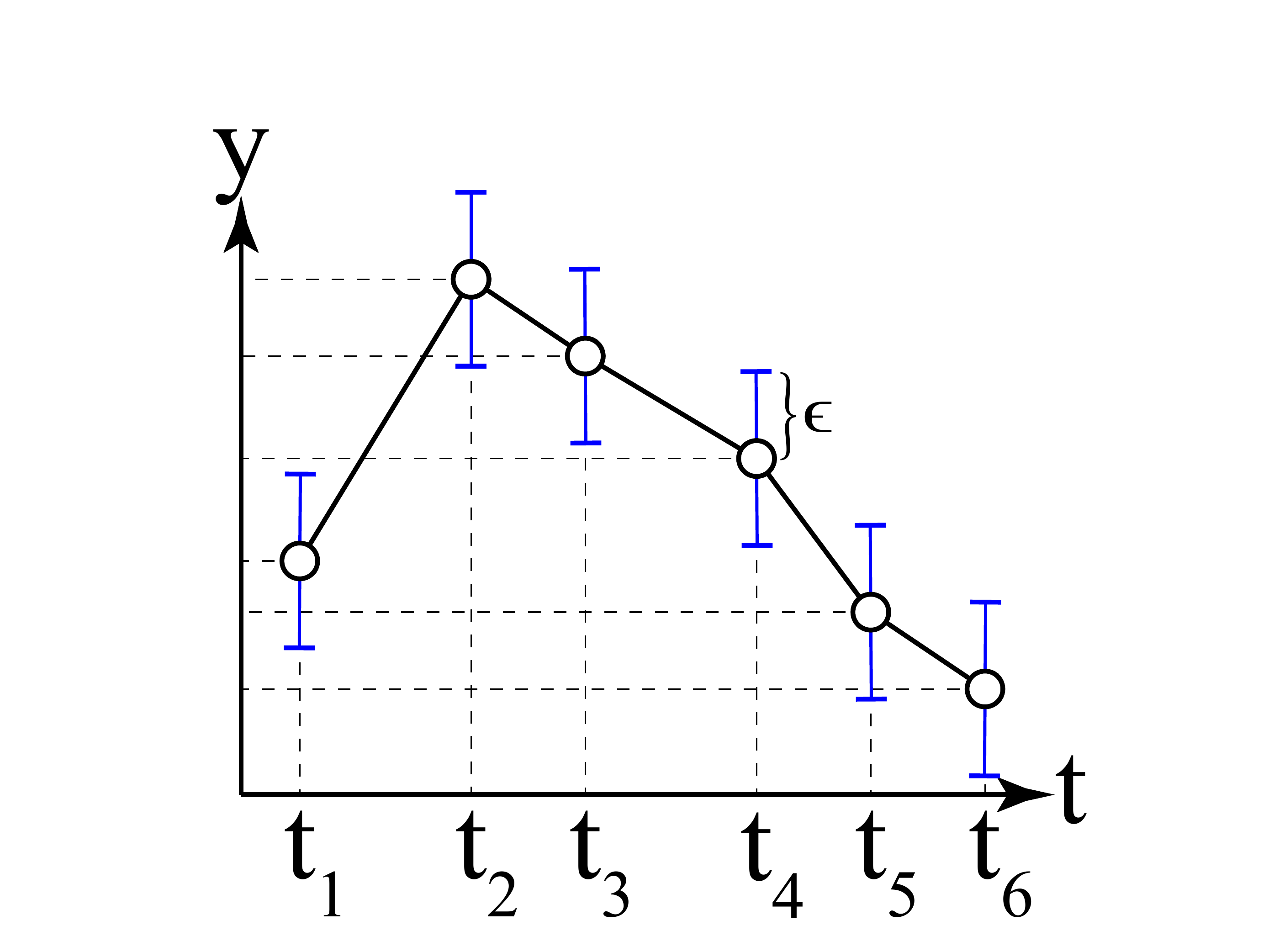}
\label{fig:ss21}}
\subfigure[Y: step $t_1$]{
\includegraphics[trim=130 0 110 70,clip,width=0.15\textwidth]{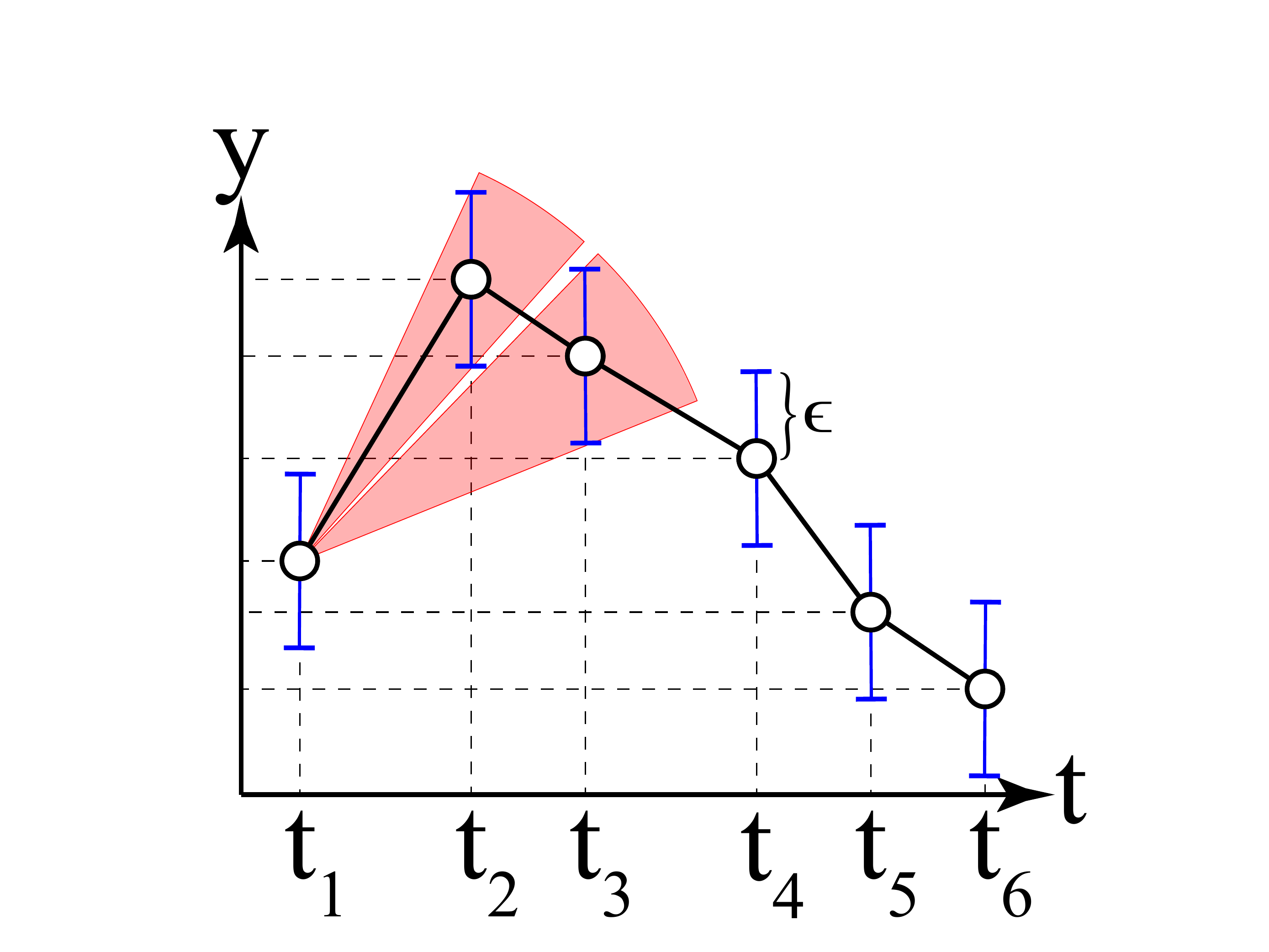}}
\subfigure[Y: step $t_2$\phantom{}$_{[1]}$]{
\includegraphics[trim=130 0 110 70,clip,width=0.15\textwidth]{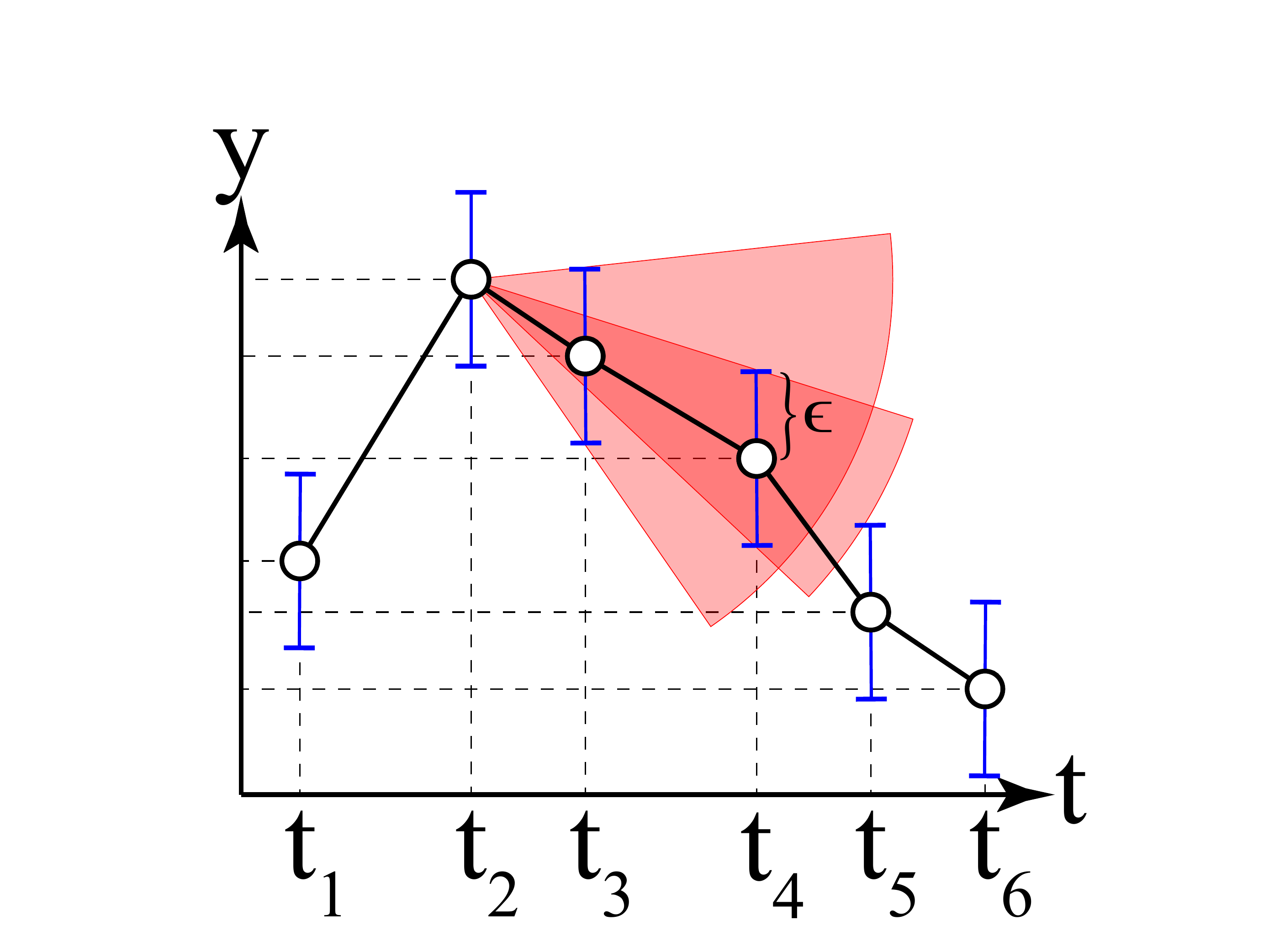}}
\subfigure[Y: step $t_2$\phantom{}$_{[2]}$]{
\includegraphics[trim=130 0 110 70,clip,width=0.15\textwidth]{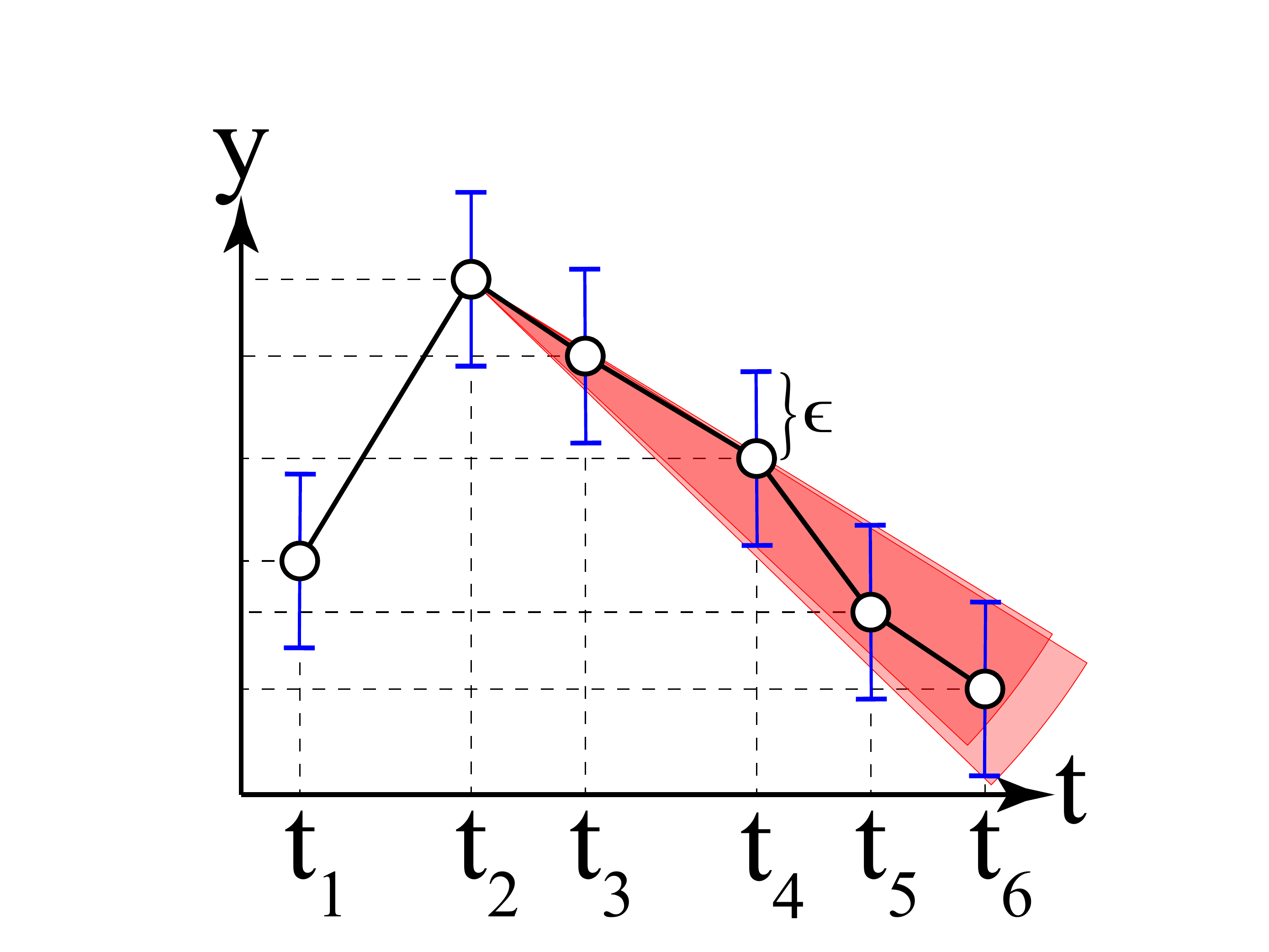}}
\subfigure[Y: step $t_2$\phantom{}$_{[3]}$]{
\includegraphics[trim=130 0 110 70,clip,width=0.15\textwidth]{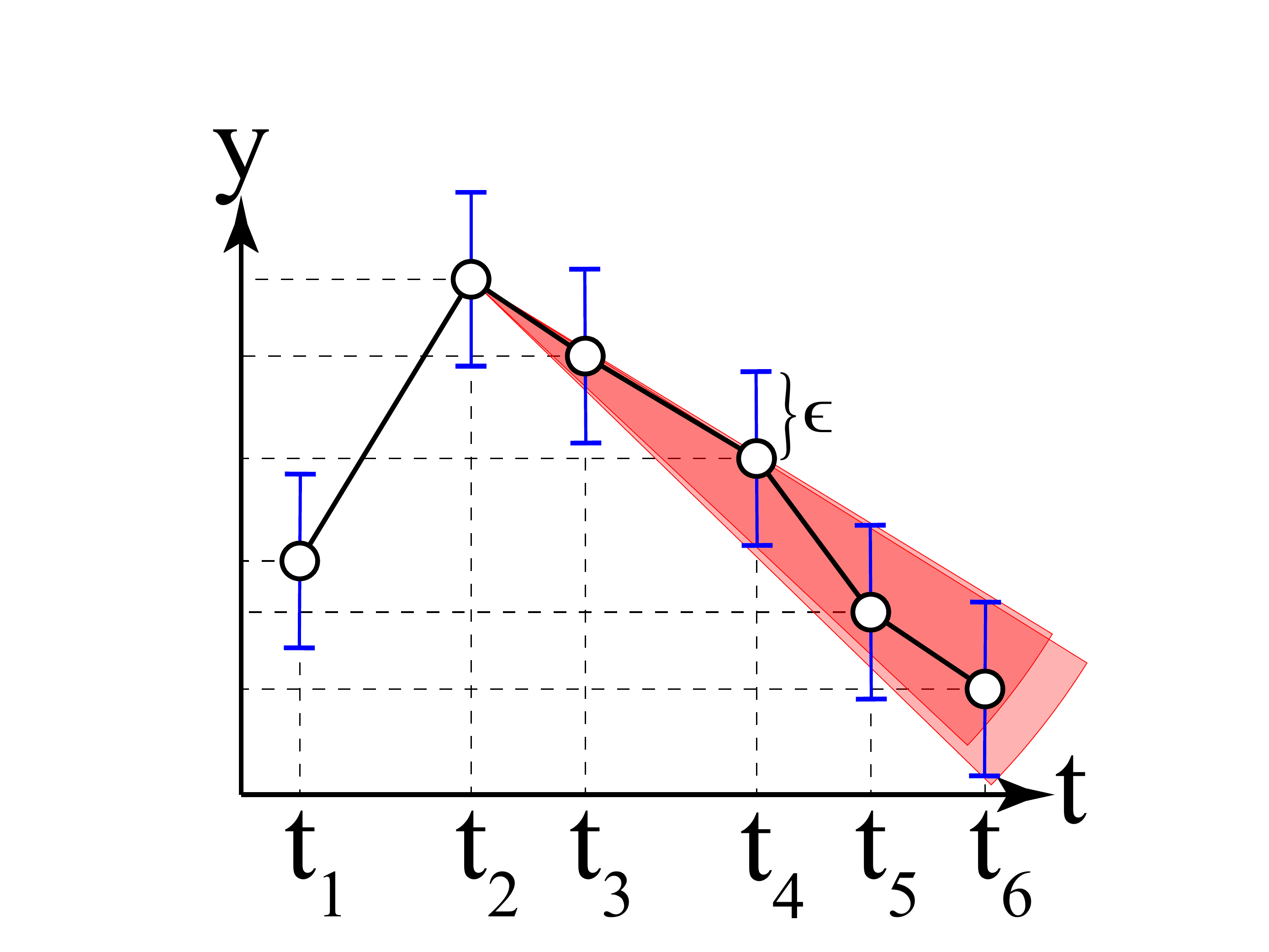}}
\subfigure[Output Y]{
\includegraphics[trim=130 0 110 70,clip,width=0.15\textwidth]{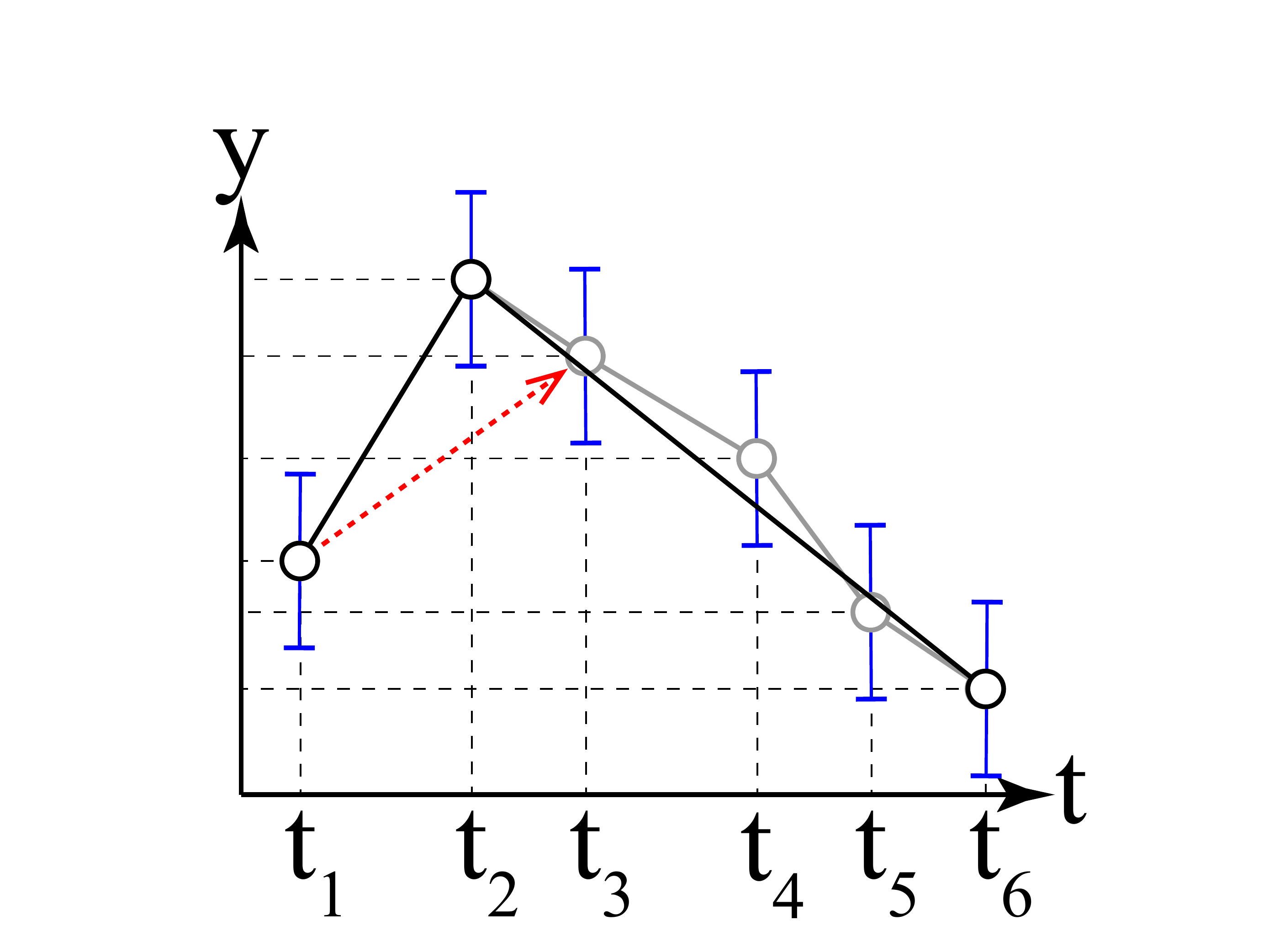}
\label{fig:ss26}}
\captionsetup{font=bf}
\caption{OPW algorithm}
\label{fig:ss}
\end{figure*}

\section{Algorithm Design}
\label{approach2}
In this section, we introduce LiMITS in detail. We first review the problem in 1-space and then discuss the generalization of existing algorithms to $m$-space. Next, we introduce multidimensional interpolation, which integrates information from all dimensions to construct simplified trajectories in $m$-space. Finally, we present a compact representation to further reduce the storage cost.

\subsection{Simplification in One Dimension}
\label{si}
We first review algorithms for trajectories in 1-space. One single number is enough to specify a location in 1-space, so trajectories are real functions of time, as well as polygonal curves in 2-space. For example, consider Figures~\ref{fig:rdp1} and~\ref{fig:rdp5}, where $X$ and $Y$ are two trajectories in 1-space and both consist of six sample points.

\subsubsection{Classic Methods}
\label{sec:ri}
RDP~\cite{douglas1973algorithms} is a well-known algorithm for polygonal curve simplification. It recursively subdivides the curve into pieces until one line segment is sufficient to simplify each piece. For example, consider the trajectory $X$ in Figures~\ref{fig:rdp1}-\ref{fig:rdp4}. Initially, the algorithm keeps the first and last point $p_1$ and $p_6$, and then finds the point that is vertically farthest from the line segment $\overline{p_1p_6}$, which is $p_3$. The vertical distance between $p_3$ and $\overline{p_1p_6}$ is greater than the error tolerance $\epsilon$ so $p_3$ is kept. Next, the algorithm subdivides $X$ into $\overline{p_1p_2p_3}$ and $\overline{p_3p_4p_5p_6}$, and recursively call itself with each piece. Note that the line segment is sufficient to simplify the corresponding piece if the deviation at the farthest point is within $\epsilon$, and hence the recursion terminates in this case. Eventually, all points in $X$ are kept and no simplification takes place here. On the other hand, RDP is able to simplify the sample trajectory $Y$ with three points $\overline{p_1p_2p_6}$ (see Figures~\ref{fig:rdp5} and~\ref{fig:rdp6}).

RDP produces a subset of the input as the result, so it belongs to strong simplification. It is easy to generalize RDP to high dimensions by simply changing the distance function. In Section~\ref{related}, we review other algorithms that can be easily extended to high dimensions. According to the existing experimental studies~\cite{zhang2018trajectory,Lin2019} and results in Section~\ref{experiment}, these algorithms are not as effective as those algorithms based on sector intersection and link distance as described in Section~\ref{sec:si} and~\ref{sec:wi}.

\subsubsection{Sector Intersection Methods}
\label{sec:si}
Effective strong simplification algorithms like OPW~\cite{meratnia2004spatiotemporal}, OPT~\cite{IMAI198631} simplify trajectories in 1-space through sector intersection. The greedy algorithm OPW checks input points one by one to determine if any can be omitted. As shown in Figures~\ref{fig:ss11}-\ref{fig:ss16}, for the trajectory $X$, OPW first keeps the starting point $p_1$, then tries to omit $p_2$. However, the deviation at $t_2$ will exceed $\epsilon$ by replacing $\overline{p_1p_2p_3}$ with $\overline{p_1p_3}$ so $p_2$ can not be omitted. In the next iteration, OPW outputs $p_2$ and checks if $p_3$ can be omitted. Since the difference between $\overline{p_2p_3p_4}$ and $\overline{p_2p_4}$ at $t_3$ is again greater than $\epsilon$, $p_3$ should be kept as well. This process goes on until reaching the last point $p_6$. Finally, the output is the same as the input and no simplification has taken place. On the other hand, OPW is able to simplify $Y$ with three points, as shown in Figure~\ref{fig:ss26}.

The time complexity of OPW is $O(n^2)$ when a linear scan is used to compute the difference between $\overline{p_ip_{i+1}\cdots p_{i+k}}$ and $\overline{p_ip_{i+k}}$. In~\cite{song2014press}, the time complexity of OPW is reduced to $O(n)$ through sector intersection. According to the problem definition, any output line $\overline{p_ip_{i+k}}$ must fall between $\overline{p_il_j}$ and $\overline{p_ih_j}$, i.e., in the angle $\angle h_jp_il_j$, where $i<j<i+k$, $l_j=(x_j-\epsilon,t_j)$ and $h_j=(x_j+\epsilon,t_j)$. Thus, $\overline{p_ip_{i+k}}$ must fall in the intersection of $\angle h_{i+1}p_il_{i+1},\cdots, \angle h_{i+k}p_il_{i+k}$. The intersections can be updated on the fly, and the current point can be omitted if and only if the intersection is not empty. For example, the intersections are all empty for $X$ while only one is empty for $Y$, as shown in Figure~\ref{fig:ss}, so $X$ remains the same while $Y$ only has three points after simplification. The overall time complexity becomes $O(n)$ as linear scans are no longer needed to determine if a point is to be omitted. OPW has optimal time complexity of $O(n)$. However, being a greedy algorithm, it does not eliminate many points.

OPT~\cite{IMAI198631} is an optimal algorithm for strong simplification, which minimizes the size of the output. It first computes the visibility between any point $p_i$ and $p_j$, i.e., whether the difference between $\overline{p_ip_{i+1}\cdots p_j}$ and $\overline{p_ip_j}$ is less than $\epsilon$. Next, it applies dynamic programming on the visibility graph to construct an optimal path in terms of the number of points. 
Similar to OPW, the time complexity of OPT can also be reduced via sector intersection (from the original complexity of $O(n^3)$ to $O(n^2)$).

The algorithms mentioned above are strong simplifications and thus their performance is limited. Although CI~\cite{Lin2019} introduces a weak simplification algorithm based on sector intersection, it has the constraint that the time stamps of output trajectories should be from 
the input. Moreover, CI cannot omit points when intersections are empty, and thus cannot simplify trajectories like $X$ as well. Sector intersection becomes complicated in higher dimensional spaces, for which we will introduce the generalization in Section~\ref{sec:g}. Compared with sector intersection, algorithms based on link distance discussed in the next subsection are more effective.

\begin{figure}
\centering
\subfigure[X boundary]{
\includegraphics[trim=130 0 110 70,clip,width=0.3\linewidth]{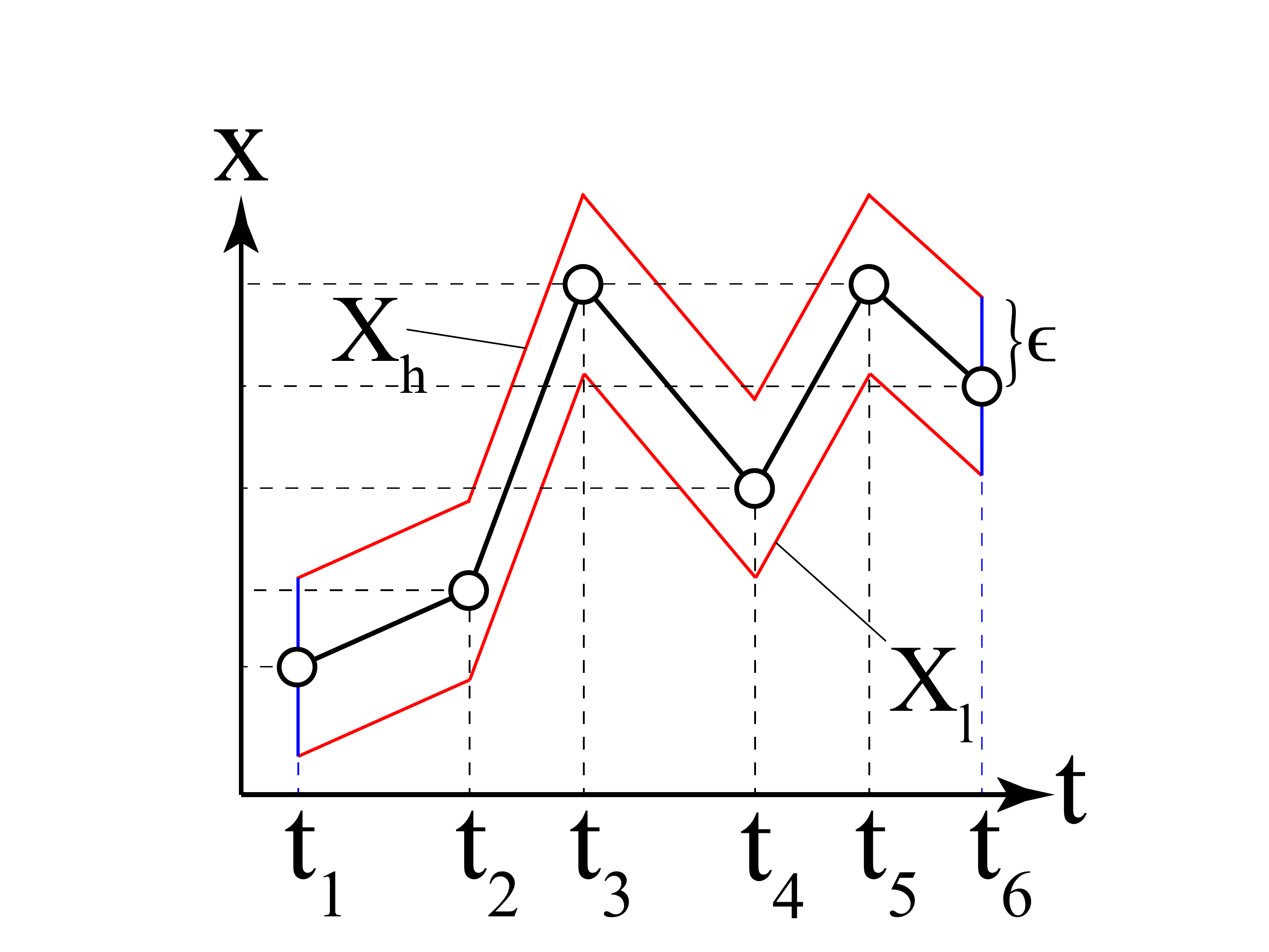}
\label{fig:w1}}
\subfigure[X windows]{
\includegraphics[trim=130 0 110 70,clip,width=0.3\linewidth]{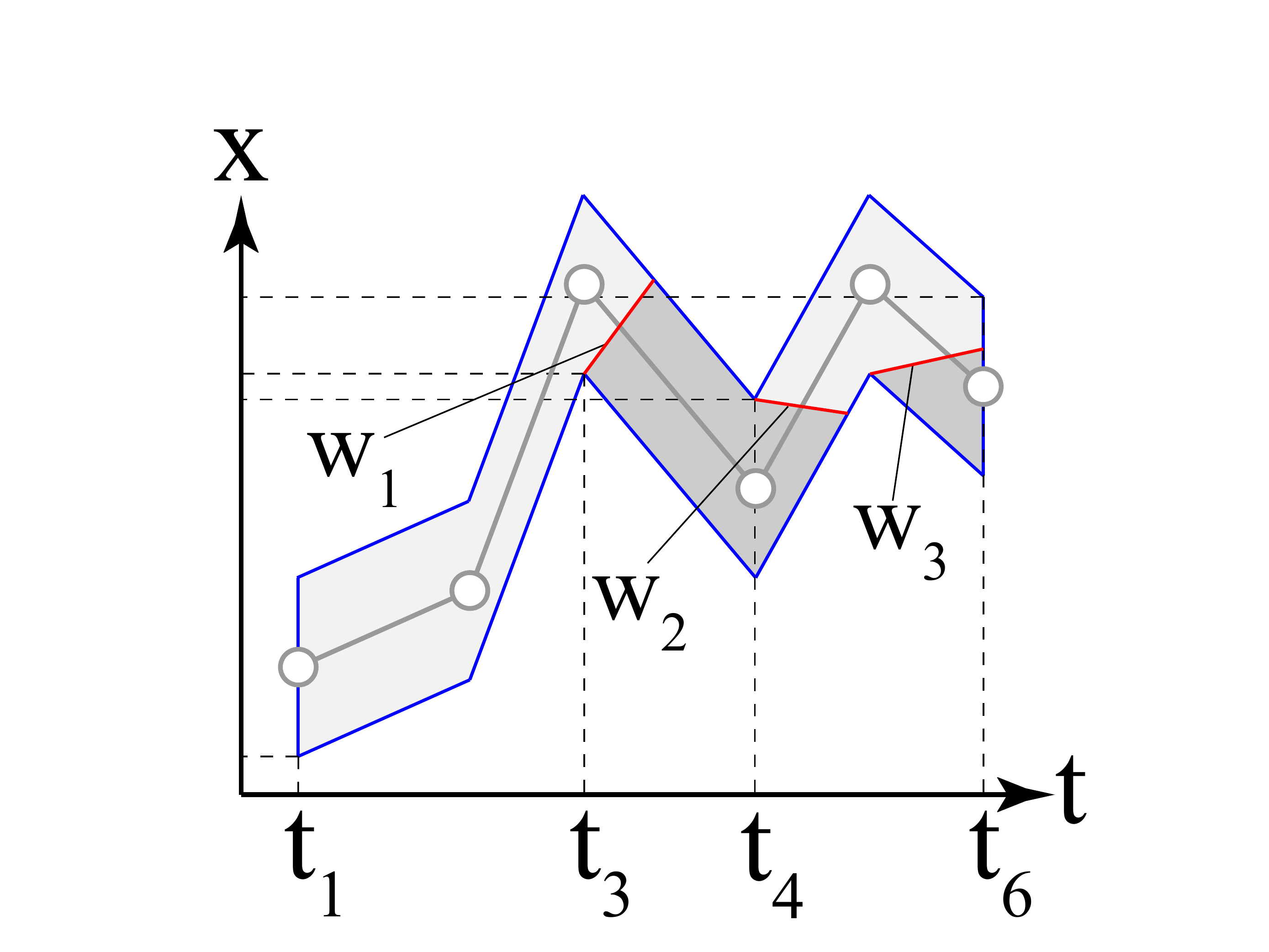}
\label{fig:w30}}
\subfigure[Simplify X]{
\includegraphics[trim=130 0 110 70,clip,width=0.3\linewidth]{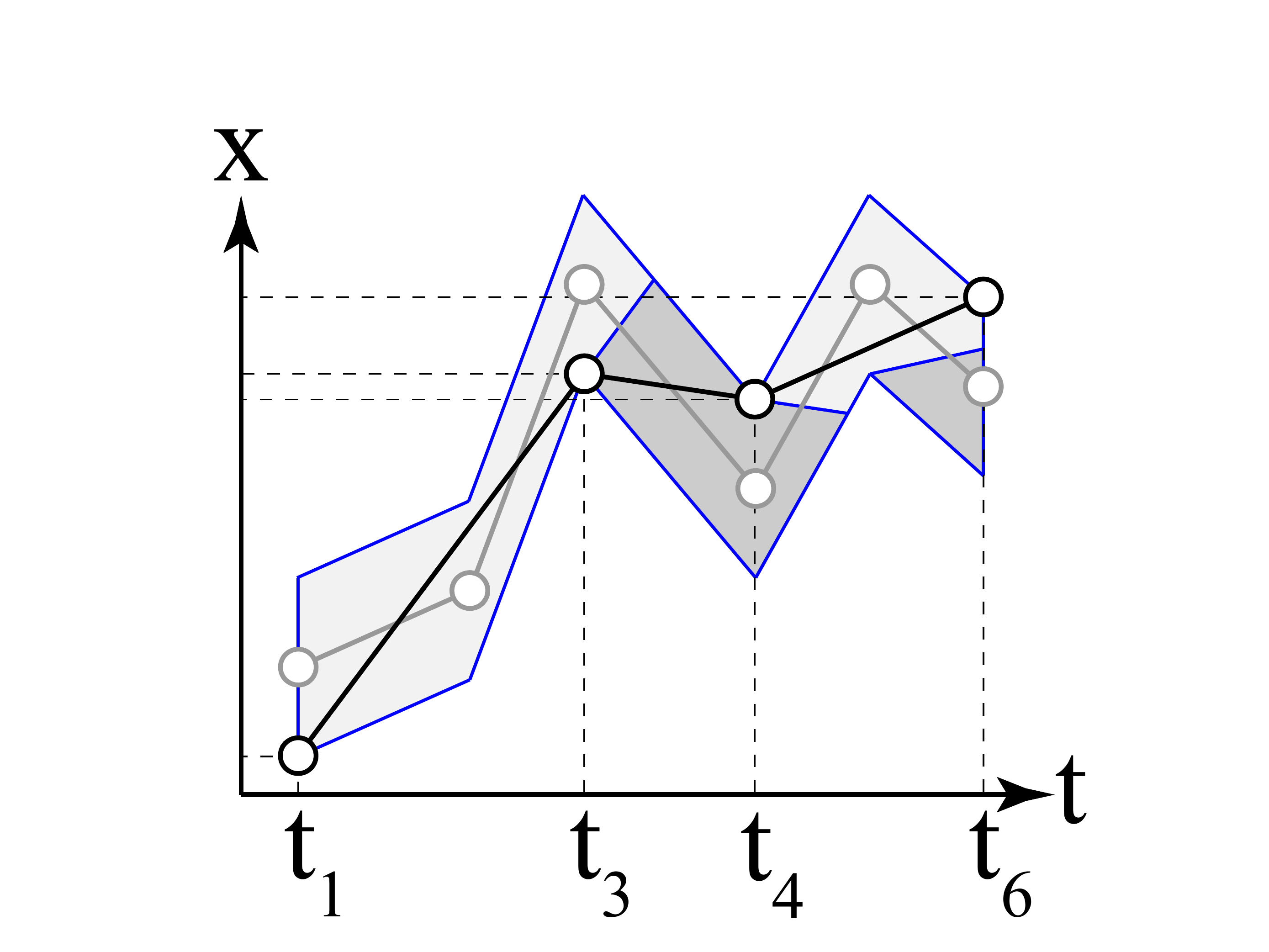}
\label{fig:w3}}\\
\subfigure[Y boundary]{
\includegraphics[trim=130 0 110 70,clip,width=0.3\linewidth]{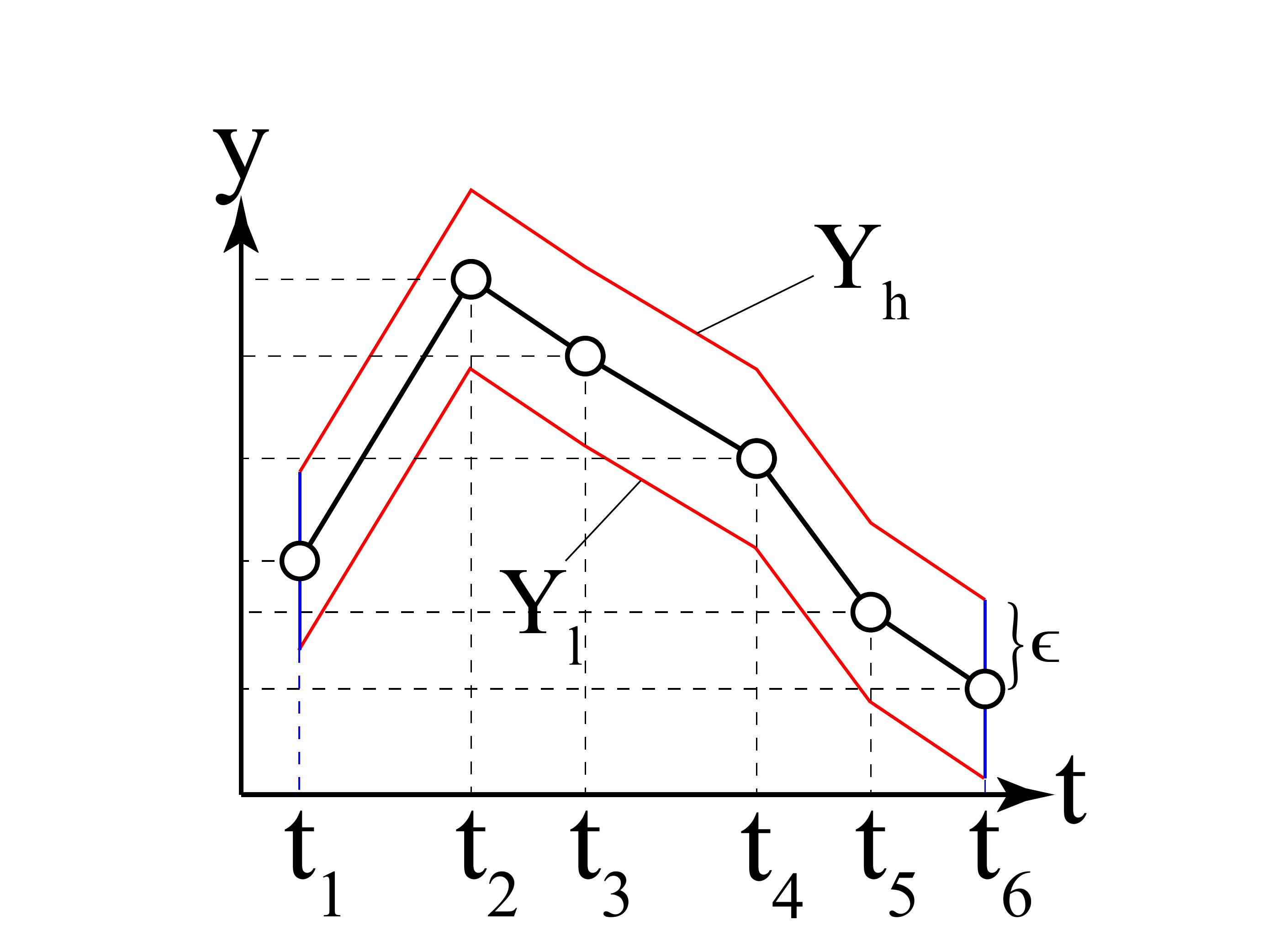}
\label{fig:w4}}
\subfigure[Y windows]{
\includegraphics[trim=130 0 110 70,clip,width=0.3\linewidth]{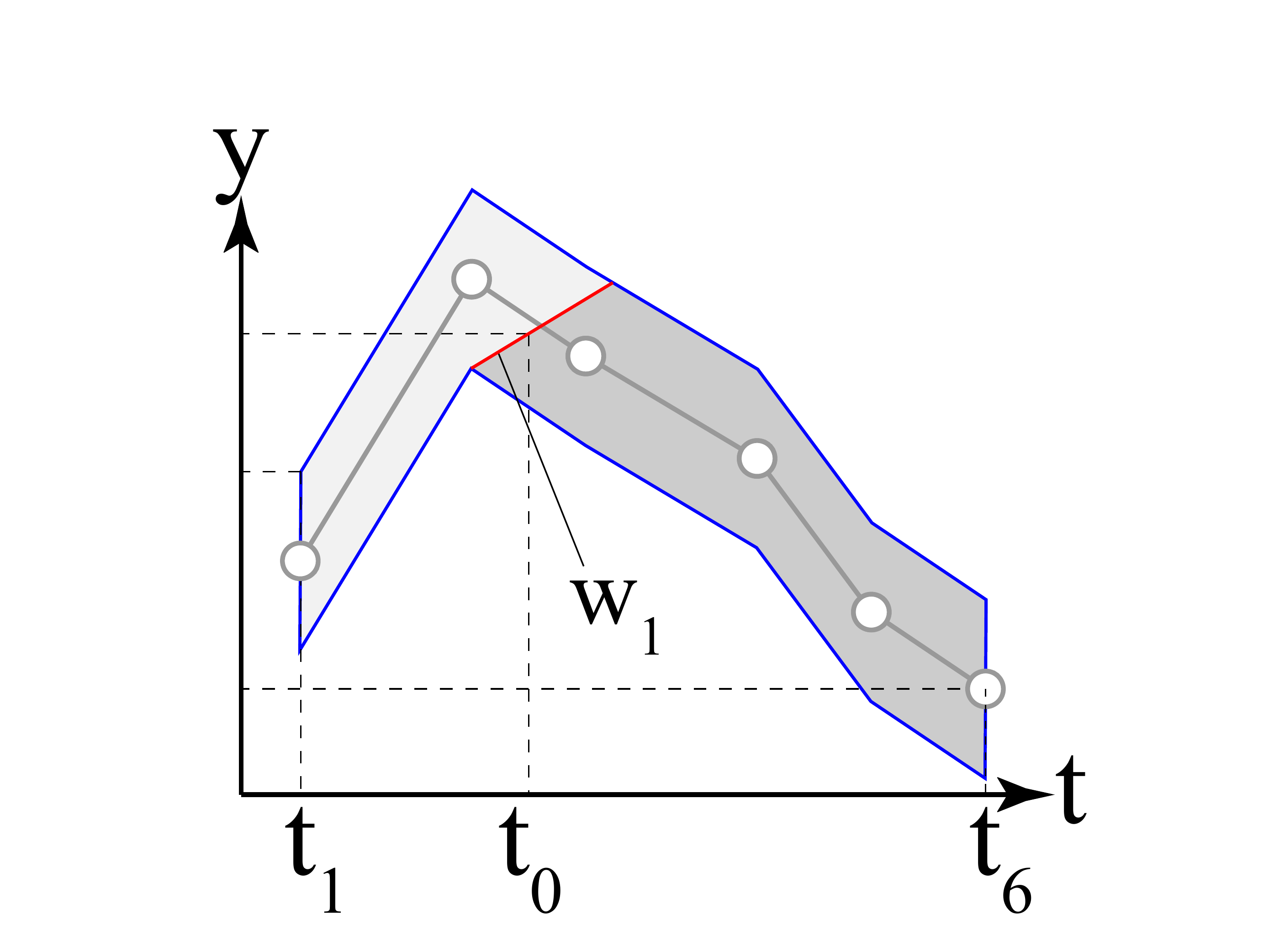}
\label{fig:w60}}
\subfigure[Simplify Y]{
\includegraphics[trim=130 0 110 70,clip,width=0.3\linewidth]{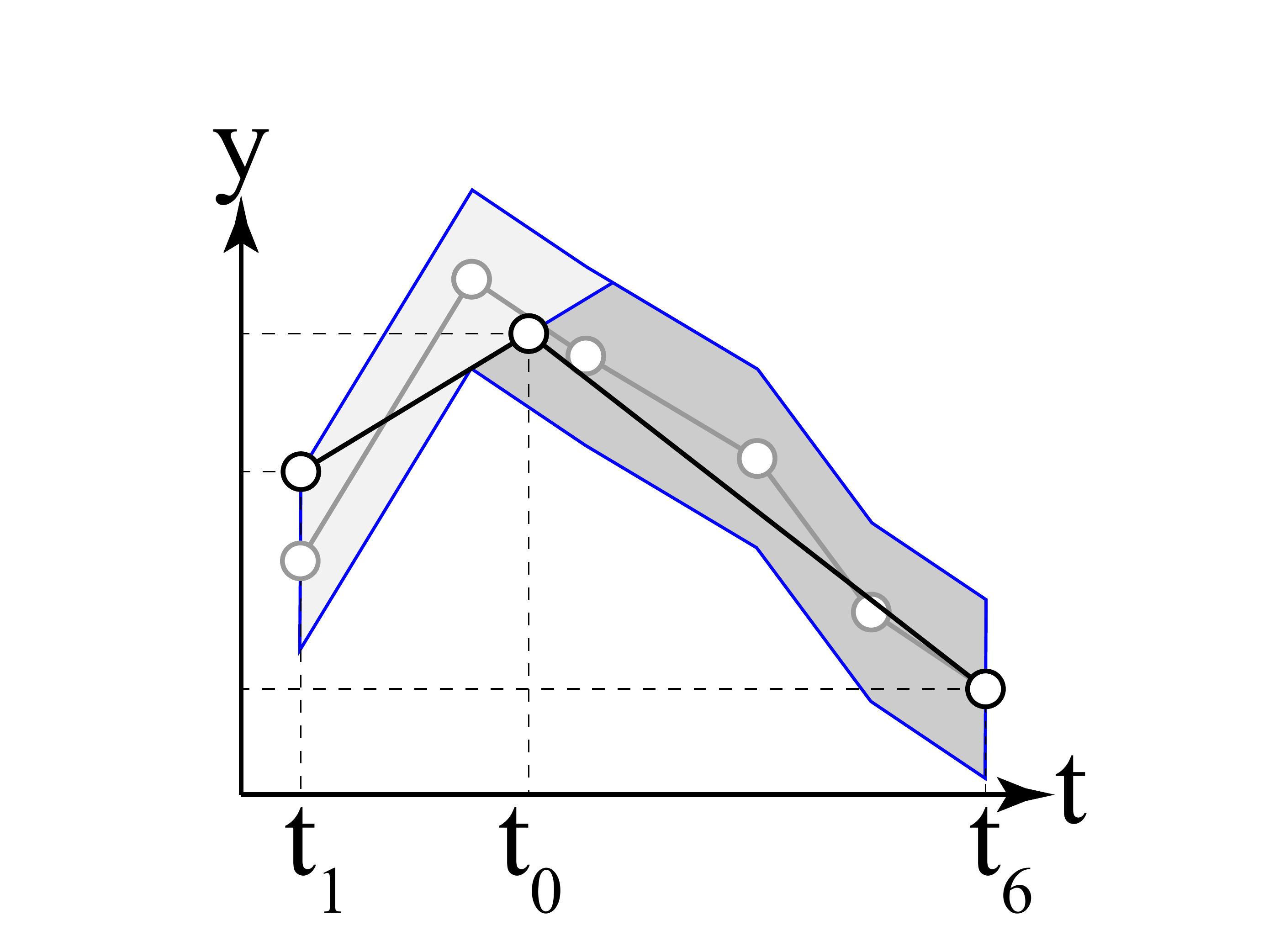}
\label{fig:w6}}
\captionsetup{font=bf}
\caption{Link distance algorithm}
\label{fig:ws}
\end{figure}
\subsubsection{Link Distance Methods}
\label{sec:wi}
The optimal simplification in 1-space is reducible to the minimum link path problem in simple polygons~\cite{han2017compress,suri1990some}. Given an error tolerance $\epsilon$, let $X_h(t) = X(t) + \epsilon$ and $X_l(t) = X(t) - \epsilon$ be two auxiliary trajectories (see Figure~\ref{fig:w1}). If a simplified trajectory $Z(t)$ falls between $X_h(t)$ and $X_l(t)$, then $X(t) -\epsilon \leq Z(t) \leq X(t) + \epsilon$ for any $t\in[t_1,t_n]$, and $Z$ is a valid solution. The two auxiliary trajectories constitute a simple polygon after we connect their starting and ending points. The polygon is a boundary of feasible trajectories, because any point $p$ outside the boundary results in the synchronized distance being greater than $\epsilon$ (see Figures~\ref{fig:w1} and~\ref{fig:w4}).

Among all feasible trajectories inside the simple polygon, we seek the ones with minimum size, i.e., the fewest vertices. As a result, the trajectory simplification problem in 1-space is reduced to the problem of computing the \textit{link distance} inside a simple polygon. The link distance between points or chords inside a polygon is defined to be the number of edges of a polyline inside the polygon connecting them. For example, the link distance between any points in a convex polygon is always one. The distance as well as the desired polyline in a polygon with $n$ vertices can be computed in $O(n)$ time through a computational geometry algorithm~\cite{suri1990some}. In this algorithm, the polygon is subdivided into pieces by chords called \textit{windows}, while the link distance from the source to the points in the same piece is constant. For example, consider $X$ in Figure~\ref{fig:ws}. Starting from the source vertical edge at $t_1$, we construct three windows $w_1$, $w_2$ and $w_3$ and subdivide the polygon into four parts until reaching the destination vertical edge at $t_6$ (see Figure~\ref{fig:w30}). In order to reach the rightmost edge, we have to go through at least two windows (i.e., $w_1$ and $w_2$), so two intermediate points are necessary to connect the leftmost and rightmost edges. Consequently, there are at least four points in the minimum link path, including the first and last points. Finally, we trace the windows back to the source to construct the optimal path with only four points (see Figure~\ref{fig:w3}). The optimal simplification of $Y$ still consists of three points, as shown in Figure~\ref{fig:w6}. Since it takes linear time to construct the simple polygon, the time complexity of optimal weak simplification in 1-space is $O(n)$.

Note that the algorithm in~\cite{suri1990some} uses polygon triangulation as its first step. Furthermore, it may re-triangulate part of the polygon during the process, so it could be time-consuming if the triangulation algorithm is not linear or not practical~\cite{chazelle1991triangulating}. Fortunately, all polygons involved in weak simplification are monotone polygons since the trajectories are monotone in the temporal dimension, and hence they can be triangulated efficiently~\cite{toussaint1984new}.

To sum up, the optimal weak simplification in 1-space is computable because we can compute the minimum link path in 2-space in $O(n)$ time. However, the link distance problem in 3-space has been proved to be NP-hard~\cite{minlinkpath2016}. Consequently, the generalization to high dimensions is not trivial. In Section~\ref{mi}, we propose an incremental approach to generalize link path construction to high dimensional space, namely multidimensional interpolation.

\subsubsection{Generalization to High Dimensions}
\label{sec:g}
\begin{figure*}
\centering
\subfigure[Input]{
\includegraphics[trim=40 0 50 0,clip,width=0.15\textwidth]{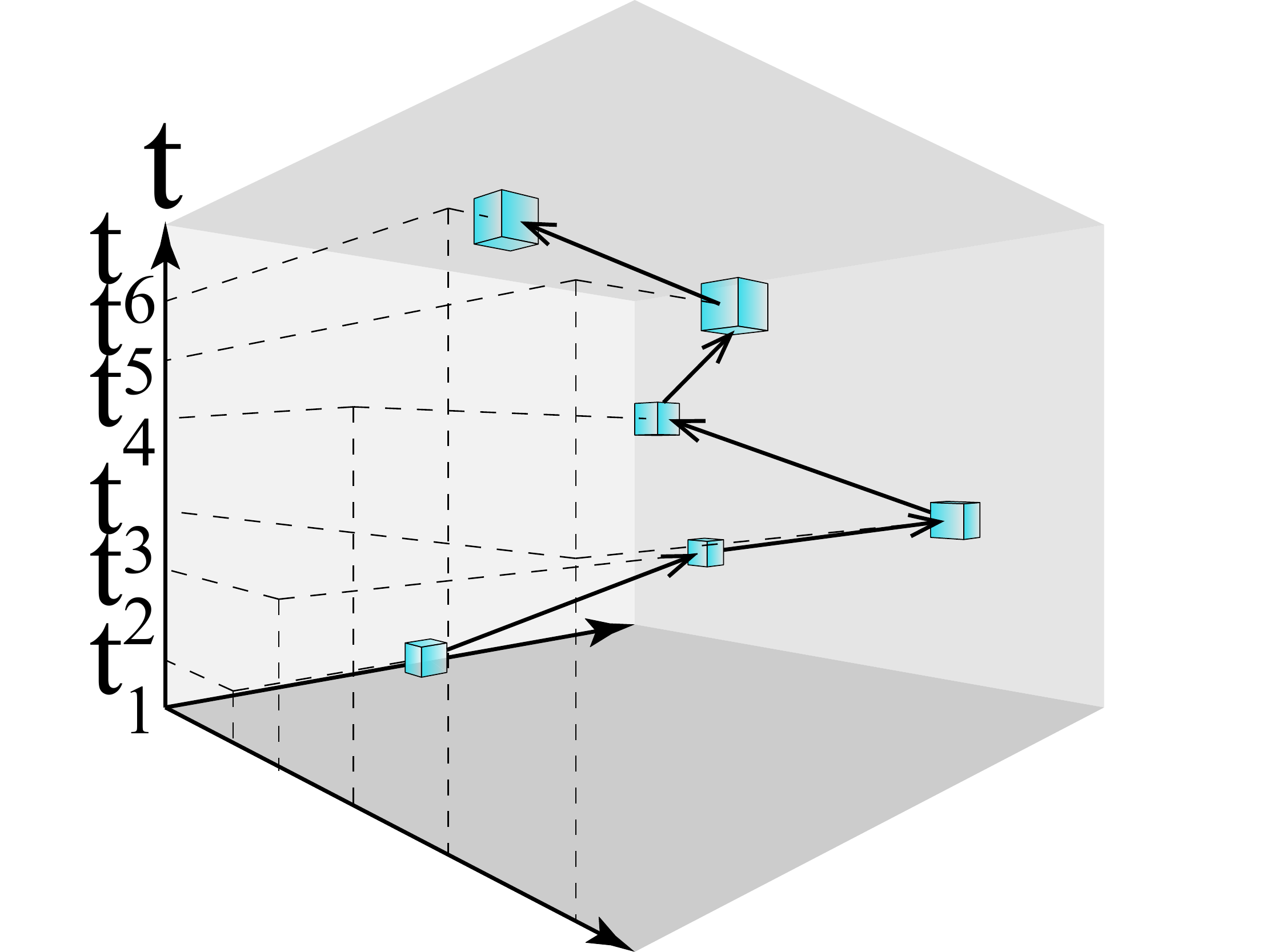}
\label{fig:traj1}}
\subfigure[Division]{
\includegraphics[trim=40 0 50 0,clip,width=0.15\textwidth]{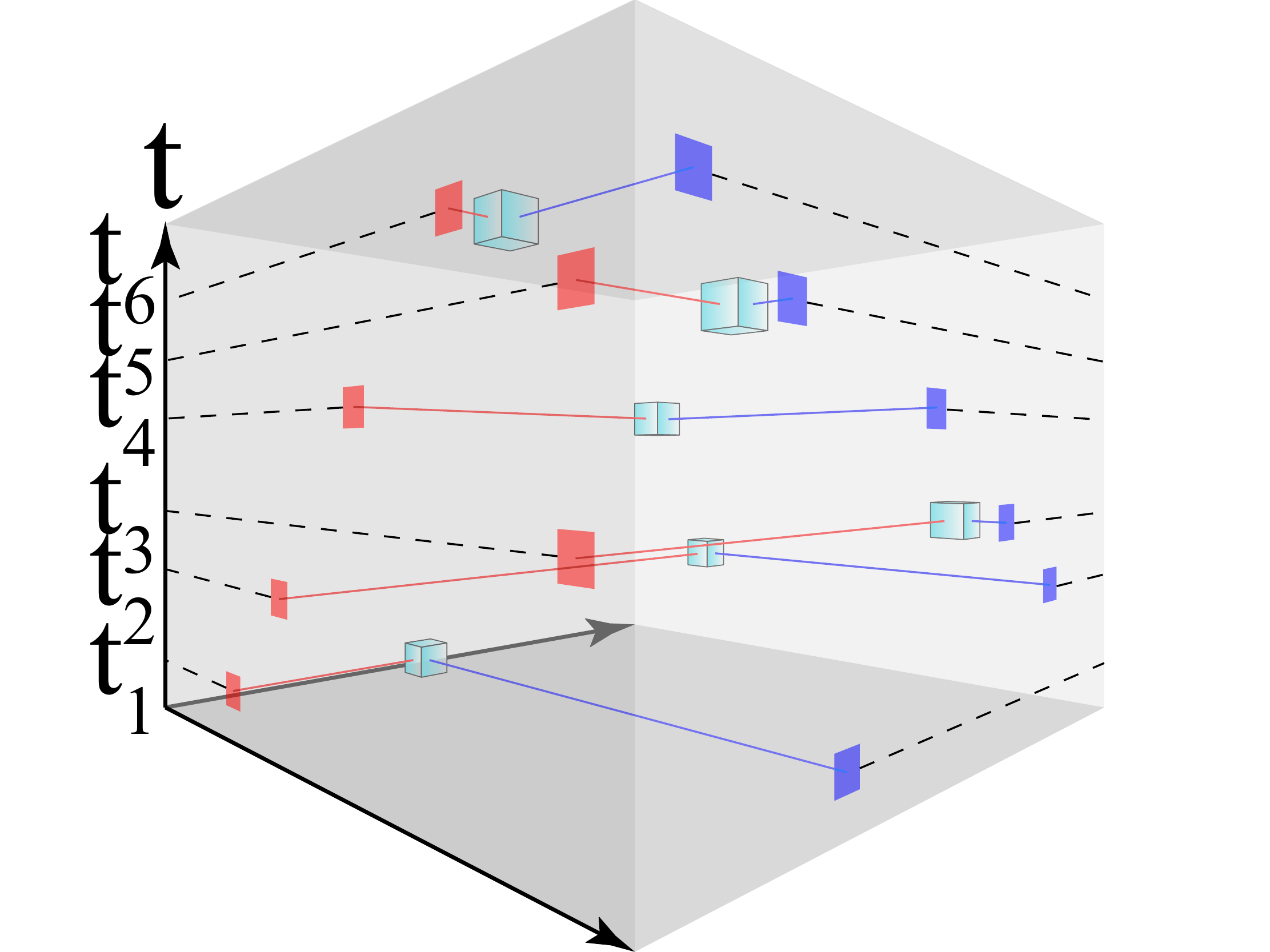}
\label{fig:traj2}}
\subfigure[Projections]{
\includegraphics[trim=40 0 50 0,clip,width=0.15\textwidth]{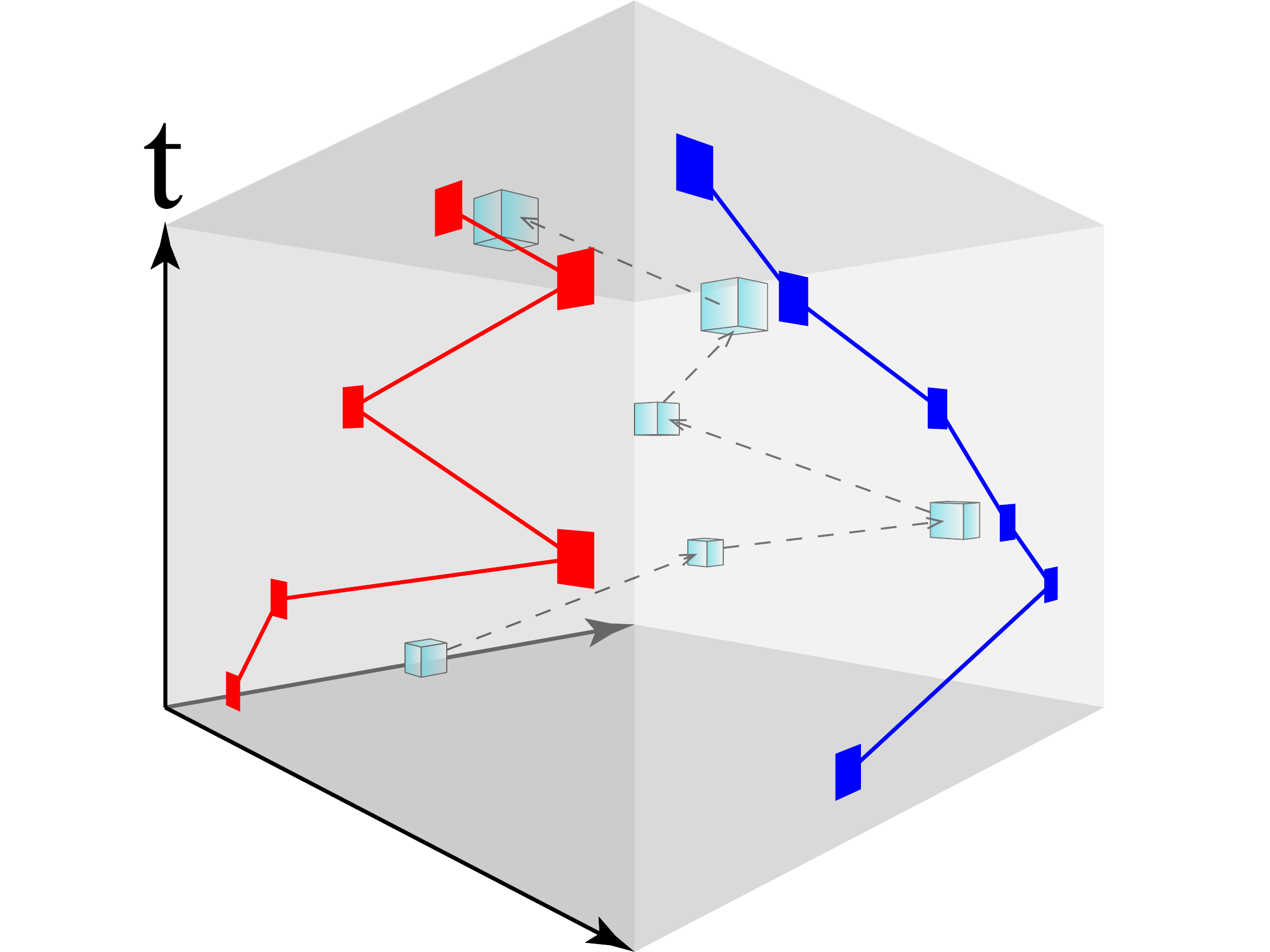}
\label{fig:traj3}}
\subfigure[Simplification$_{[1]}$]{
\includegraphics[trim=40 0 50 0,clip,width=0.15\textwidth]{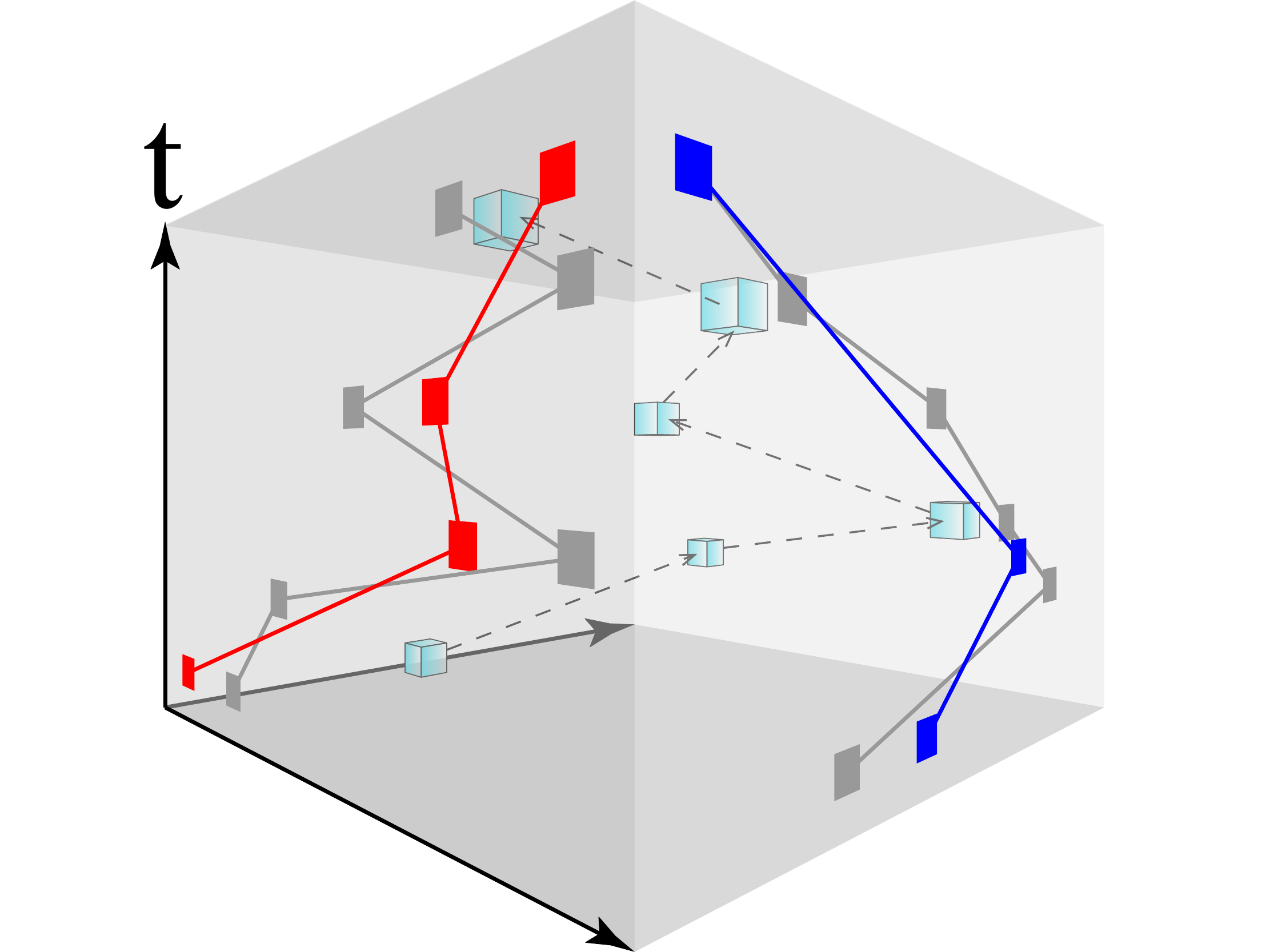}
\label{fig:traj11}}
\subfigure[Interpolation$_{[1]}$]{
\includegraphics[trim=40 0 50 0,clip,width=0.15\textwidth]{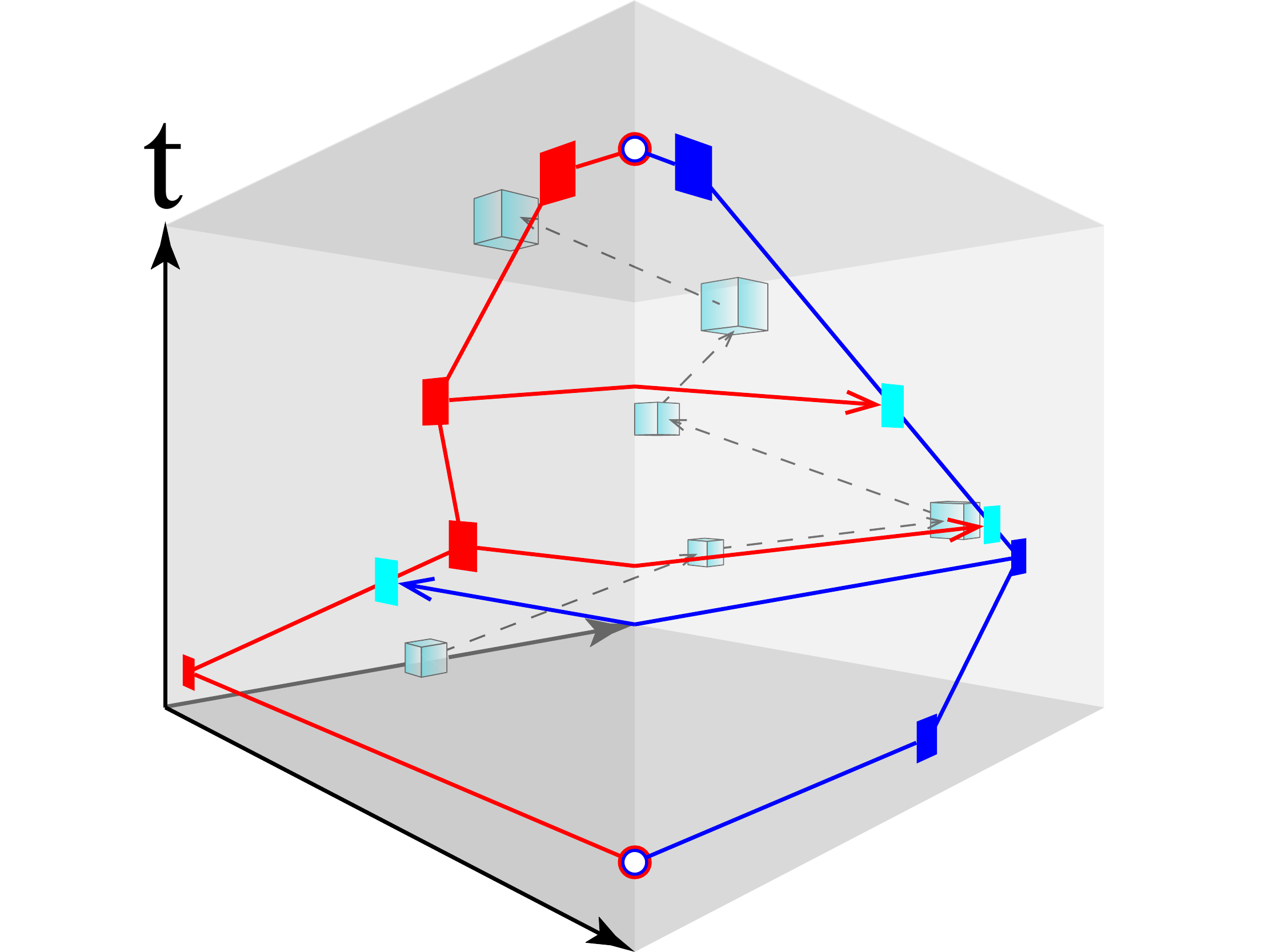}
\label{fig:traj12}}\\
\subfigure[Combination$_{[1]}$]{
\includegraphics[trim=40 0 50 0,clip,width=0.15\textwidth]{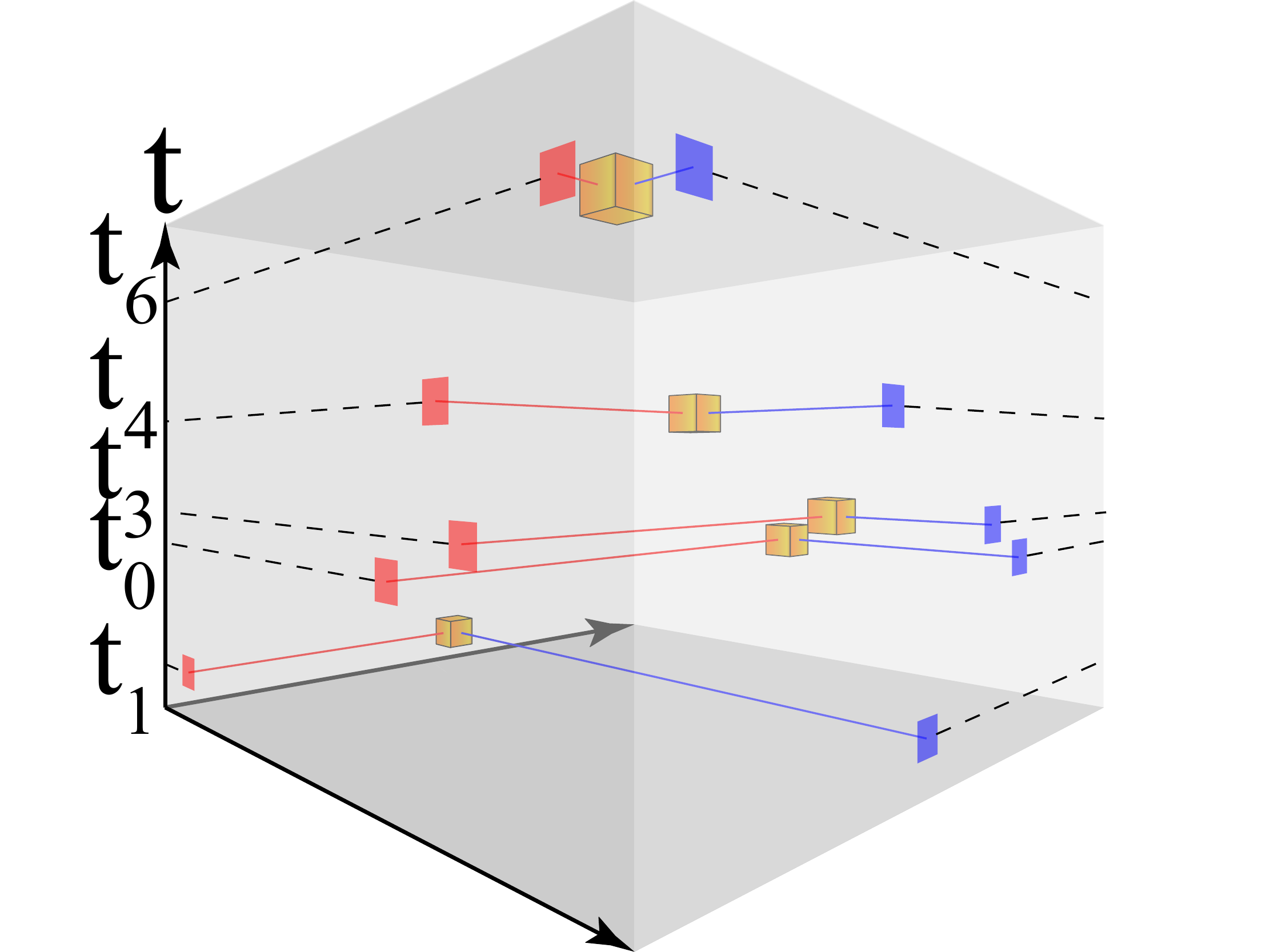}
\label{fig:traj13}}
\subfigure[Output$_{[1]}$]{
\includegraphics[trim=40 0 50 0,clip,width=0.15\textwidth]{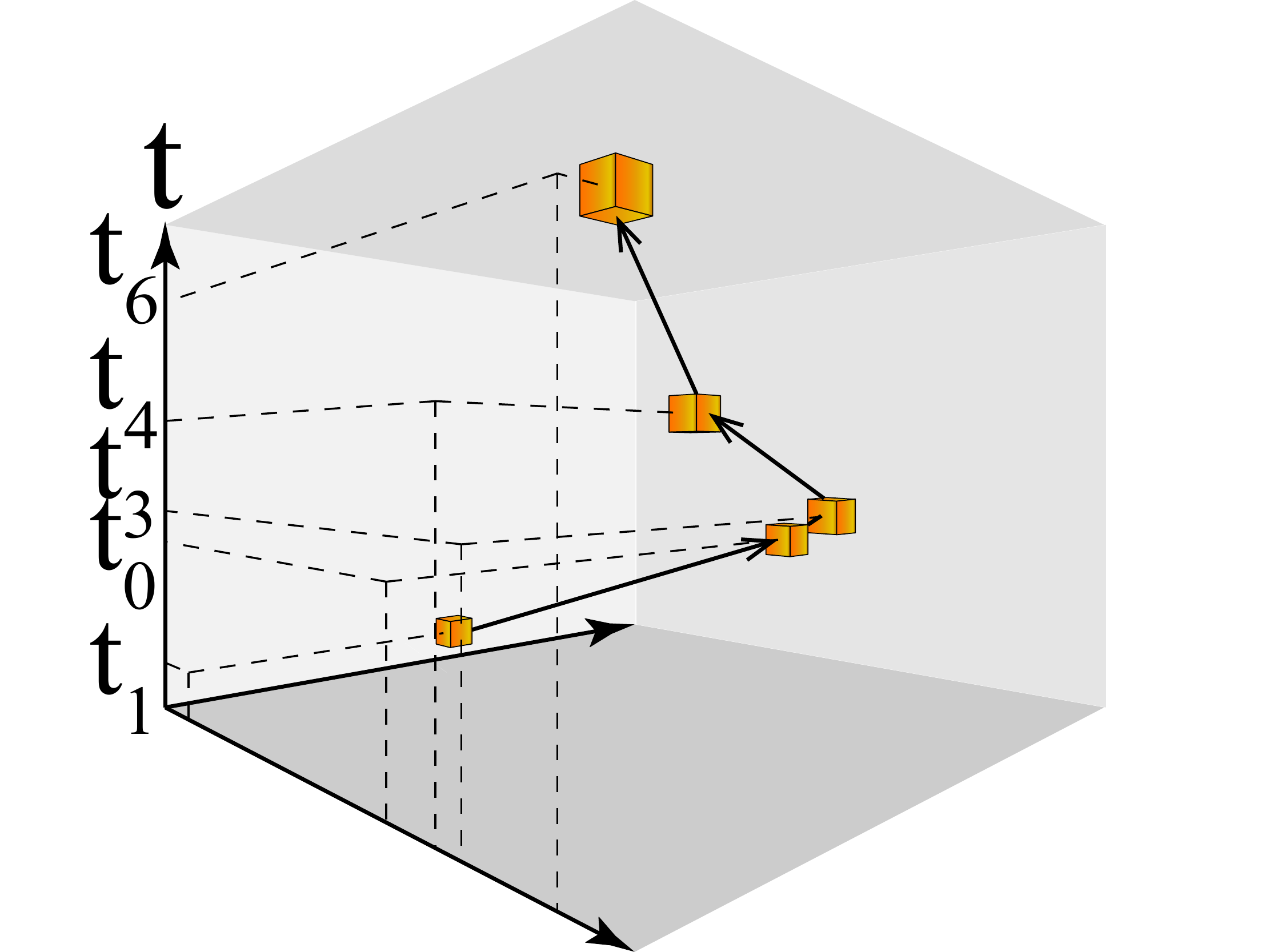}
\label{fig:traj14}}
\subfigure[Simplification$_{[2]}$]{
\includegraphics[trim=40 0 50 0,clip,width=0.15\textwidth]{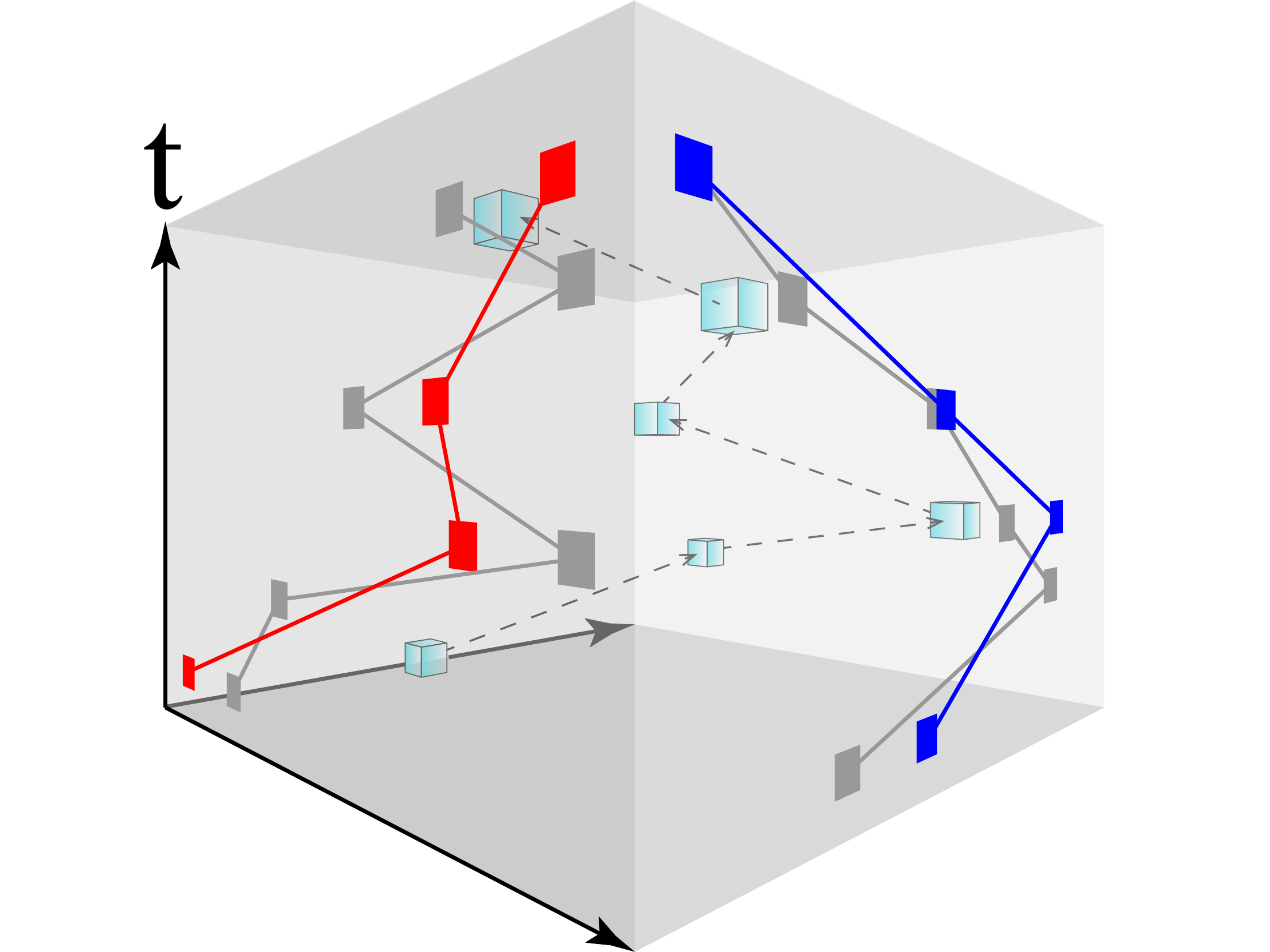}
\label{fig:traj21}}
\subfigure[Interpolation$_{[2]}$]{
\includegraphics[trim=40 0 50 0,clip,width=0.15\textwidth]{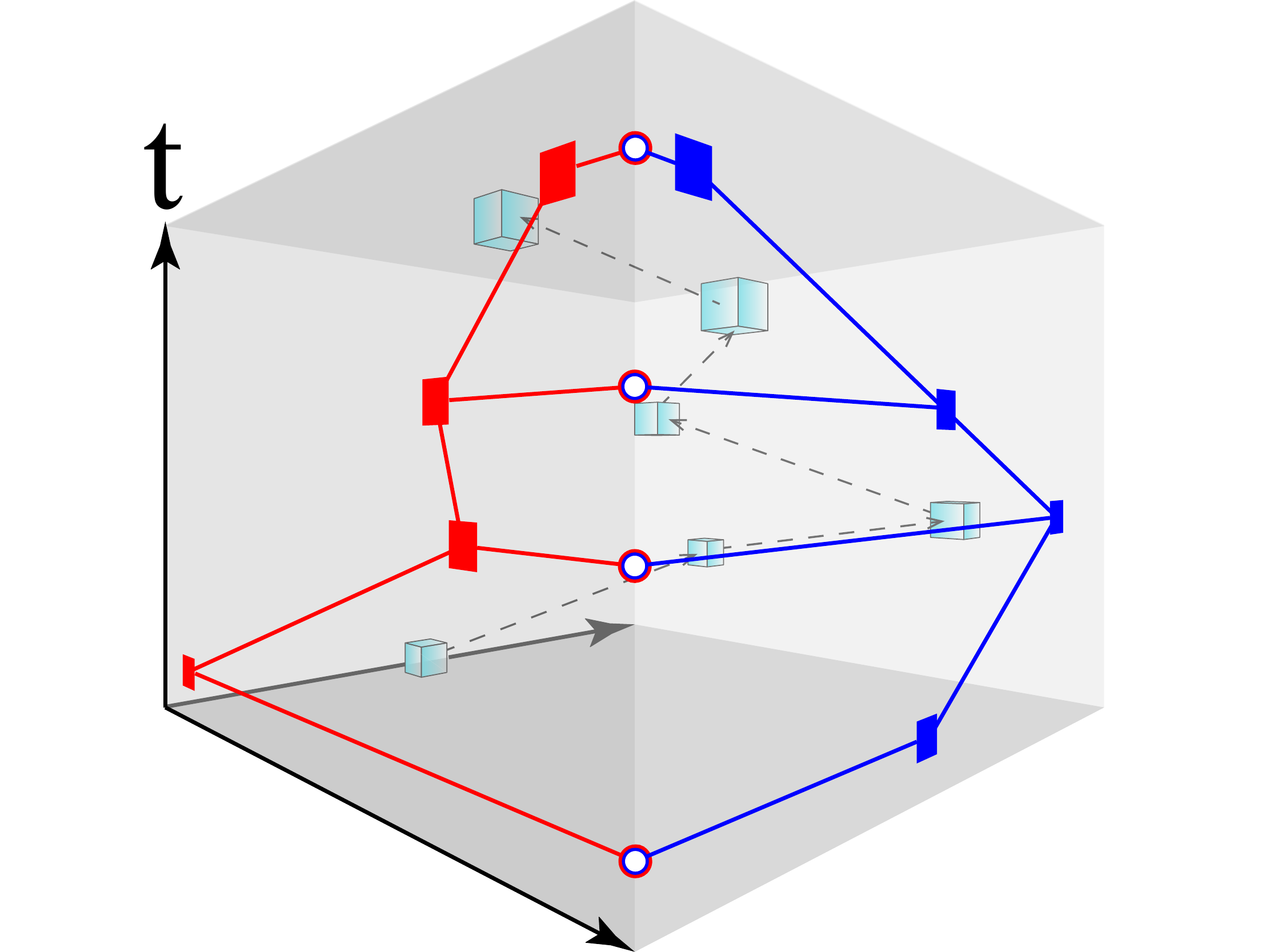}
\label{fig:traj22}}
\subfigure[Combination$_{[2]}$]{
\includegraphics[trim=40 0 50 0,clip,width=0.15\textwidth]{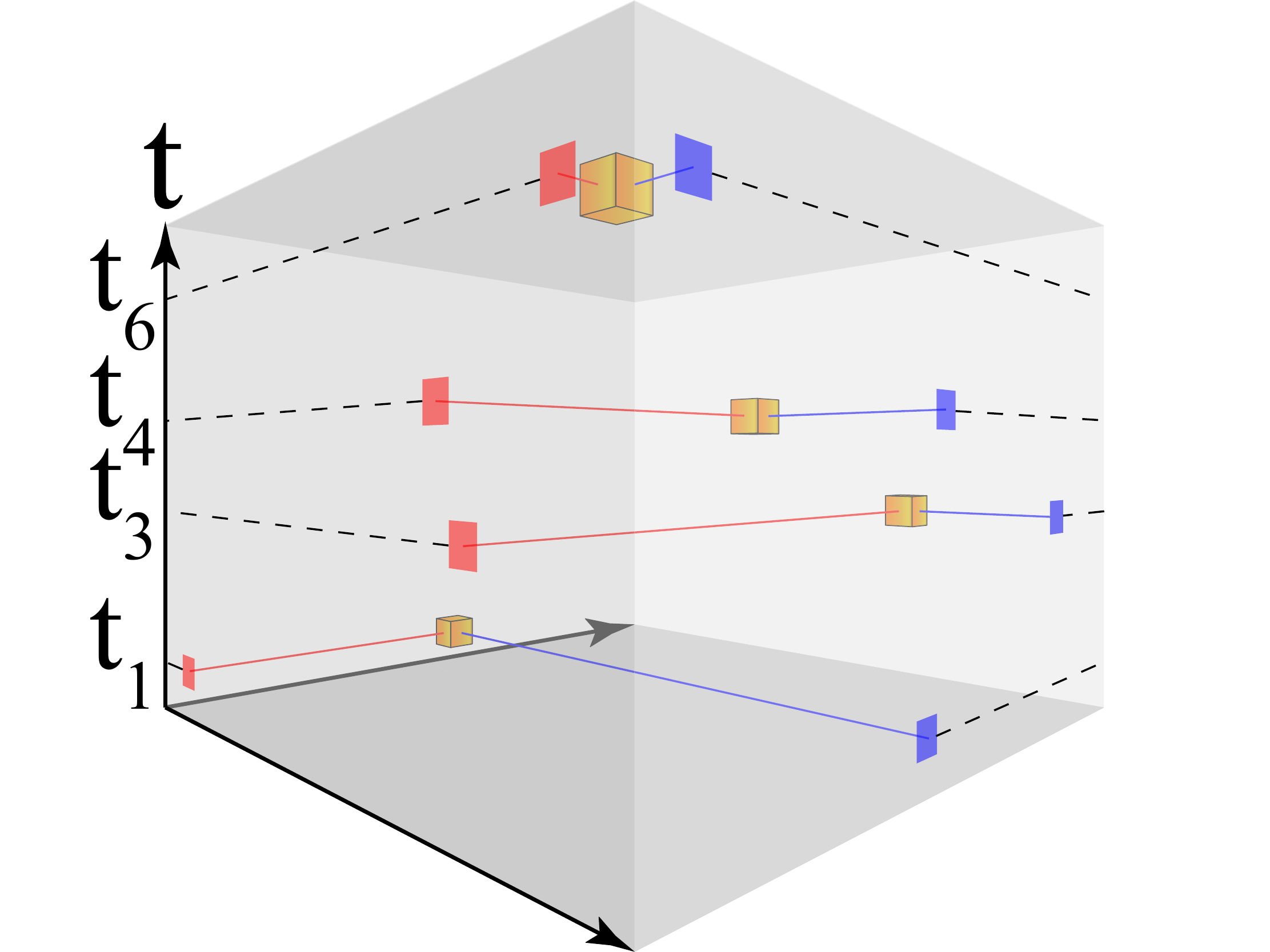}
\label{fig:traj23}}
\subfigure[Output$_{[2]}$]{
\includegraphics[trim=40 0 50 0,clip,width=0.15\textwidth]{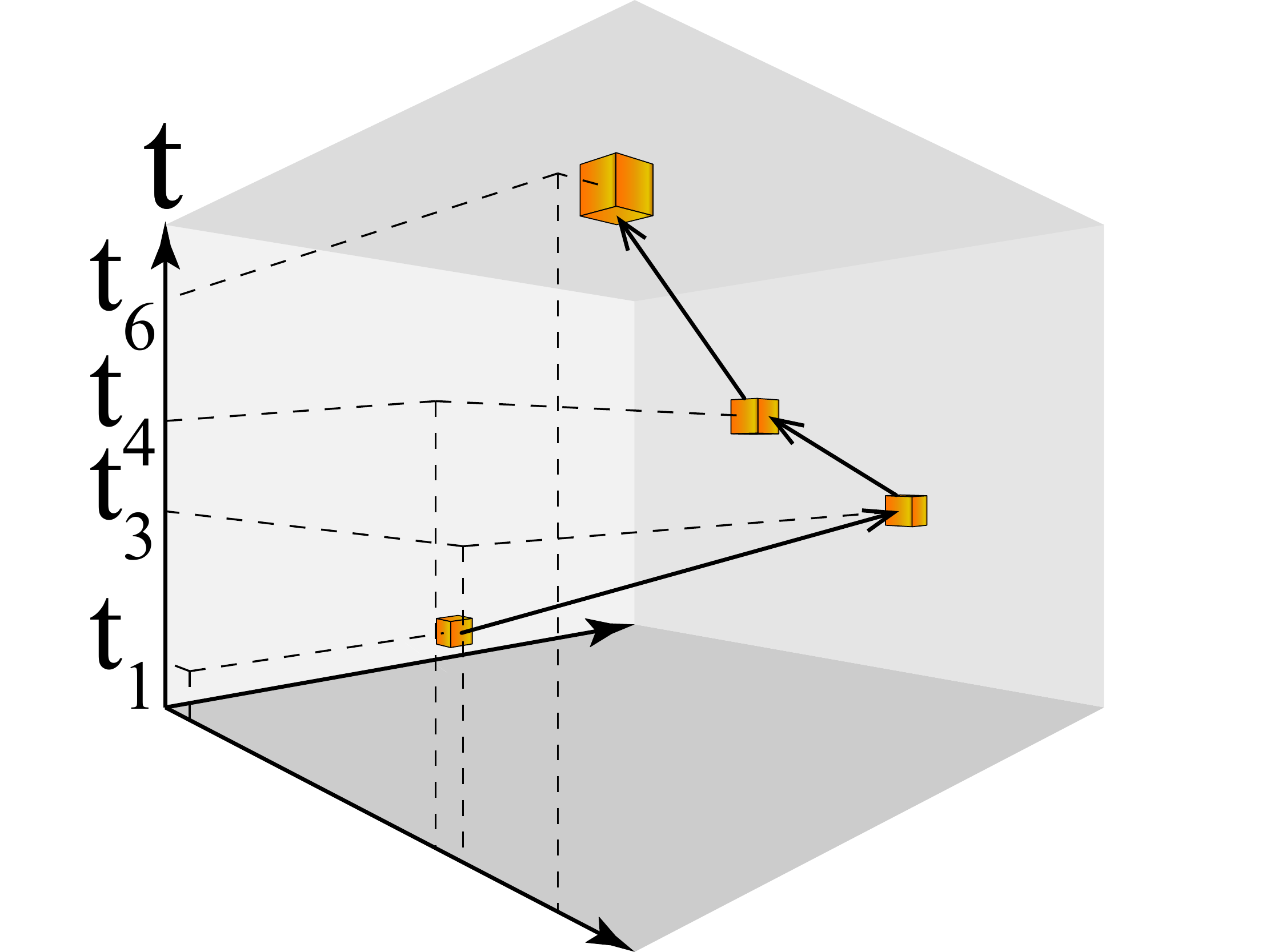}
\label{fig:traj24}}
\captionsetup{font=bf}
\caption{Multidimensional interpolation}
\label{sec:main}
\label{fig:traj}
\end{figure*}

On the basis of existing algorithms in 1-space, we discuss their generalizations to high dimensions under $L_\infty$. In order to simplify the explanation, we illustrate the algorithms with example trajectory data in 2-space. For example, consider Figure~\ref{fig:traj1}, where the input trajectory consists of six sample points in 2-space, whose x-coordinate and y-coordinate values are $X$ and $Y$ in Figures~\ref{fig:rdp} and~\ref{fig:ss}, respectively.

As mentioned in Section~\ref{sec:ri}, the generalization of algorithms such as RDP is straightforward, while it takes some effort to generalize sector intersection algorithms in Section~\ref{sec:si}. It is simple to perform sector intersection in 1-space, because the sectors are equivalent to vertical line segments---the unit ball of all metrics in 1-space. However, the unit balls become complicated in high dimensions. For example, it takes $\Theta(n \log n)$ time to compute the intersection of $n$ unit balls of the $L_p$ metric in 2-space or 3-space (e.g., disks or balls) and the complexity of the intersection by $n$ unit balls in $m$-space could be up to $O(n^{\lfloor m/2\rfloor})$~\cite{mcmullen1970maximum}. Consequently, sector intersection becomes difficult in high dimensional space. Fortunately, the unit balls of $L_\infty$ are hypercubes, and it only takes $O(n)$ time to compute the intersection of $n$ hypercubes in any space. Therefore, all algorithms based on sector intersection can be perfectly generalized to high dimensional space under the $L_\infty$ metric.

The performance of the generalized algorithms is limited by their performance in 1-space. For algorithms like RDP, the $L_\infty$ distance over all dimensions is less than $\epsilon$ if and only if the distance in each dimension is less than $\epsilon$. For those algorithms based on sector intersection, the intersection is not empty if and only if all intersections in lower dimensions are not empty. Consequently, if the projection of a point in $T_i$ is kept by the original algorithm in 1-space, then the point must be kept by the generalized algorithm, too. Thus,
$|G(\mathbf{T})| \geq \max_{i=1}^m |g(T_i)|$, where $G(\cdot)$ is the generalized simplification algorithm in $m$-space while $g(\cdot)$ is its original version in 1-space. For example, the algorithms discussed in Sections~\ref{sec:ri} and~\ref{sec:si} cannot simplify the input trajectory in Figure~\ref{fig:traj1}, because the simplification of one of its projections $X$ (see Figures~\ref{fig:rdp} and~\ref{fig:ss}) consists of all points. In order to achieve a higher compression ratio, we resort to the generalization of optimal weak simplification based on link distance.

$L_\infty$ also makes it possible to generalize the algorithm in Section~\ref{sec:wi}, which is based on link distance, to higher dimensions. Here we illustrate the idea from the perspective of intrinsic properties of distance metrics. The synchronized distance with $L_\infty$ is 
\begin{align*}
    d(\mathbf{S}, \mathbf{T})&=\max_{t\in[t_1,t_n]}\|\mathbf{S}(t) - \mathbf{T}(t)\|_\infty\\
    &=\max_{t\in[t_1,t_n]}\max_{i=1,2,\cdots,m}{|S_i(t) - T_i(t)|}.
\end{align*}
In this case, we can swap the maximum operators without changing the result, and hence the overall error under $L_\infty$ is determined by the error in each dimension:
\begin{align}
    d(\mathbf{S}, \mathbf{T})&=\max_{i=1}^m\max_{t\in[t_1,t_n]}{|S_i(t) - T_i(t)|}\nonumber\\
    &=\max_{i=1}^m~d(S_i,T_i)\label{swap}
\end{align}
where $d(S_i, T_i)$ is the synchronized distance in 1-space. It is therefore possible to break down the problem into simplification in 1-space and then combine the results to form a solution. On the contrary, for any finite $p$, the synchronized $L_p$ distance could be greater than $\epsilon$ even if synchronized distances in all dimensions are bounded by $\epsilon$. On the basis of this property, we propose the new approach, namely multidimensional interpolation, to generalize the algorithm in Section~\ref{sec:wi}.

\subsection{Multidimensional Interpolation}
\label{mi}
In this section, we first present the three implementations of LiMITS, including Direct Interpolation (DI), Midpoint Correlated Interpolation (MCI) and Various Interpolation (VI), which are all based on optimal simplification in 1-space introduced in Section~\ref{sec:wi}. Then we introduce a compact representation of the output of LiMITS to further improve the compression ratio.

\subsubsection{Direct interpolation}
The first implementation is Direct Interpolation (DI). It directly combines results from all dimensions. Given a trajectory $\mathbf{T}$ in $m$-space, we first project it onto the planes spanned by each spatial dimension and the temporal dimension. Let the projection in the $i$-th plane be $T_i$, where the $j$-th sample point in $T_i$ is $\langle T_i(\mathbf{T}.t_j),\mathbf{T}.t_j \rangle$. Then we simplify all projected trajectories in 1-space, after which we obtain $m$ simplifications $S_1,S_2,\cdots,S_m$. Finally, we combine all $S_i$ by interpolation. Given any time $t$, the point in the output trajectory $\mathbf{S}$ is $\langle S_1(t), S_2(t),\cdots, S_{m}(t)\rangle$. Since $T_i(t) -\epsilon \leq S_i(t) \leq T_i(t) + \epsilon$ for all $i$, the $L_\infty$ distance between $\mathbf{T}(t)$ and $\mathbf{S}(t)$ should be within $\epsilon$, so it is a feasible simplification. In practice, we only need to store points at time stamps of points from $m$ simplifications and other points can be obtained through linear interpolation, as introduced in Section~\ref{sec:pre}.

\begin{algorithm}
\SetKwFunction{PR}{project}
\SetKwFunction{P}{polygon}
\SetKwFunction{MLP}{link}
\SetKwFunction{IN}{interpolation}
\SetKwInOut{Input}{input}
\SetKwInOut{Output}{output}
\SetAlgoLined
\Input{input trajectory $\mathbf{T}$ and tolerance $\epsilon$}
\Output{output trajectory $\mathbf{S}$}
 \For{$i\gets 1$ \KwTo  $m$}{
 $T_i \gets$ $\mathbf{T}$.\PR{$\mathcal{P}_{i}$}\;\label{alg1:pr}
  $P_{T_i} \gets$ \P{$T_i,\epsilon$}\;\label{alg1:p}
  $R_i  \gets$ $P_{T_i}$.\MLP{$ T_i(\mathbf{T}.t_1)\pm\epsilon, T_i(\mathbf{T}.t_n)\pm\epsilon$}\;\label{alg1:lk}
 \lIf{$i=1$}{ 
  $\mathbf{S}_1 \gets R_i$
 }\lElse{
  $\mathbf{S}_{i}\gets$\IN{$\mathbf{S}_{i-1}, R_i$}\label{alg1:it}
  }
 }
 \Return $\mathbf{S}_m$\;
 \caption{Direct Interpolation (DI)}
 \label{MI1}
\end{algorithm}
For the input trajectory in Figure~\ref{fig:traj1}, the projections are $X$ and $Y$ respectively (see Figures~\ref{fig:traj2} and~\ref{fig:traj3}).
According to the computation in Section~\ref{sec:wi}, the simplification of $X$ consists of four points at $t_1$, $t_3$, $t_4$ and $t_6$, while that of $Y$ only consists of three points at $t_1$, $t_0$ and $t_6$, as shown in Figure~\ref{fig:traj11}. Afterwards, one interpolation point is inserted in $X$ at $t_0$ while two are inserted in $Y$ at $t_3$ and $t_4$, as Figure~\ref{fig:traj12} shows. Finally, we take the x-coordinate and y-coordinate values from simplified $X$ and $Y$ at all time stamps to construct a simplification containing five points (see Figures~\ref{fig:traj13} and~\ref{fig:traj14}).

The pseudo-code is given by Algorithm~\ref{MI1}. In each iteration, the algorithm first projects the input trajectory in the plane $\mathcal{P}_i$ spanned by the $i$-th spatial dimension and the temporal dimension (line~\ref{alg1:pr}) and then constructs a polygon with respect to the projection $T_i$ (line~\ref{alg1:p}). Next, the polygon along with the vertical edges at $t_1$ and $t_n$ is the input of the link distance algorithm in 1-space, which computes the link distance between two edges $T_i(\mathbf{T}.t_1)\pm\epsilon$ and $T_i(\mathbf{T}.t_n)\pm\epsilon$, where $p\pm\epsilon$ represents the vertical line segment of length $2\epsilon$ centered at a point $p$. The algorithm generates a feasible trajectory $R_i$ for every $T_i$ (line~\ref{alg1:lk}). Moreover, it integrates $\mathbf{S}_{i-1}$ (the combination of $R_1,R_2,\cdots,R_{i-1}$) with $R_i$ through interpolation and produces a trajectory $\mathbf{S}_{i}$ in $i$-space (line~\ref{alg1:it}). The steps in each iteration take $O(n)$ time, so the time complexity is $O(nm)$. In practice, the dimension $m$ is a constant and the algorithm is linear.

\begin{figure}
\centering
\subfigure[Division]{
\includegraphics[trim=130 0 110 70,clip,width=0.3\linewidth]{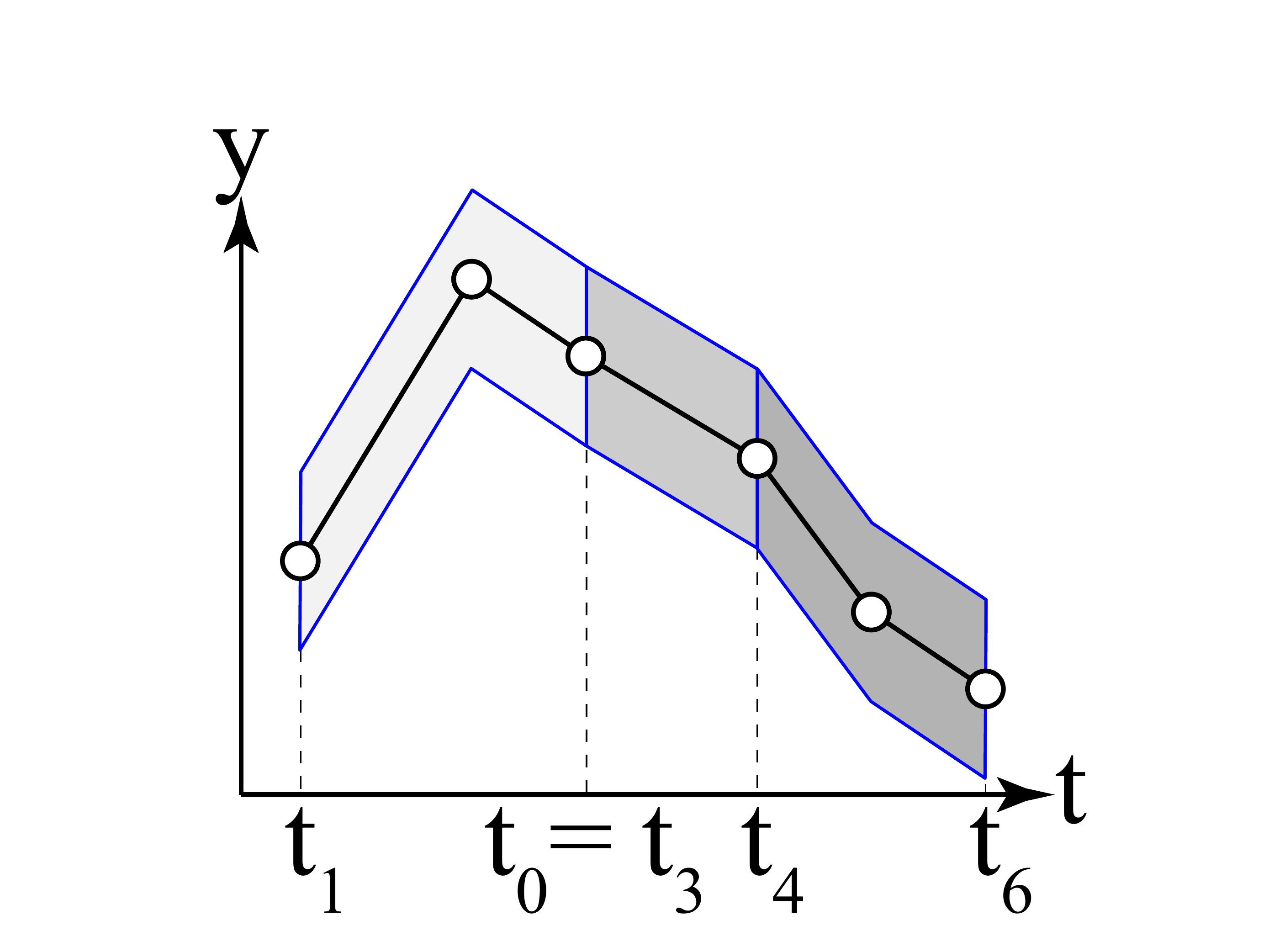}
\label{fig:w7}}
\subfigure[Visibility]{
\includegraphics[trim=130 0 110 70,clip,width=0.3\linewidth]{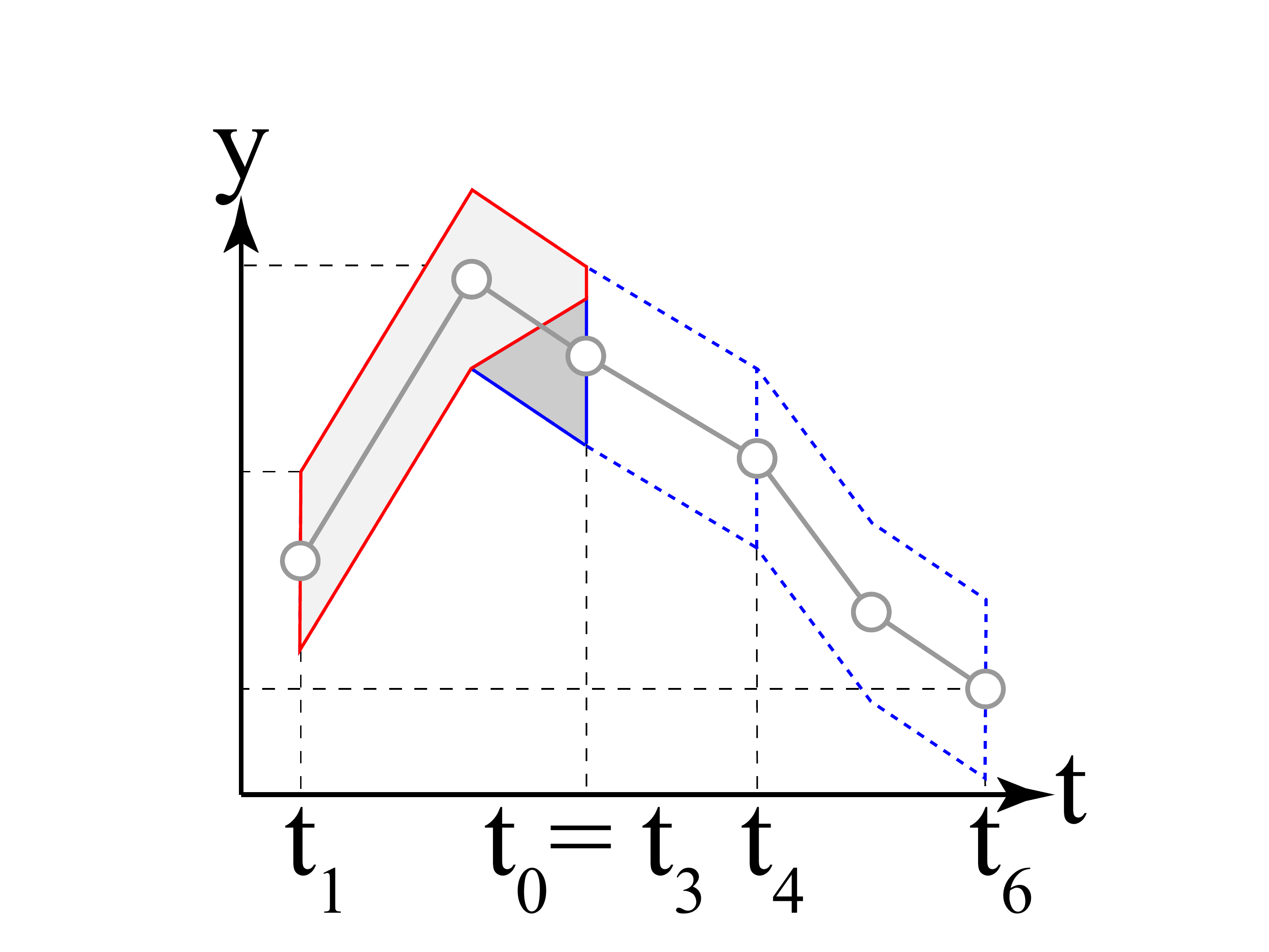}
\label{fig:w8}}
\subfigure[Simplify Y]{
\includegraphics[trim=130 0 110 70,clip,width=0.3\linewidth]{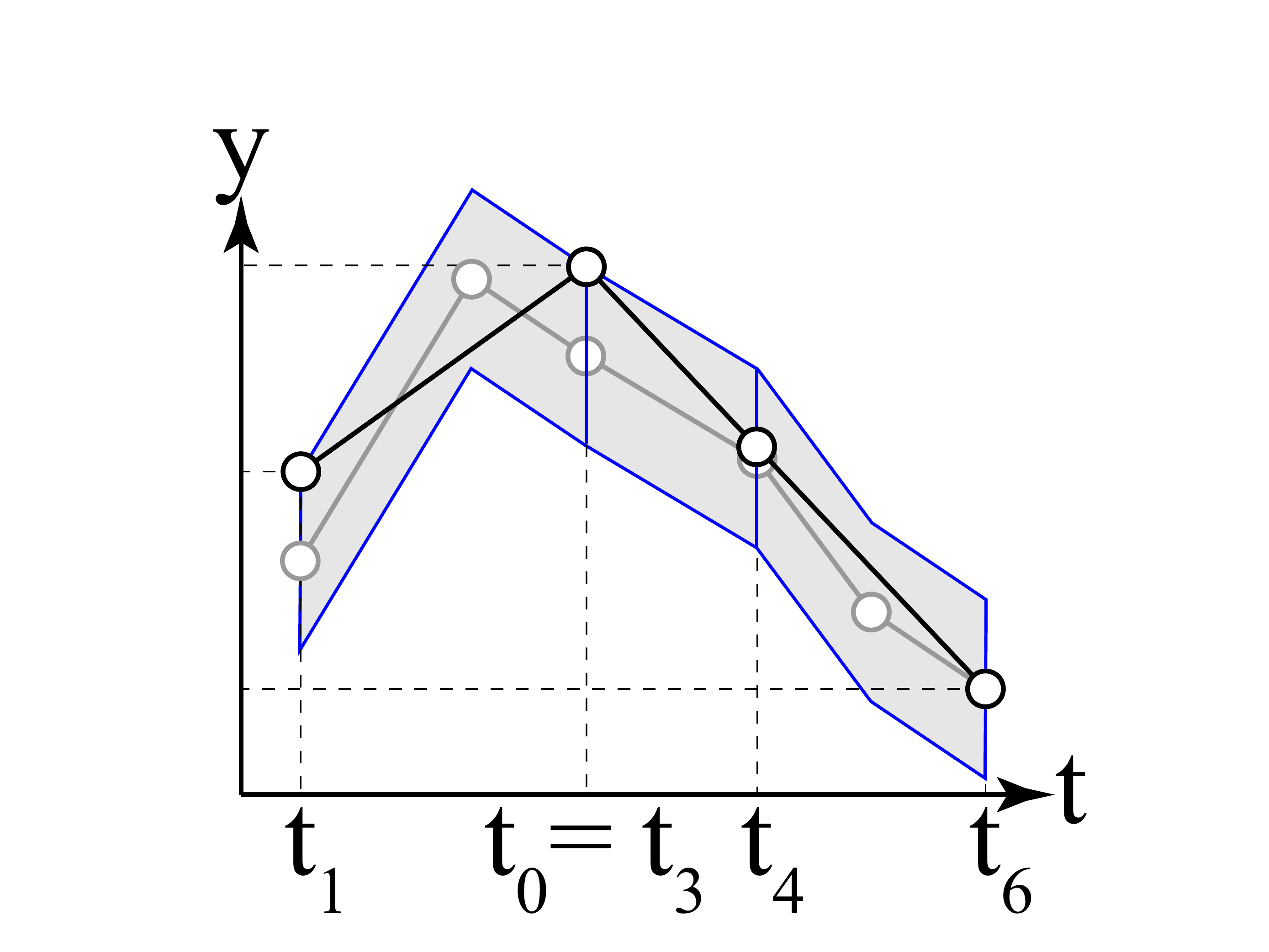}
\label{fig:w9}}
\captionsetup{font=bf}
\caption{Correlated simplification}
\label{fig:ws2}
\end{figure}

\begin{figure}
\centering
\subfigure[Projection]{
\includegraphics[trim=0 90 0 90,clip,width=0.48\linewidth]{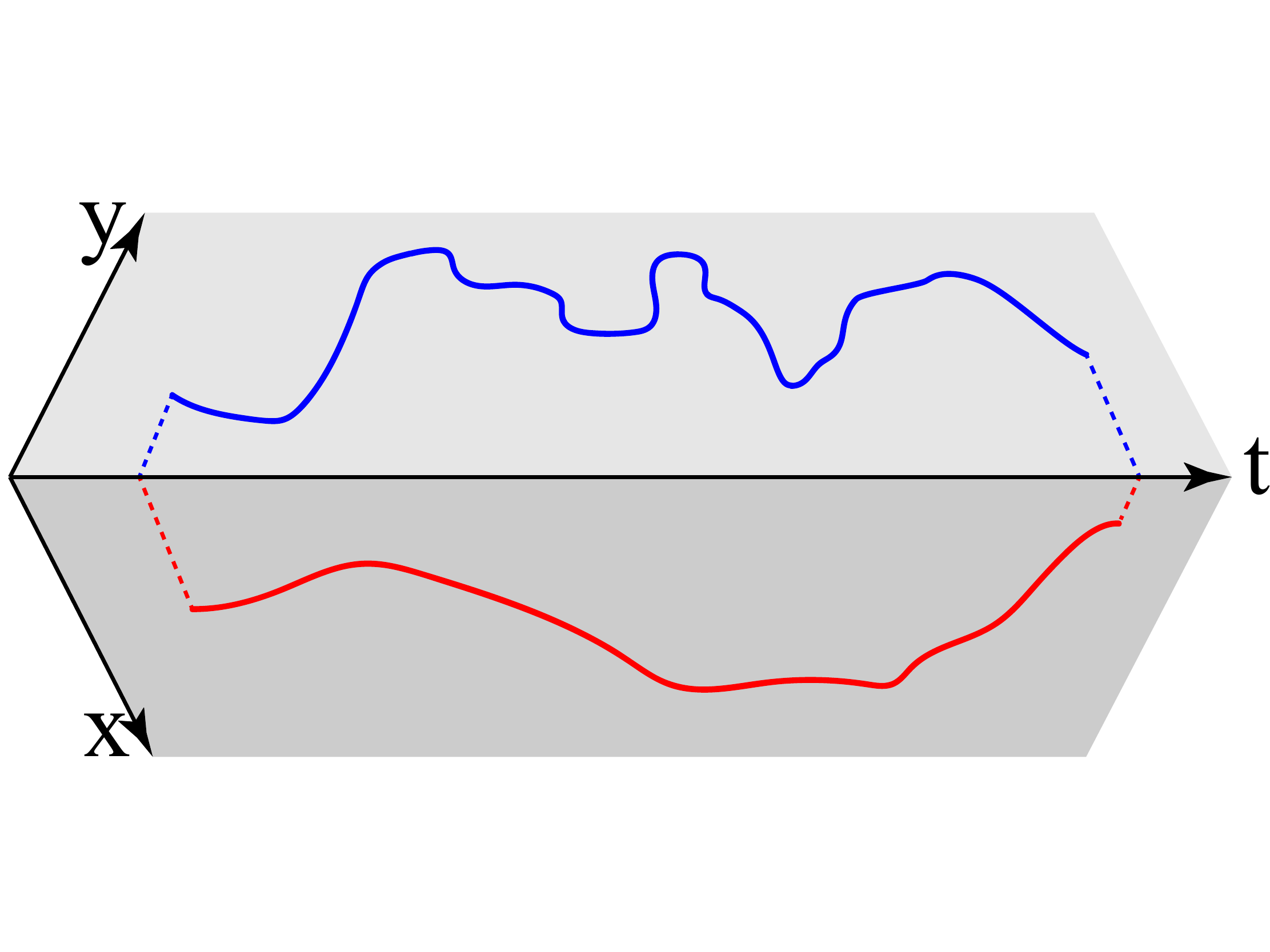}}
\subfigure[Simplify X(t)]{
\includegraphics[trim=0 90 0 90,clip,width=0.48\linewidth]{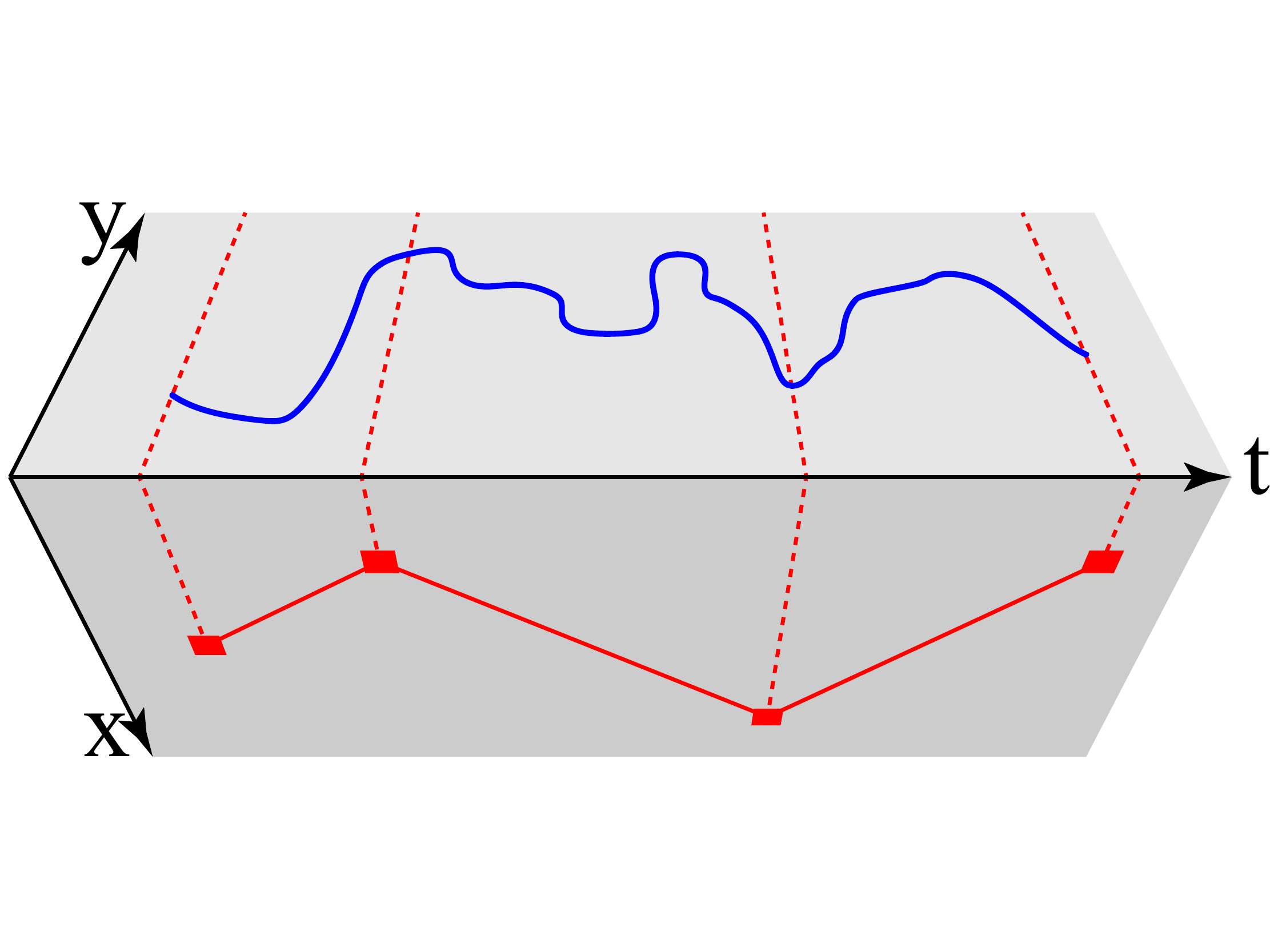}}
\subfigure[Simplify Y(t)$_{[1]}$]{
\includegraphics[trim=0 90 0 90,clip,width=0.48\linewidth]{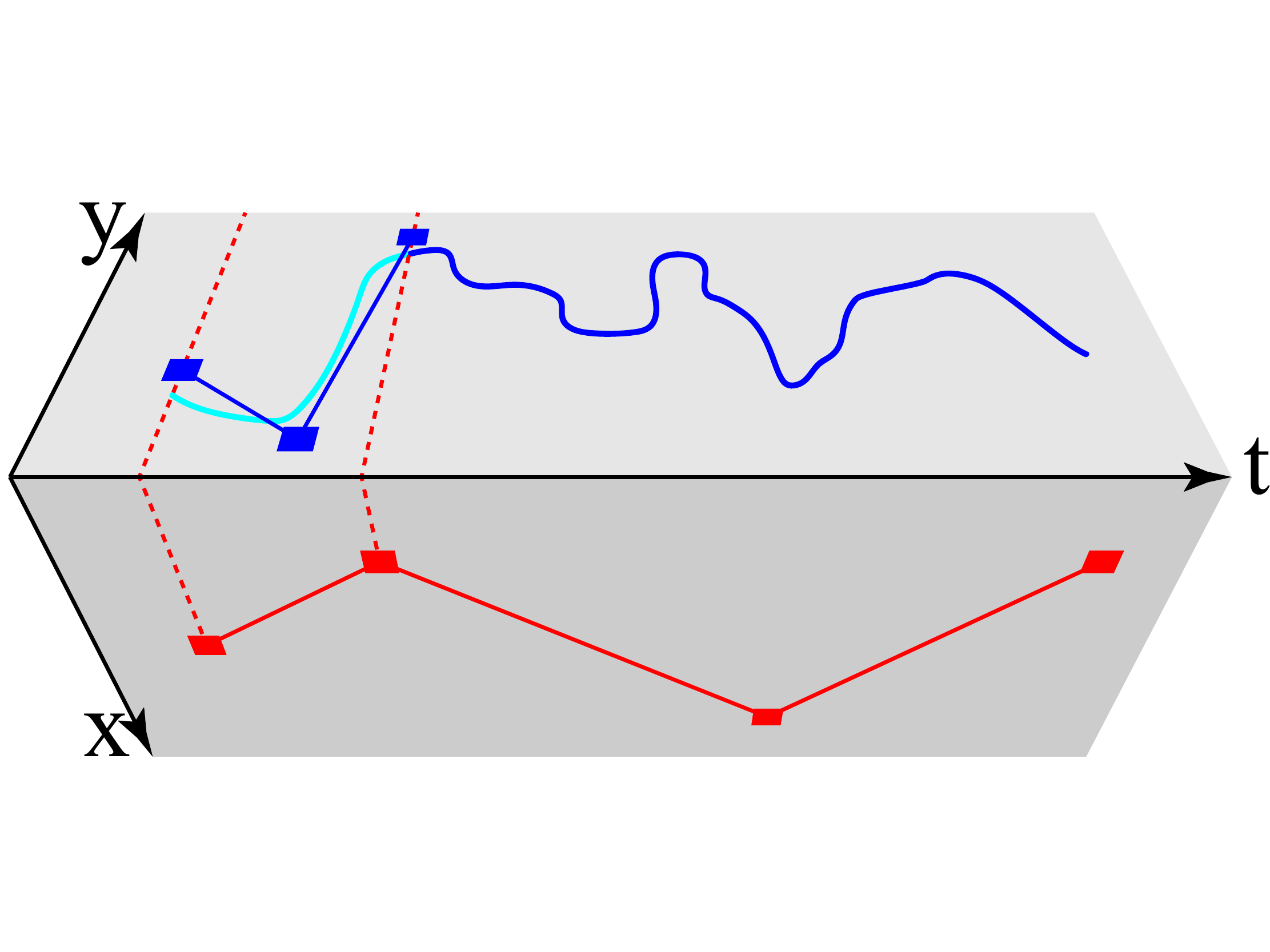}}
\subfigure[Simplify Y(t)$_{[2]}$]{
\includegraphics[trim=0 90 0 90,clip,width=0.48\linewidth]{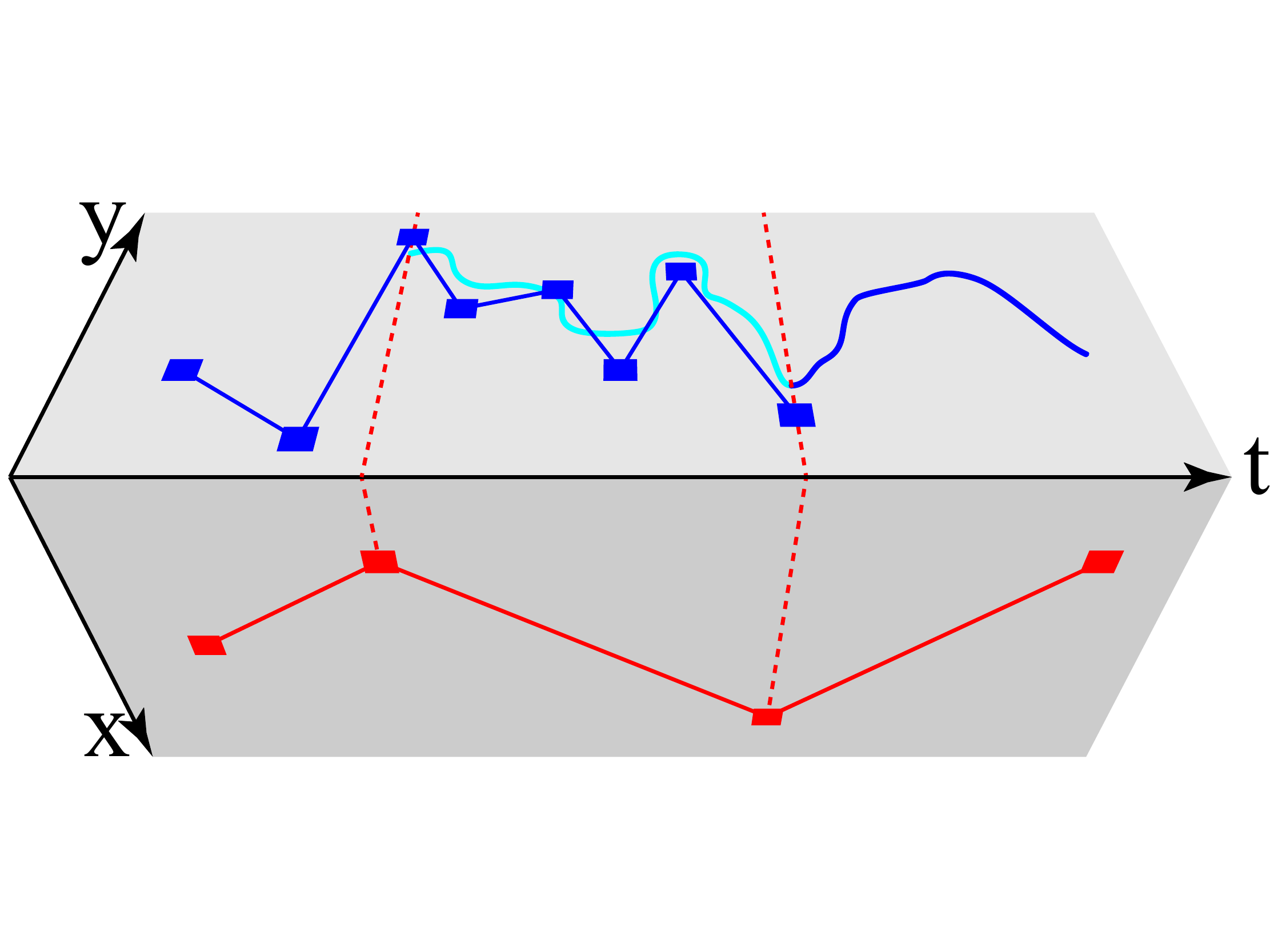}}
\subfigure[Simplify Y(t)$_{[3]}$]{
\includegraphics[trim=0 90 0 90,clip,width=0.48\linewidth]{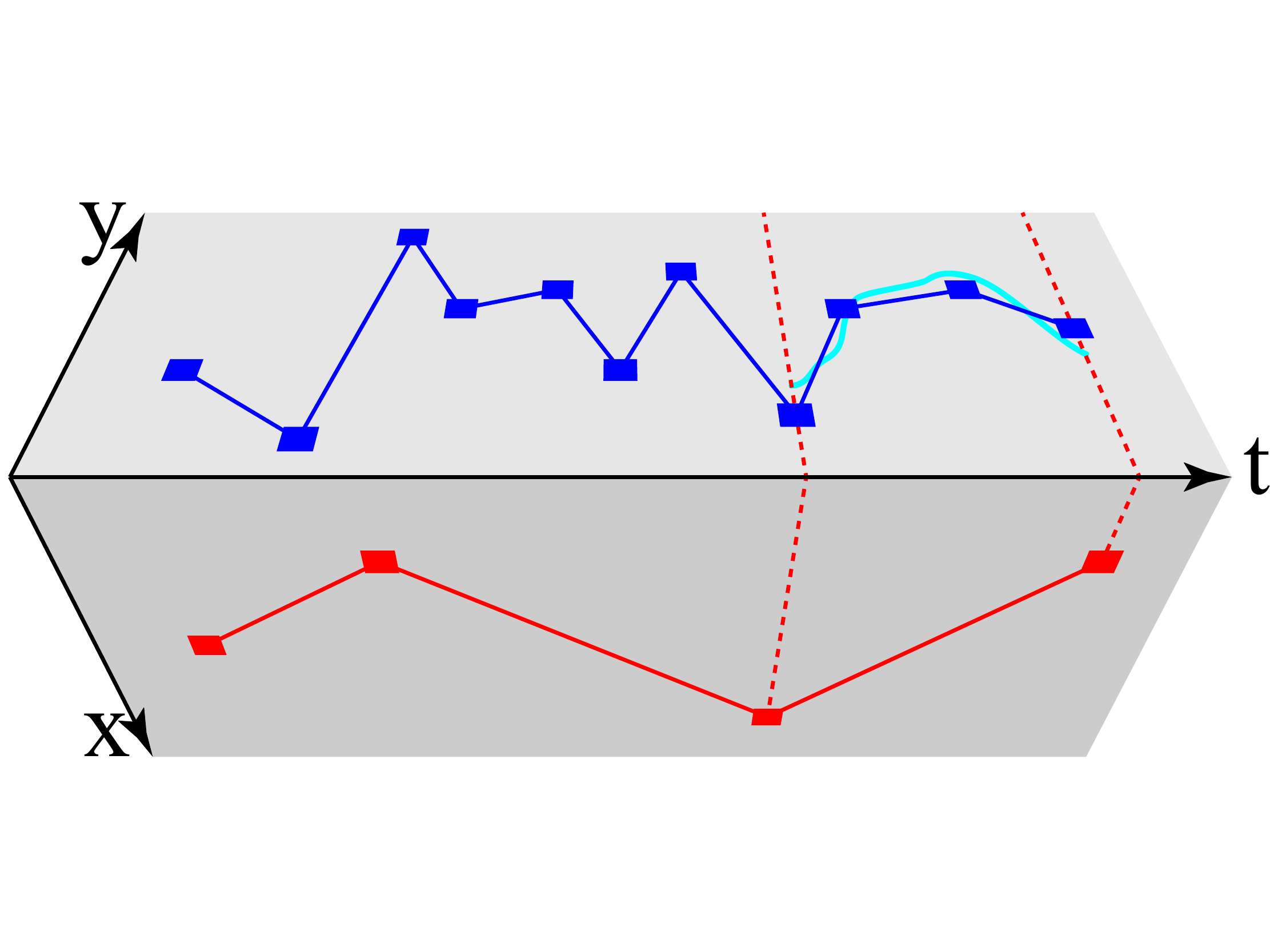}}
\subfigure[Interpolation]{
\includegraphics[trim=0 90 0 90,clip,width=0.48\linewidth]{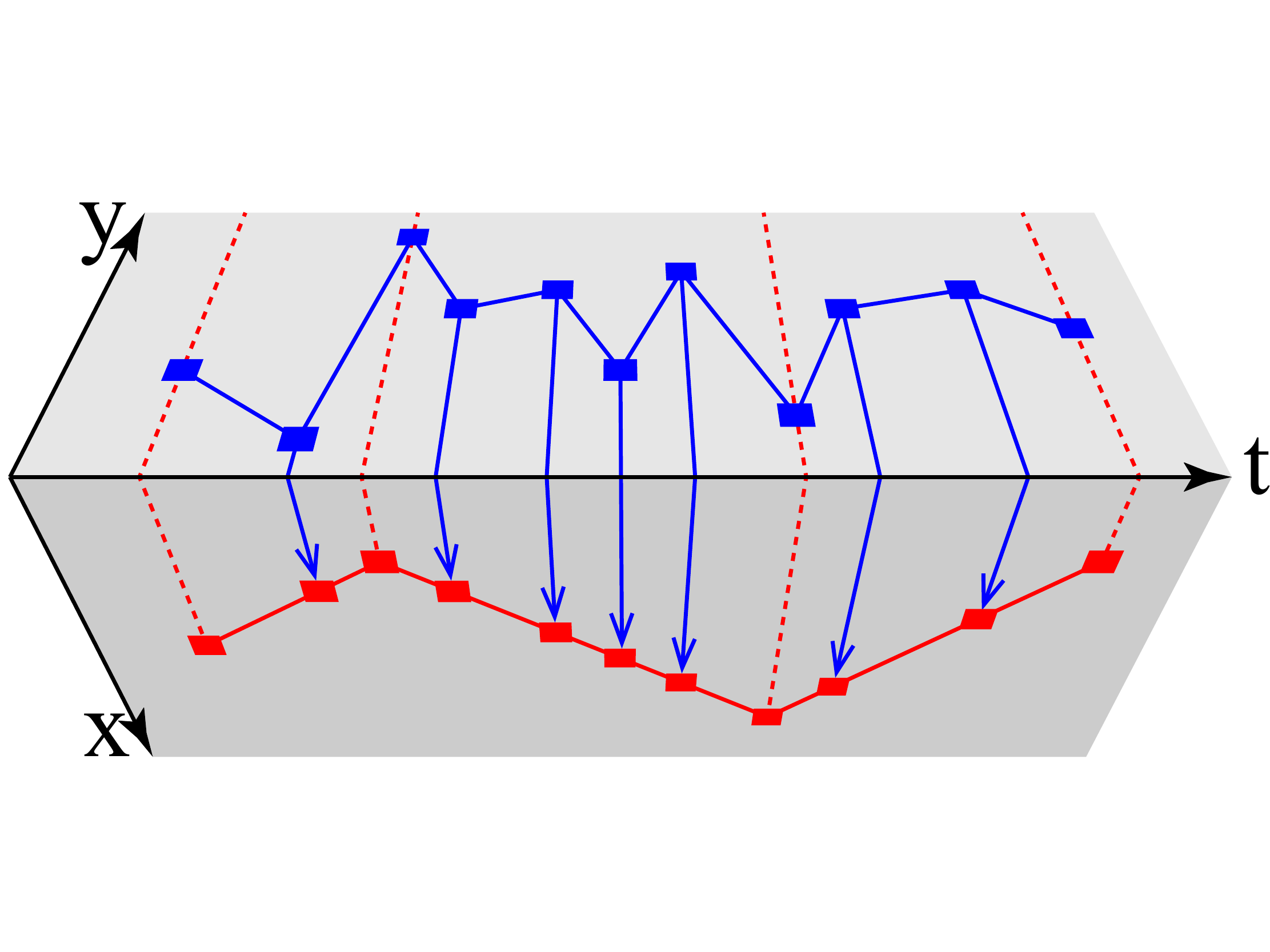}}
\captionsetup{font=bf}
\caption{Correlated interpolation}
\label{fig:i2}
\end{figure}

\begin{algorithm}
\SetKwFunction{PR}{project}
\SetKwFunction{P}{polygon}
\SetKwFunction{MLP}{link}
\SetKwFunction{IN}{interpolation}
\SetKwInOut{Input}{input}
\SetKwInOut{Output}{output}
\SetAlgoLined
\Input{input trajectory $\mathbf{T}$ and tolerance $\epsilon$}
\Output{output trajectory $\mathbf{S}$}
 \For{$i\gets 1$ \KwTo  $m$}{
 $T_i \gets$ $\mathbf{T}$.\PR{$\mathcal{P}_{i}$}\;
 \eIf{$i=1$}{
  $P_{T_1} \gets$ \P{$T_1,\epsilon$}\;
  $\mathbf{S}_1 \gets$ $P_{T_i}$.\MLP{$ T_1(\mathbf{T}.t_1)\pm\epsilon, T_1(\mathbf{T}.t_n)\pm\epsilon$}\label{alg2:lk1}
 }
 {
  $R_i\gets\varnothing$\;
  $k\gets T_i(\mathbf{T}.t_1)\pm\epsilon$\tcp*{first k is an edge}
  \For {$j\gets 2$ \KwTo $|\mathbf{S}_{i-1}|$} {\label{alg2:inner1}
    $T'_j \gets$ $T_i \cap [\mathbf{S}_{i-1}.t_{j-1},\mathbf{S}_{i-1}.t_j]$\;\label{alg2:dv}
    $P_{T'_j} \gets$ \P{$T'_j,\epsilon$}\; \label{alg2:pl}
    $S'_j \gets$ $P_{T'_j}$.\MLP{$k,T'_j(\mathbf{S}_{i-1}.t_j)\pm\epsilon$}\;\label{alg2:lk2}
    $R_i\gets R_i\cup S'_j$\;\label{alg2:cb}
    $k\gets S'_j(\mathbf{S}_{i-1}.t_j)$\tcp*{other k are points}\label{alg2:st}
  }\label{alg2:inner2}
  $\mathbf{S}_{i}\gets$\IN{$\mathbf{S}_{i-1}, R_i$}\;
 }
 }
 \Return $\mathbf{S}_m$\;
 \caption{Midpoint Correlated Interpolation (MCI)}
 \label{MI2}
\end{algorithm}

\subsubsection{Correlated Interpolation}
DI is based on optimal simplification that is more effective than any 1-dimensional algorithm $g(\cdot)$ in Section~\ref{sec:g}, so it could produce better results than existing algorithms. However, it simply gathers information from all dimensions and ignores the correlation among the projections, so it is still ineffective. If two points from different projections coincide in time, only one point should be generated in the output, e.g., at the starting and ending times. Nonetheless, DI will still output two points even if their time stamps are very close. For example, the second and third points in the simplified trajectory in Figure~\ref{fig:traj14} are redundant, because the second point in simplified $X$ and the second point in simplified $Y$ are close in time. In general, the size of the output could be up to $n\cdot m$ when $\epsilon$ is small. Consequently, further improvement is necessary.

Taking into account such correlation among projections, we propose the second implementation, namely Midpoint Correlated Interpolation (MCI), as shown in Algorithm~\ref{MI2}. Similar to the first implementation, the algorithm also generates $m$ simplifications in the planes spanned by each spatial dimension and the temporal dimension. It produces a simplification for the projection of the input trajectory in the $i$-space spanned by the first $i$ spatial dimensions in the $i$-th iteration. Eventually, the algorithm produces an output trajectory in $m$-space after $i=m$. In the first iteration, the algorithm is equivalent to the optimal algorithm in 1-space, and generates a trajectory $\mathbf{S}_1$ for the projection $T_1$ (line~\ref{alg2:lk1}). In the second iteration, it subdivides the projection $T_2$ into $|\mathbf{S}_1| - 1$ pieces according to the time stamps of $\mathbf{S}_1$ (line~\ref{alg2:dv}). Here the link distance algorithm in Section~\ref{sec:wi} is applied in the polygons subdivided by the time stamps of $\mathbf{S}_1$ (lines~\ref{alg2:pl}-\ref{alg2:lk2}). After that, the algorithm simplifies each piece separately and connects the results together to form $R_2$, a simplification for $T_2$ (line~\ref{alg2:cb}).  At the end of the iteration, it combines $\mathbf{S}_{1}$ with $R_2$ to form the new trajectory $\mathbf{S}_2$. In the third iteration, it subdivides $T_3$ according to $\mathbf{S}_2$ and then simplifies it with $R_3$. By combining $R_3$ and $\mathbf{S}_2$ it produces $\mathbf{S}_3$. The process continues until $i>m$, and the trajectory $\mathbf{S}_m$ is the simplification for $\mathbf{T}$. Note that the 1-space simplification in Algorithm~\ref{MI2} is slightly different from that in Algorithm~\ref{MI1}, because it involves the link distance between a point, i.e., $k$ in line~\ref{alg2:lk2}, and an edge, while only link distance between edges is used in Algorithm~\ref{MI1}.

For the same input trajectory in Figure~\ref{fig:traj1}, the algorithm first simplifies $X$, and the result remains the same (Figure~\ref{fig:w3}). Next, it subdivides $Y$ according to the time stamps of the simplified $X$, i.e., $t_1$, $t_3$, $t_4$ and $t_6$, as shown in Figure~\ref{fig:w7}. Although there is a window in the first polygon between $t_1$ and $t_3$, i.e., the maximum link distance in the polygon could be two, the window intersects the vertical edge at $t_3$, as shown in Figure~\ref{fig:w8}. Hence the link distance between the source and destination is only one, i.e., the simplification of the first piece of $Y$ only has two points. The simplification of the other two pieces is similar and the final results for $Y$ are shown in Figures~\ref{fig:w9} and~\ref{fig:traj21}. Note that now the final result, i.e., the combination of simplified $X$ and $Y$, only contains four points, though the simplification of $Y$ is not optimal (see Figures~\ref{fig:traj23} and~\ref{fig:traj24}). Another example of a longer trajectory is shown in Figure~\ref{fig:i2}. Note that we use curves to demonstrate polygonal curves containing lots of sample points in this example.

MCI outperforms DI because it exploits the correlation among projections. Algorithm~\ref{MI1} simplifies projections separately, while Algorithm~\ref{MI2} simplifies later projections on the basis of the current simplification $\mathbf{S}_{i-1}$. In Algorithm~\ref{MI1}, we construct the simplified trajectory incrementally, and the trajectory $\mathbf{S}_{i}$ contains all points from $\mathbf{S}_{i-1}$. No matter how we simplify $T_i$, the interpolation points at time stamps from $\mathbf{S}_{i-1}$ are always kept. Consequently, we construct the simplification of $T_i$ according to the time stamps from $\mathbf{S}_{i-1}$, and all interpolation points at these time stamps will not increase the size of $\mathbf{S}_{i}$. For example, $Y$ can be simplified with three points in Algorithm~\ref{MI1} while it is four points in Algorithm~\ref{MI2}, but the result is even better because the extra point in $Y$ and the third point in simplified $X$ coincide (see Figure~\ref{fig:traj22}). Furthermore, it makes the simplification of $T_i$ easier, as the link distance between two consecutive time stamps could be very short. For example, all simplifications of $Y$ pieces only consist of one line segment, as Figure~\ref{fig:w9} shows.

In the $j$-th piece of $T_i$, there could be multiple optimal simplifications and their last vertices form a continuous vertical line segment at $t_{j+1}$. In practice, we take the one whose last vertex is the midpoint of the line segment, and this is the reason why we call the algorithm MCI. The steps in the inner loop (lines~\ref{alg2:inner1}-\ref{alg2:inner2}) takes $O(n_j)$ time for the $j$-th piece of $T_i$ whose size is $n_j$. In total, the sum of $n_j$ is $O(n)$, so the time complexity of the second implementation is $O(mn)$ as well.

\begin{figure}
\centering
\subfigure[Projection]{
\includegraphics[trim=0 0 0 0,clip,width=0.45\linewidth]{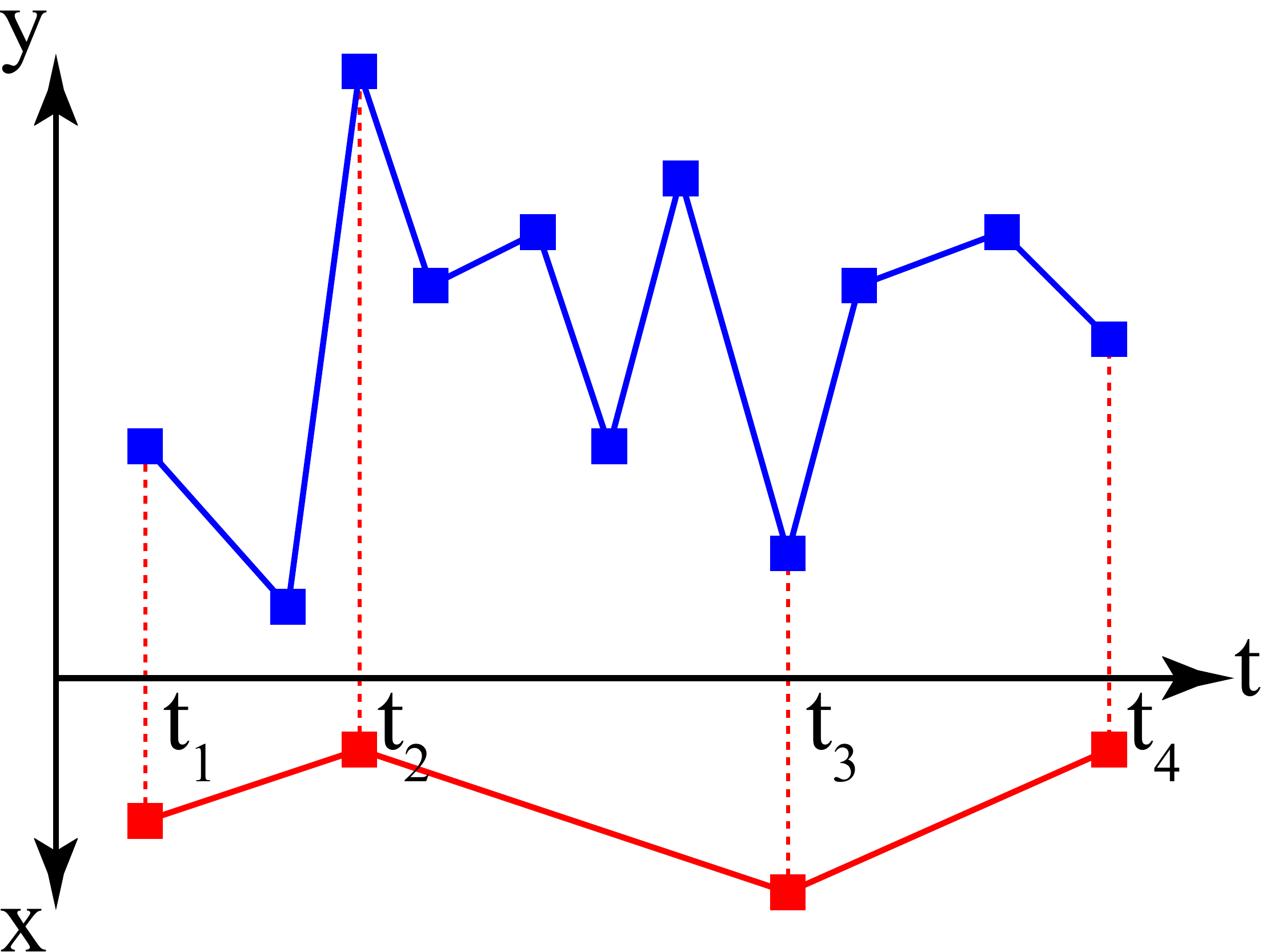}
\label{fig:i31}}
\subfigure[Link distance]{
\includegraphics[trim=0 0 0 0,clip,width=0.45\linewidth]{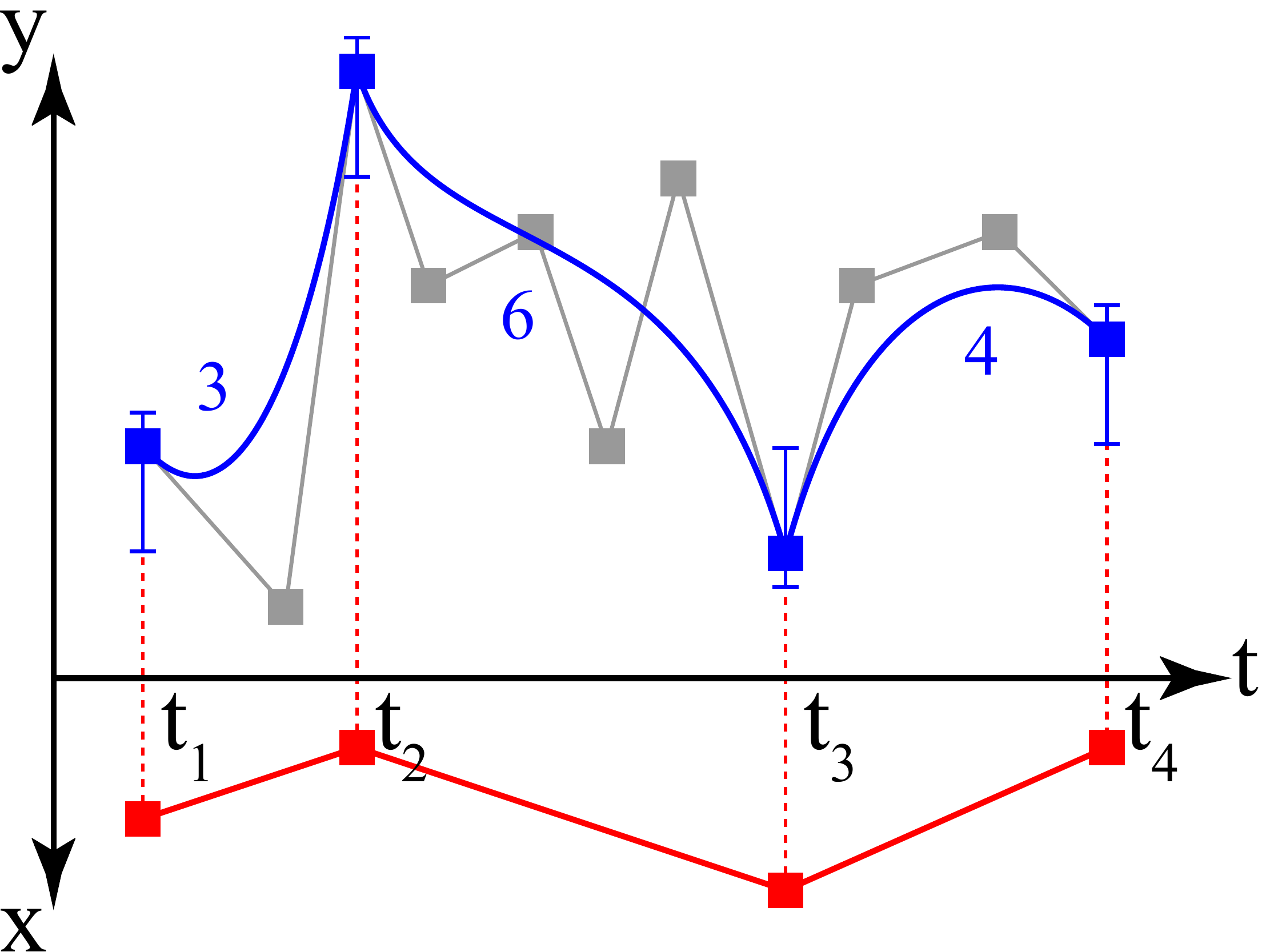}
\label{fig:i32}}
\subfigure[Another path]{
\includegraphics[trim=0 0 0 0,clip,width=0.45\linewidth]{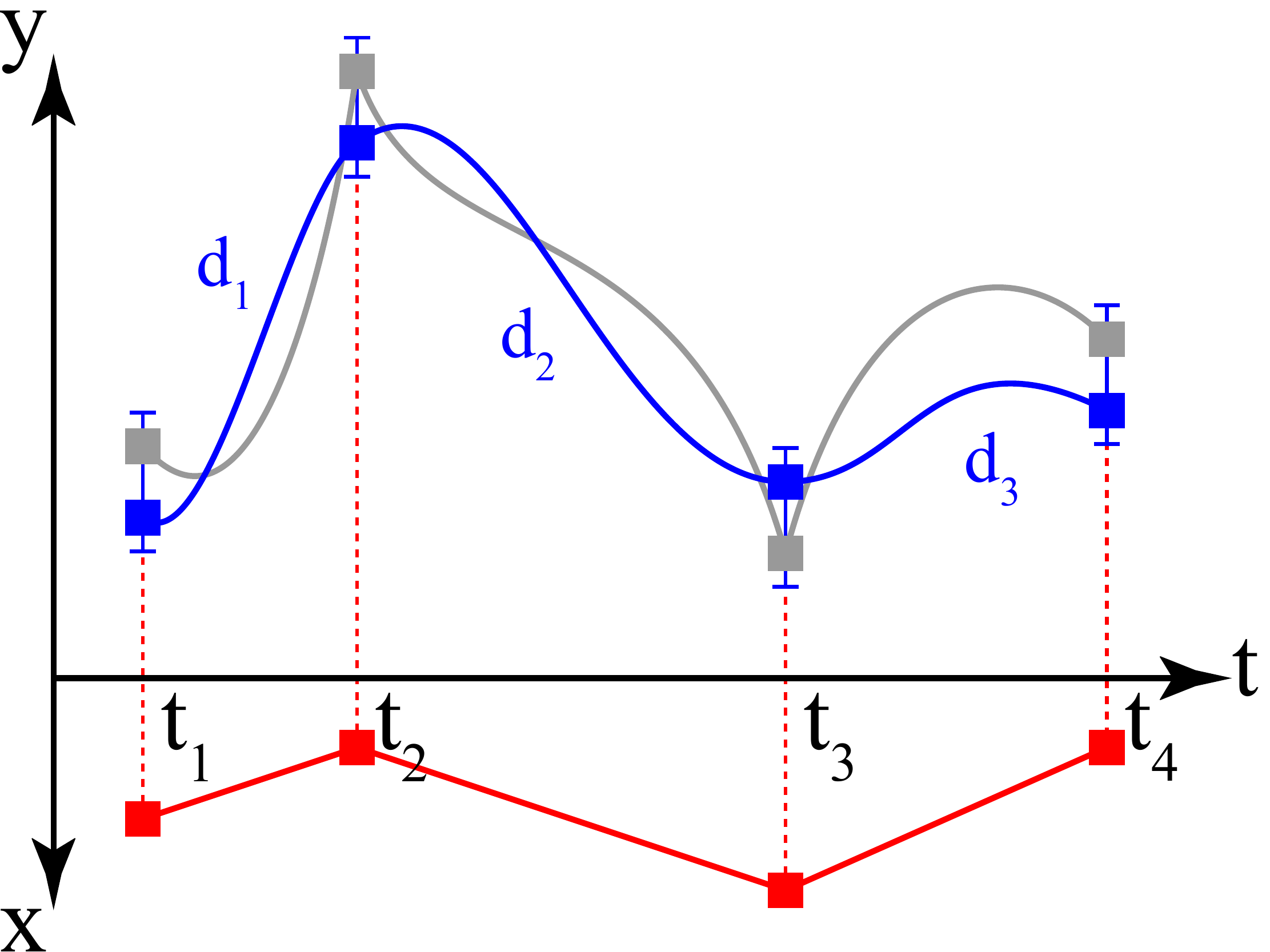}
\label{fig:i33}}
\subfigure[Link paths]{
\includegraphics[trim=0 0 0 0,clip,width=0.45\linewidth]{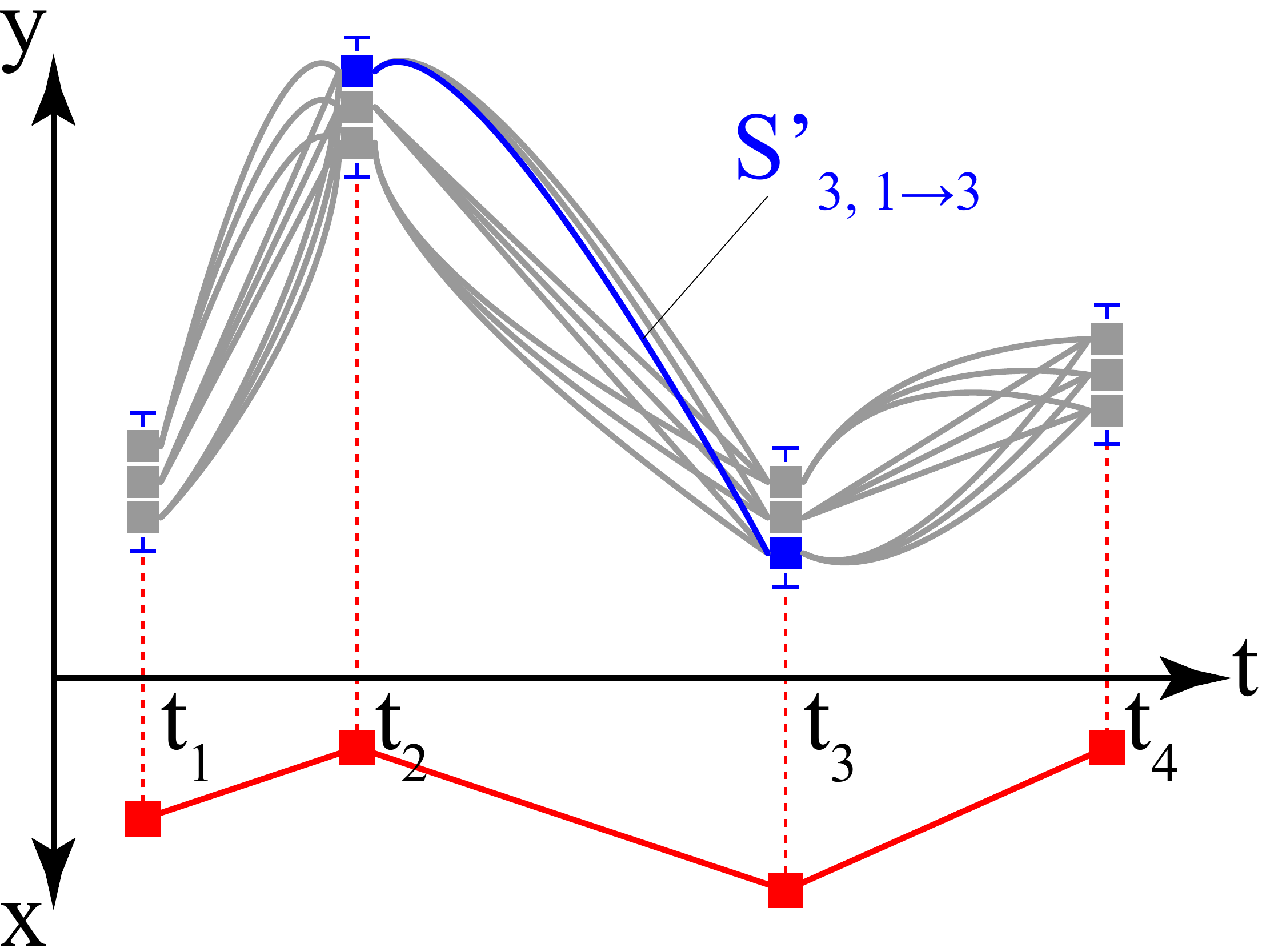}
\label{fig:i34}}
\captionsetup{font=bf}
\caption{Various interpolation}
\label{fig:i3}
\end{figure}

\begin{algorithm}
\SetKwFunction{PR}{project}
\SetKwFunction{P}{polygon}
\SetKwFunction{MLP}{link}
\SetKwFunction{IN}{interpolation}
\SetKwInOut{Input}{input}
\SetKwInOut{Output}{output}
\SetAlgoLined
\Input{input trajectory $\mathbf{T}$,\\ tolerance $\epsilon$ and sampling rate $r$}
\Output{output trajectory $\mathbf{S}$}
 \For{$i\gets 1$ \KwTo  $m$}{
 $T_i \gets$ $\mathbf{T}$.\PR{$\mathcal{P}_{i}$}\;
 \eIf{$i=1$}{
  $P_{T_1} \gets$ \P{$T_1,\epsilon$}\;
  $\mathbf{S}_1 \gets$ $P_{T_i}$.\MLP{$ T_1(\mathbf{T}.t_1)\pm\epsilon, T_1(\mathbf{T}.t_n)\pm\epsilon$}
 }
 {
 \lFor {$q \gets 0$ \KwTo $r$}
        {$v_{1,q}\gets1$}
  \For{$j\gets 2$ \KwTo $|\mathbf{S}_{i-1}|$} {
        $T'_j \gets$ $T_i \cap [\mathbf{S}_{i-1}.t_{j-1},\mathbf{S}_{i-1}.t_j]$\;\label{alg3:dv}
        $P_{T'_j} \gets$ \P{$T'_j,\epsilon$}\;\label{alg3:pl}
        \For {$q \gets 0$ \KwTo $r$} {
        \For {$p \gets 0$ \KwTo $r$} {
            $k_q \gets T_j'(\mathbf{S}_{i-1}.t_{j})+ (2q/r-1)\cdot\epsilon$\;\label{alg3:s1}
            $k_p \gets T_j'(\mathbf{S}_{i-1}.t_{j-1}) + (2p/r-1)\cdot\epsilon$\;\label{alg3:s2}
            $S'_{j,p\to q} \gets$ $P_{T'_j}$.\MLP{$k_p,k_q$}\;\label{alg3:s3}
        }
        $v_{j,q}\gets\min_{p}{\{v_{j-1,p}+|S'_{j,p\to q}| - 1\}}$\;\label{alg3:dp}
        }
  }
   $R_i\gets\varnothing$;\label{alg3:r1}
   $q\gets \argmin_{p}{v_{|\mathbf{S}_{i-1}|,p}}$\;
  \For{$j\gets |\mathbf{S}_{i-1}|$ $\mathbf{downto}$ $2$} {
  $p\gets\argmin_{p}{\{v_{j-1,p}+|S'_{j,p\to q}| - 1\}}$\;
  $R_i\gets R_i\cup S'_{j,p\to q}$;
  $q\gets p$\;
  }\label{alg3:r2}
  $\mathbf{S}_{i}\gets$\IN{$\mathbf{S}_{i-1}, R_i$}\;
 }
 }
 \Return $\mathbf{S}_m$\;
 \caption{Various Interpolation (VI)}
 \label{MI3}
\end{algorithm}

\subsubsection{Various Interpolation}
In the correlated interpolation algorithm, the simplified trajectory $S'_j$ ends with the point $k$, which becomes the starting point of $S'_{j+1}$. The point $k$ could be any one on the edge $T'_j(\mathbf{S}_{i-1}.t_j)\pm\epsilon$ as long as the local link distance inside the polygon is minimized. In practice, we take the average position of all possible $k$, i.e., the midpoint. However, such greedy strategy to select $k$ can be further improved and we should take into account all points on the vertical edges, including those resulting in non-optimal local link distance. We therefore select connecting points $k$ through a better strategy to reduce the size of $R_i$ as well as $\mathbf{S}_{i}$. In this section, we propose the third implementation of LiMITS, namely Various Interpolation (VI).

For the input trajectory in Figure~\ref{fig:i2}, the simplification in the x dimension consists of four points, while the simplification in the y dimension consists of 11 points, as shown in Figure~\ref{fig:i31}. More precisely, the projection in the y dimension is subdivided into three pieces, because the simplification in the x dimension has four points. When we apply the correlated interpolation, the link distances in the three pieces are three, six and four respectively (see Figure~\ref{fig:i32}). Nevertheless, a feasible simplification may pass through the vertical edges at any point (see Figure~\ref{fig:i33}), and the link distances could be $d_1$, $d_2$ and $d_3$ respectively. Consequently, there are an infinite number of possible positions of $k$ on each vertical edge. In order to select $k$ effectively, we propose what we call Various Interpolation (VI). We first sample $r+1$ uniformly distributed points on each vertical edge as candidates. Next, we construct $O(r^2)$ minimum link paths between candidates on starting and ending edges in each piece. After that, an optimal simplification through candidates can be computed via dynamic programming. Let $v_{j,q}$ be the minimum number of vertices to reach the $q$-th candidate at the $j$-th vertical edge, which can be computed recursively: 
\begin{equation}
v_{j,q} =
    \begin{cases}
      min_{p}{\{v_{j-1,p}+|S'_{j,p\to q}| - 1\}} & j > 1,\\
      1 & j = 1,
   \end{cases} 
   \label{eq:dp}
\end{equation}
where $|S'_{j,p\to q}|$ is the link distance between the candidates $p$ and $q$ on the consecutive vertical edges (see Figure~\ref{fig:i34}).

The pseudo-code is shown in Algorithm~\ref{MI3}. Again, we subdivide the projection into pieces and construct corresponding polygons in each iteration. However, different from the correlated interpolation, we then sample candidates (lines~\ref{alg3:s1}-\ref{alg3:s2}) and construct link paths (line~\ref{alg3:s3}). Meanwhile, we compute $v_{j,q}$ according to Equation~\ref{eq:dp} (line~\ref{alg3:dp}). 
Next, we construct an optimal simplification $R_i$ through candidates according to $v_{j,q}$ values (line~\ref{alg3:r1}-\ref{alg3:r2}). The time complexity of the algorithm is $O(r^2mn)$. In Section~\ref{experiment}, we show that a small value of $r$ is sufficient to achieve better performance than the correlated interpolation. In practice, the link distance algorithm is only applied $r+1$ times in each iteration of the inner loop and thus we do not need to perform the link distance computation between all $O(r^2)$ pairs. Furthermore, we do not need to store all link paths and only construct those that form the optimal simplification. Regardless, these minor improvements do not affect the compression ratio as well as the time complexity. For more details of the implementation, please refer to the source code available online (see Section~\ref{experiment}).

\begin{figure}
\centering
\includegraphics[trim=0 40 0 40,clip,width=0.35\textwidth]{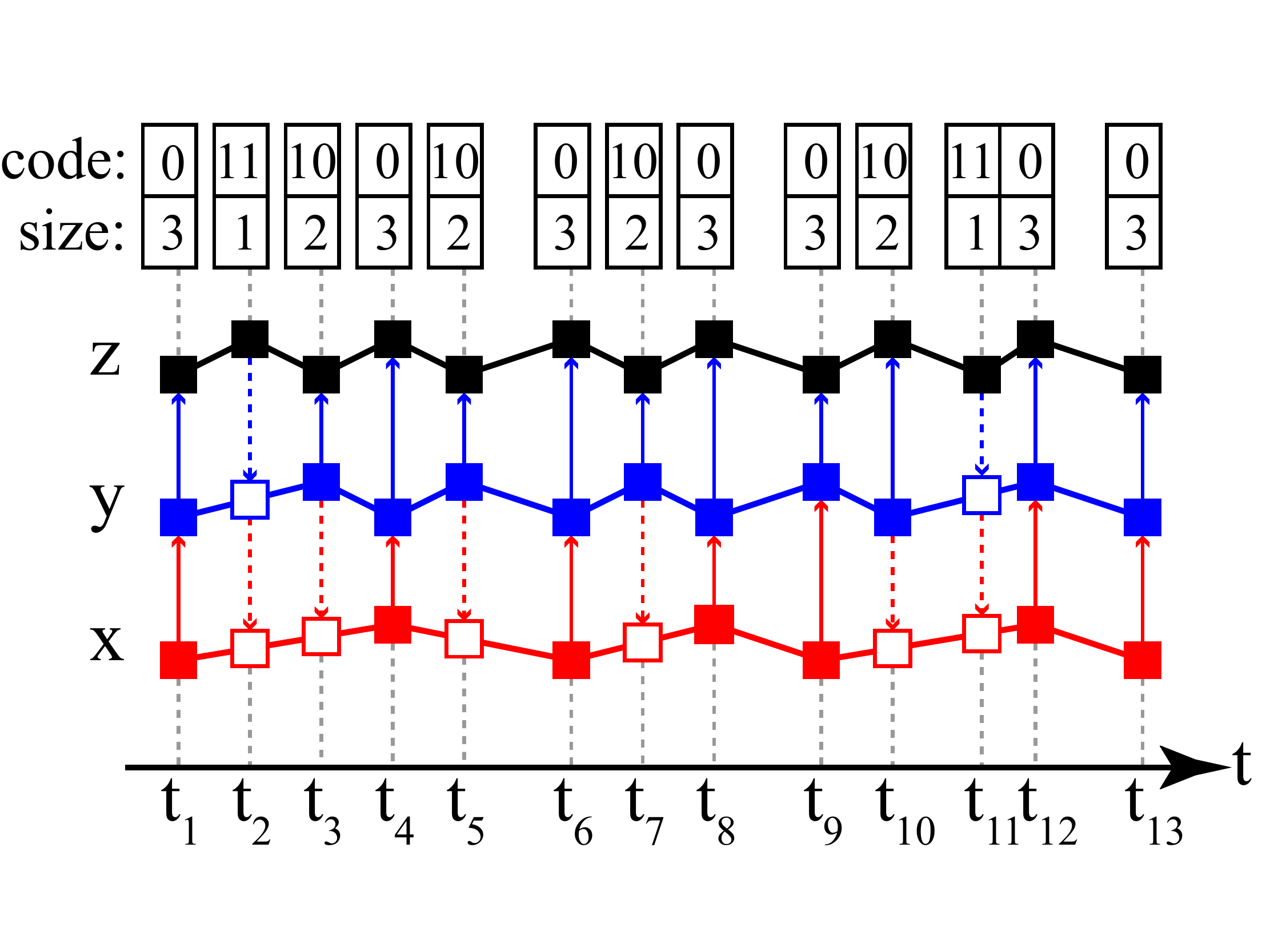}
\captionsetup{font=bf}
\caption{Compact representation}
\label{fig:cp}
\end{figure}

\subsubsection{Compact Representation}
Besides the normal format where a trajectory is represented by series of points, LiMITS also allows a compact representation which leads to even better performance. Although there are no consecutive collinear points in the results after simplification, some points could still be collinear in one or more dimensions. For example, consider the output trajectory in Figure~\ref{fig:cp}, where the first four points are collinear in the x dimension because the x-coordinate values of the second and third points are interpolation values. Similarly, the first three points are collinear in the y dimension. In the compact representation, we do not store these interpolation values so that the storage cost is reduced.

In LiMITS, an output point in $m$-space contains $0$ to $m-1$ interpolation values. Furthermore, the coordinate values of the first $k$ dimensions are interpolation values while the remaining coordinate values are not, if the point contains $k$ interpolation values. We therefore only store the coordinate values of the last $m-k$ dimensions. For example, we store all coordinate values of $\mathbf{p_1}$, only the z-coordinate value of $\mathbf{p_2}$ and both the y-coordinate value and the z-coordinate value of $\mathbf{p_3}$ for the trajectory in Figure~\ref{fig:cp}. As a result, the number of coordinate values in a point varies in the compact format while it is fixed in the normal format. Consequently, we append an extra codeword to each point to indicate the number of coordinate values, where the an entropy encoding like Huffman coding~\cite{huffman1952method} may be used.

For example, 2-bit codewords are sufficient for trajectories in 3-space. In the experiments, we use the codewords $\underline{11}$, $\underline{10}$ and $\underline{0}$ to indicate one, two and three coordinate values respectively, because the frequencies of points containing all three coordinate values are the highest. The compact representation of the example trajectory in Figure~\ref{fig:cp} is 
$
    \langle \underline{0x_1y_1z_1t_1},\underline{11z_2t_2},\underline{10y_3z_3t_3},\cdots,\underline{0x_{13}y_{13}z_{13}t_{13}}\rangle
$,
where $x_i$, $y_i$, $z_i$ and $t_i$ are floating point numbers which require 64 bits if double precision is chosen. There are $7$ points containing all three coordinate values, $4$ points containing two and $2$ points containing one, so the size of the compact representation is $7\times(1+4\times64) + 4 \times(2+3\times64) + 2\times(2+2\times64) = 2,835$ bits, while that of the normal representation is $13 \times4\times 64 = 3,328$ bits.

The conversion between compact and normal representations is lossless and only takes linear time. In particular, for each point, we first read the codeword to obtain its size, and then read the corresponding number of floating-point numbers. This process continues until we have decoded all points in the compact representation. After decoding all points, we compute the remaining coordinate values through interpolation. Note that the compact representation is not applicable to other conventional algorithms, because the colinearity in the individual dimensions is not guaranteed.

\section{Experiments}
\label{experiment}
In this section, we compare LiMITS with existing algorithms using real and generated data, in terms of both effectiveness and efficiency. All algorithms are implemented in Java and run on a computer with Intel Xeon CPU E5-2620 @ 2.10GHz and 64GB memory.
We select OPT~\cite{IMAI198631} and CIS~\cite{Lin2019} as the representatives of strong simplification, and CIW~\cite{Lin2019} as the representative of weak simplification, because their compression ratios outperform other algorithms~\cite{zhang2018trajectory,Lin2019}. RDP~\cite{douglas1973algorithms} is a classic algorithm for the problem so we also implemented it as a baseline. The source code is publicly available online.\footnote{\url{https://github.com/yhhan19/LiMITS}}

\begin{table}
\centering
\captionsetup{font=bf}
\caption{Dataset information}
\begin{tabular}{  |c|c|c|c|c| }
\hline
 Dataset & Source & Records & Points & Dimension  \\
  \hline
   Wiener & Random & 100 & 200,000 & 2D/3D   \\
 \hline
  Bee & Animal & 876 & 2,523,180 & 2D \\  
\cline{1-1}\cline{3-5}
   Danio&  & 42 & 64,586 & 3D   \\
  \hline
 Beijing & Human & 18,655 & 24,106,047 & 2D/3D \\  
\cline{1-1}\cline{3-5}
   Mopsi & & 6,779 & 7,612,499 & 2D/3D   \\
 \hline
\end{tabular}
\label{fig:rdi}
\end{table}

\subsection{Datasets}
In order to evaluate the performance of algorithms, we use following trajectory datasets in our experiments: Beijing (Geolife)~\cite{zheng2010geolife}, Mopsi~\cite{2017mopsi}, Bee~\cite{bozek2018towards}, Danio~\cite{stienessen2013effect} and Wiener~\cite{mandrekar1995mathematical}. The detailed information including the numbers of trajectories and sample points is given in Table~\ref{fig:rdi}.

The first two datasets consist of trajectories collected by human users. These human trajectories record wide ranges of activities, such as walking, hiking and traveling via vehicles. A sample point in the human trajectory data usually consists of a latitude, a longitude, an altitude and a time stamp. The majority of the human trajectories contain valid altitudes and hence are used as 3D data. We also use these trajectories as 2D data by ignoring the altitudes. The other two datasets from animals are collected from video data through object tracking techniques. The bees are on a planar bee hive so their trajectories are 2D, while the danios are in a fish tank so the trajectories are 3D. Besides the real datasets, we also generated random trajectories simulating Brownian motion according to the standard Wiener process~\cite{mandrekar1995mathematical}.

\begin{figure*}
\centering
\subfigure[Beijing 2D]{
\includegraphics[width=0.24\textwidth]{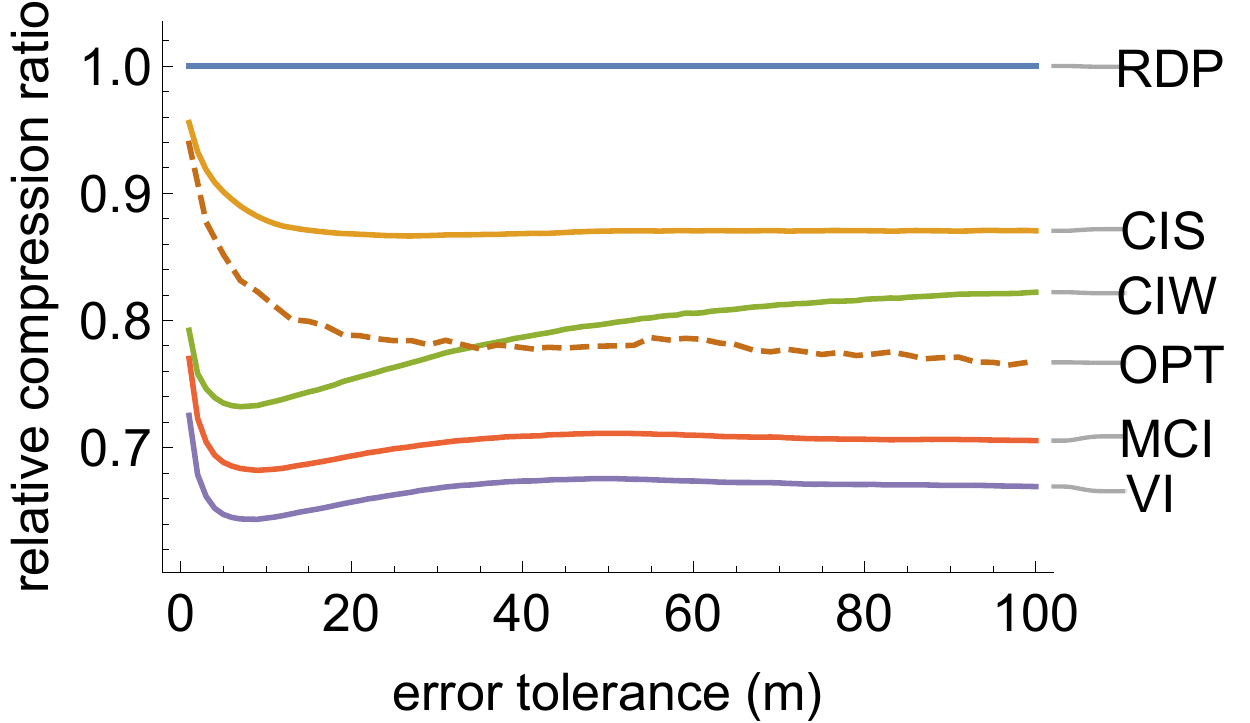}
\label{fig:effectrbj2}}
\subfigure[Beijing 3D]{
\includegraphics[width=0.24\textwidth]{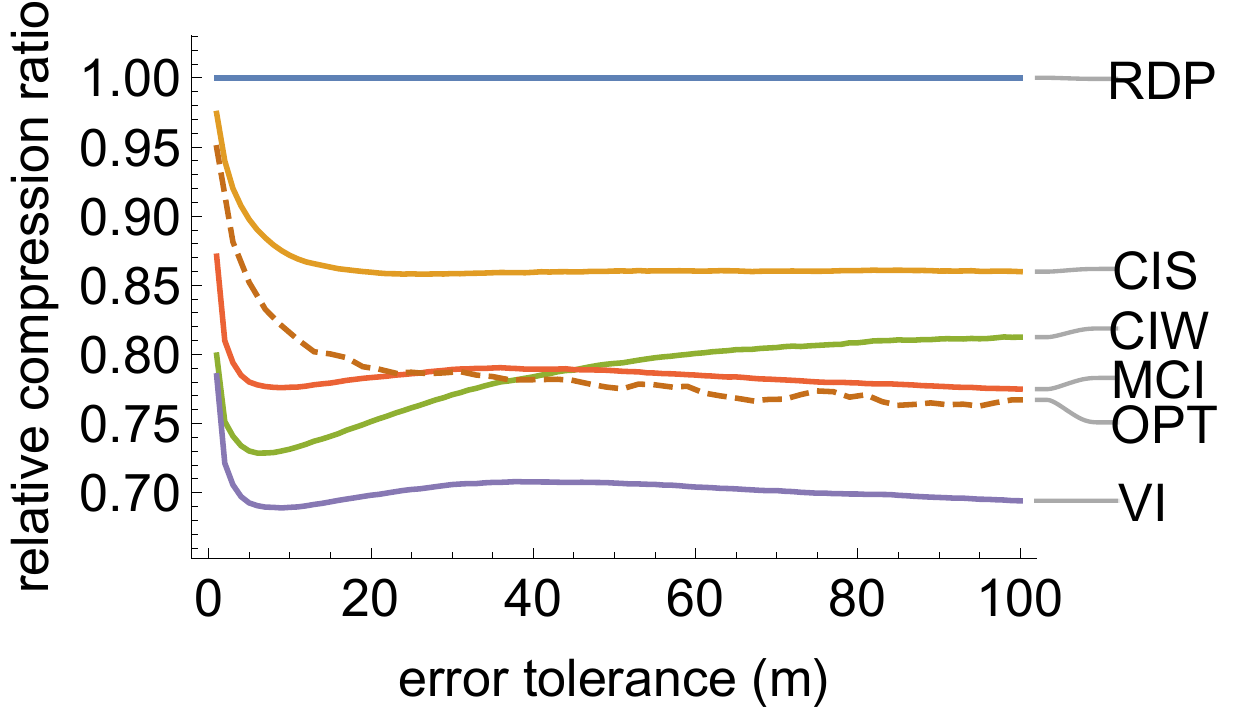}
\label{fig:effectrbj3}}
\subfigure[Mopsi 2D]{
\includegraphics[width=0.24\textwidth]{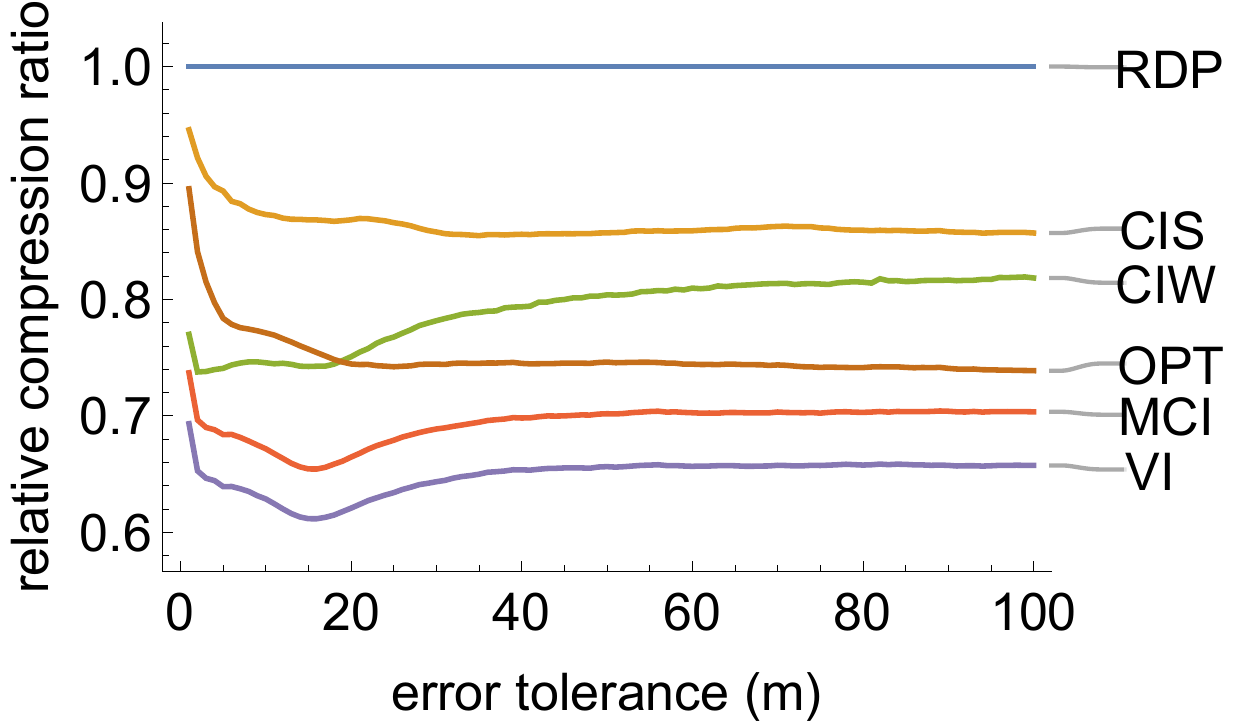}}
\subfigure[Mopsi 3D]{
\includegraphics[width=0.24\textwidth]{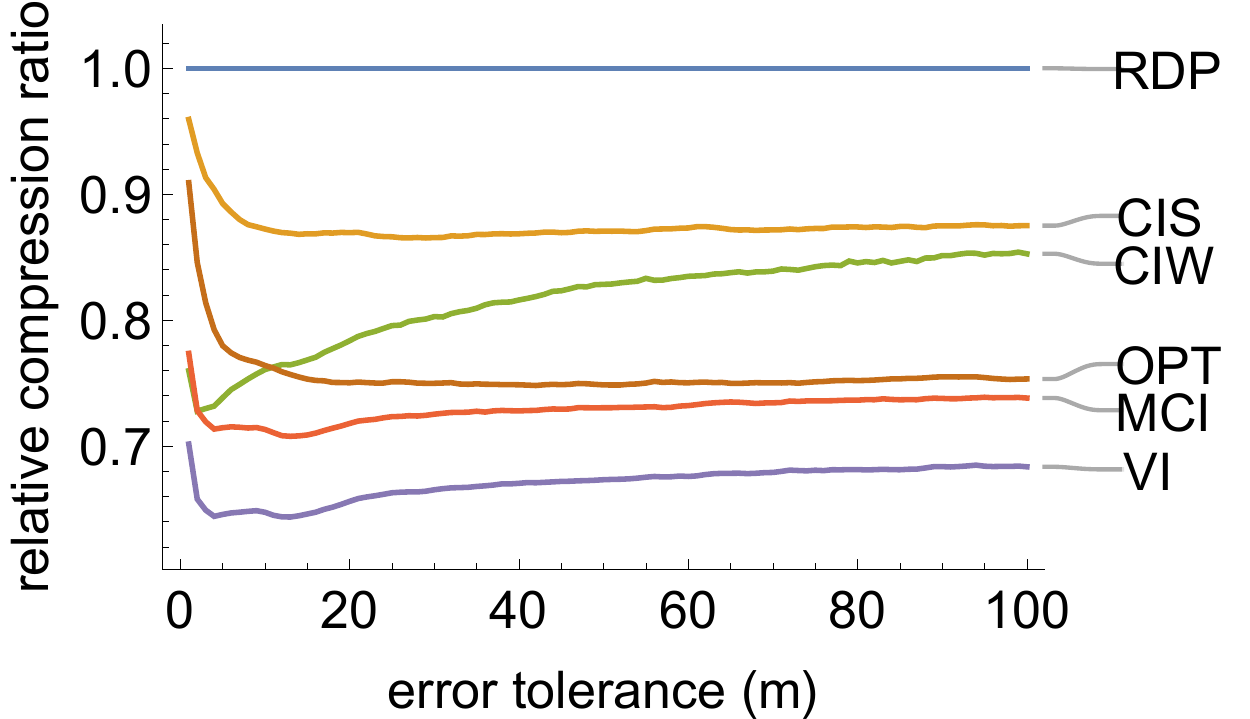}}
\subfigure[Bee 2D]{
\includegraphics[width=0.24\textwidth]{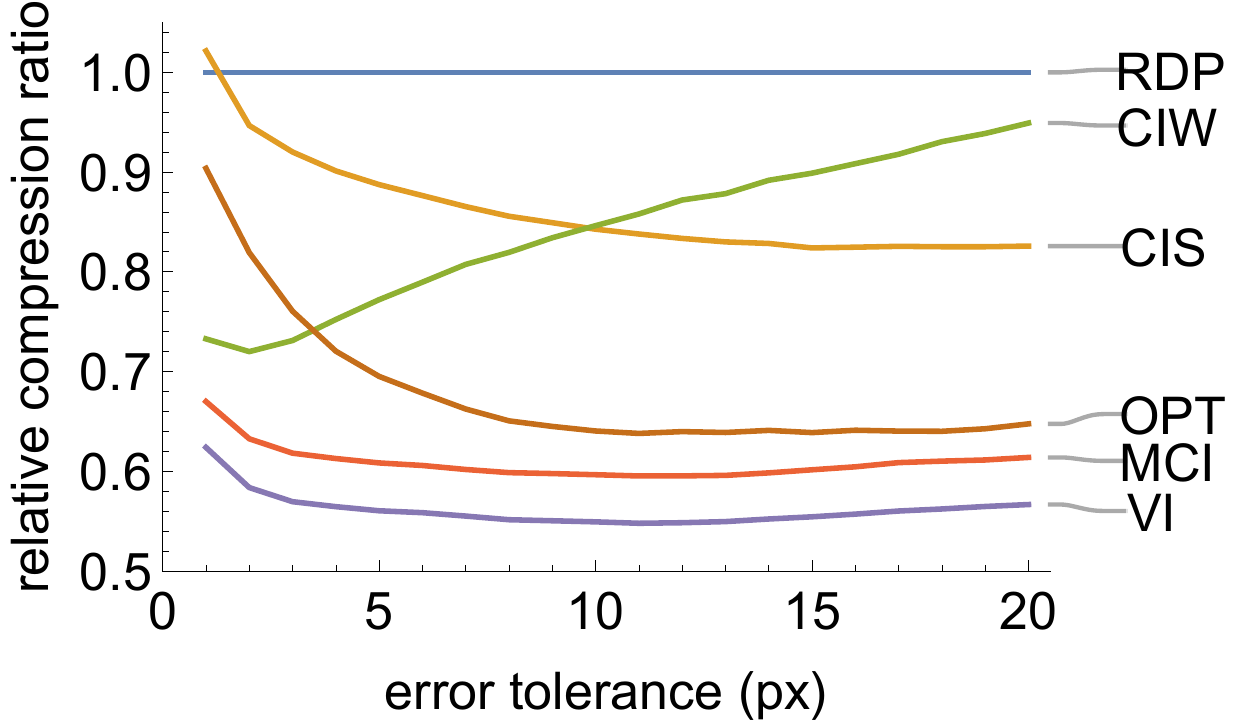}}
\subfigure[Danio 3D]{
\includegraphics[width=0.24\textwidth]{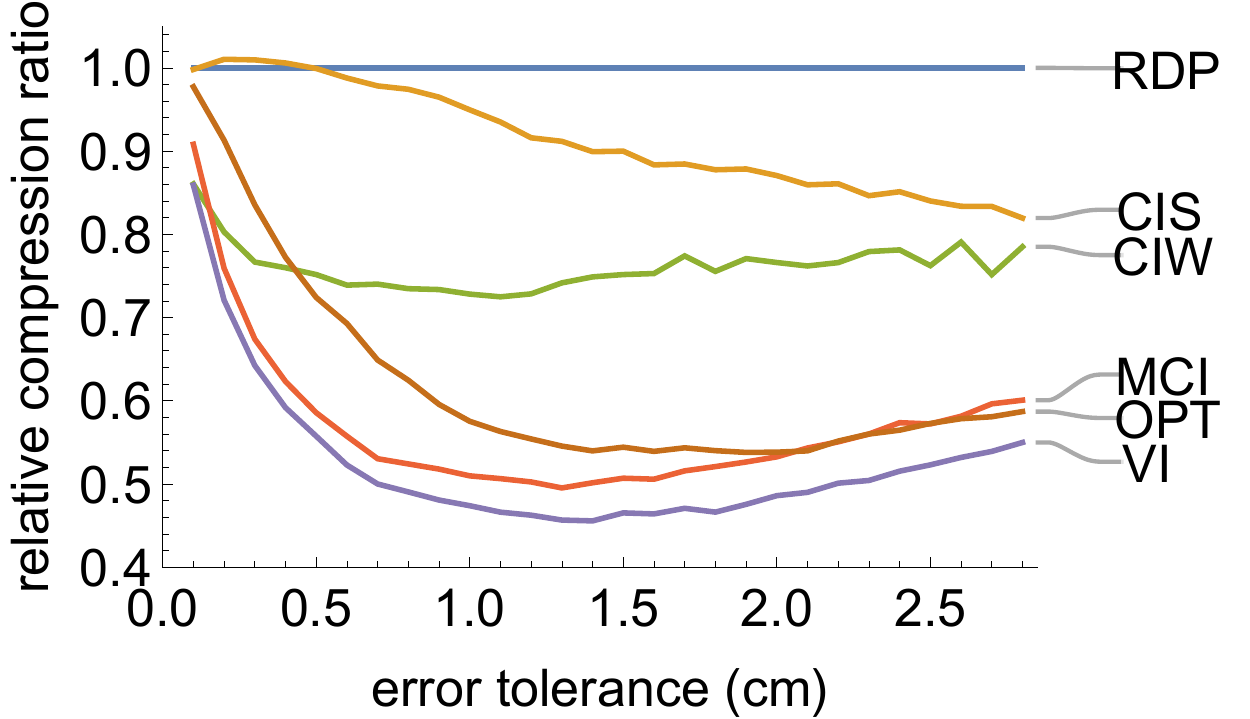}}
\subfigure[Wiener 2D]{
\includegraphics[width=0.24\textwidth]{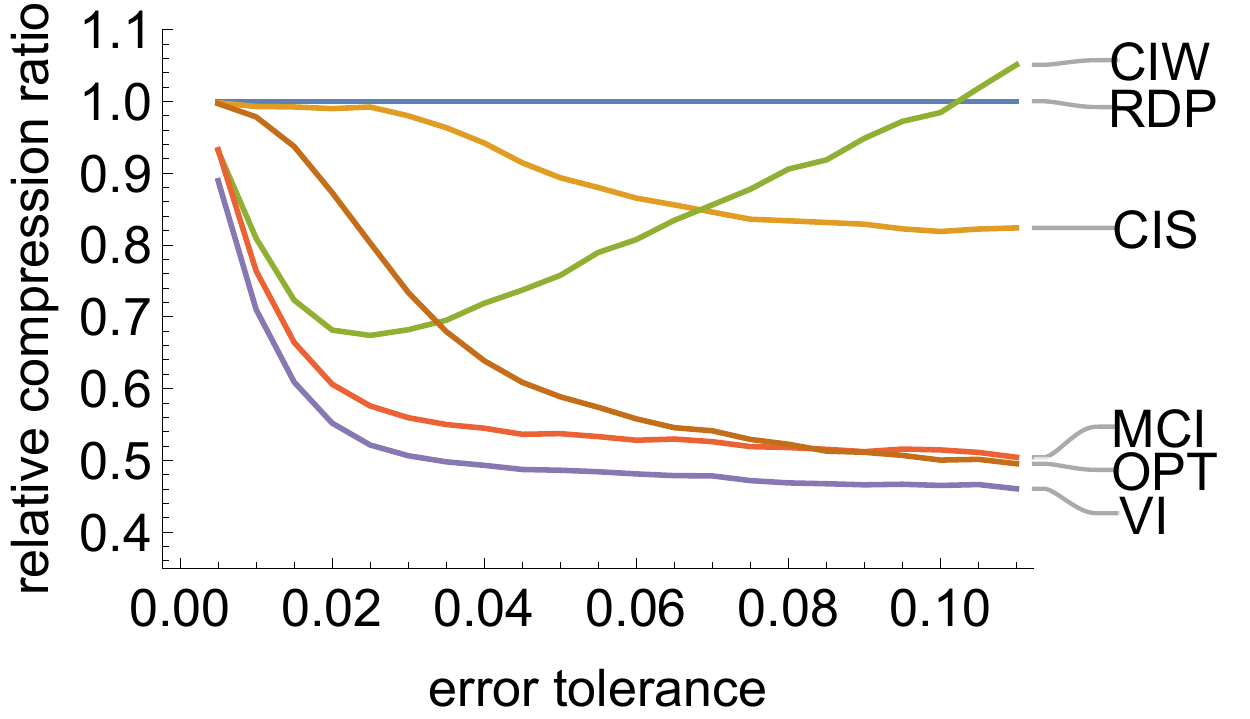}}
\subfigure[Wiener 3D]{
\includegraphics[width=0.24\textwidth]{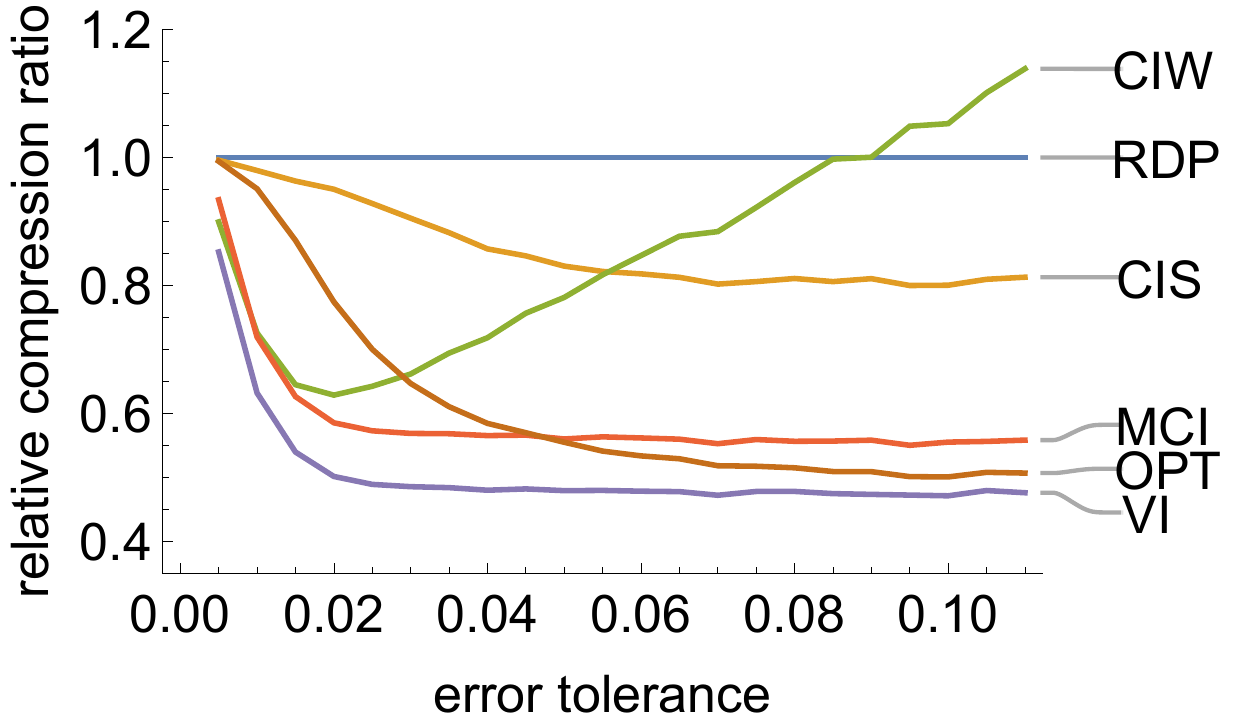}}
\captionsetup{font=bf}
\caption{Relative compression ratio}
\label{fig:effectr}
\end{figure*}
\begin{figure*}
\centering
\subfigure[Beijing 2D]{
\includegraphics[width=0.24\textwidth]{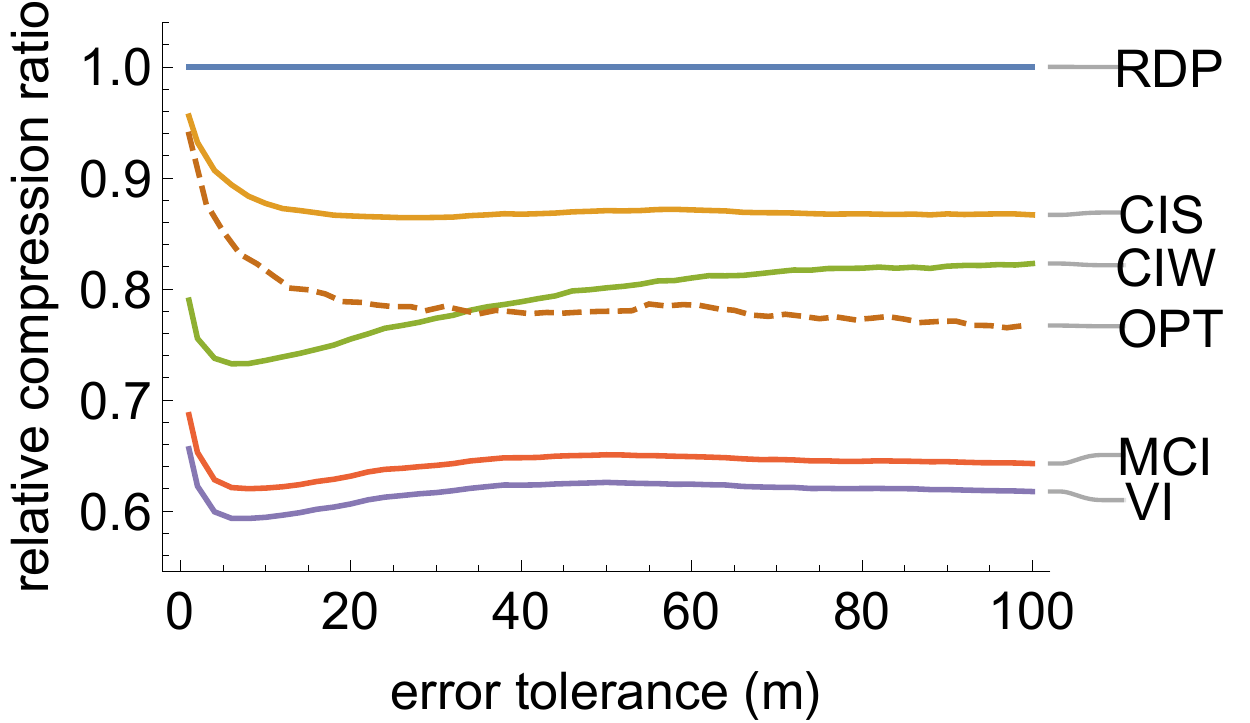}}
\subfigure[Beijing 3D]{
\includegraphics[width=0.24\textwidth]{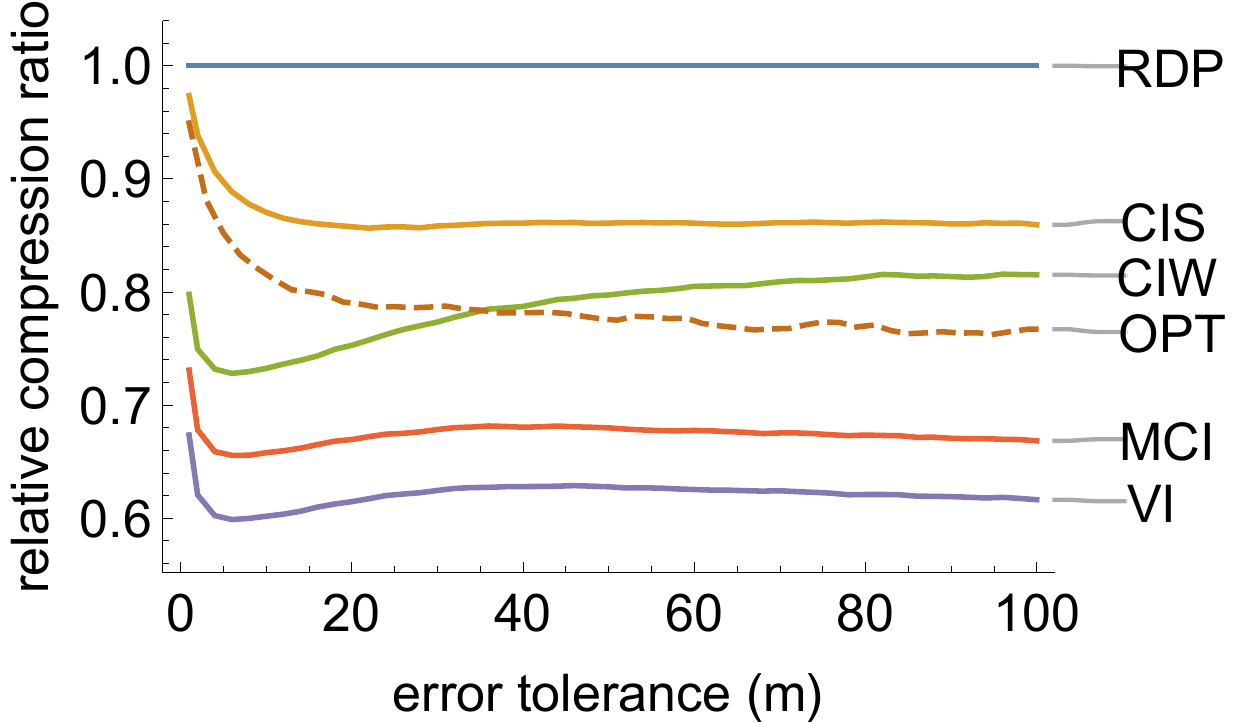}}
\subfigure[Mopsi 2D]{
\includegraphics[width=0.24\textwidth]{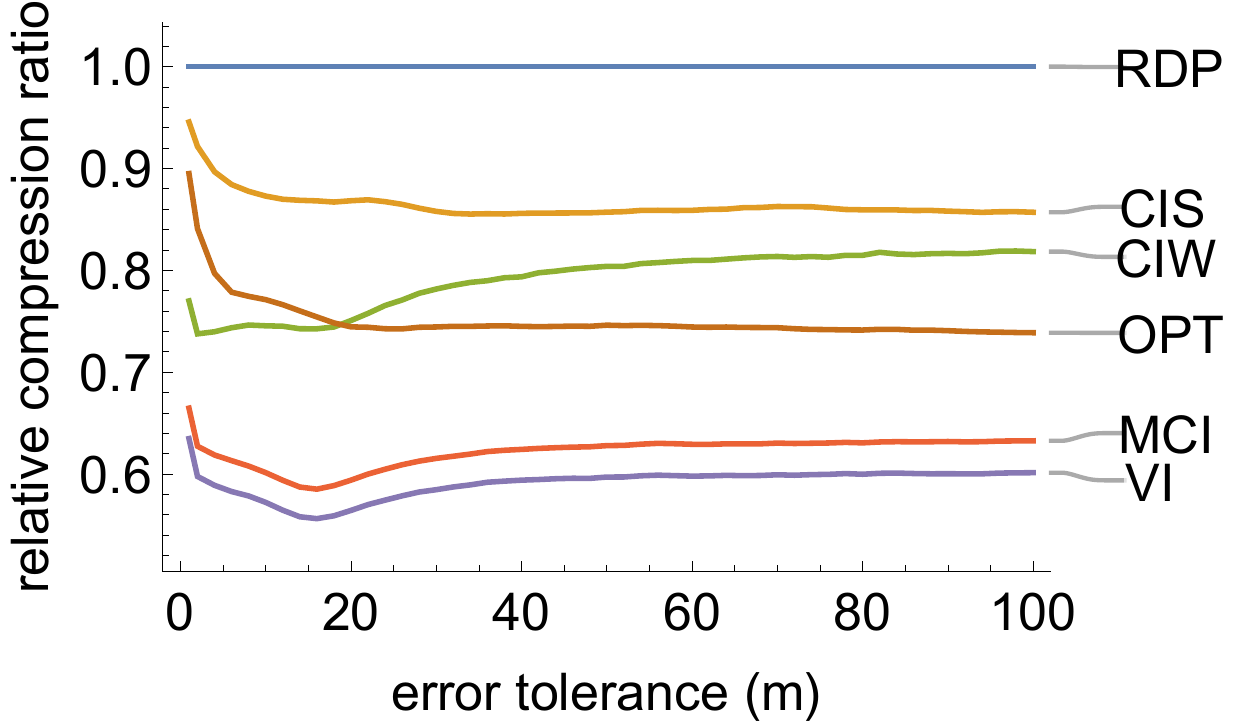}}
\subfigure[Mopsi 3D]{
\includegraphics[width=0.24\textwidth]{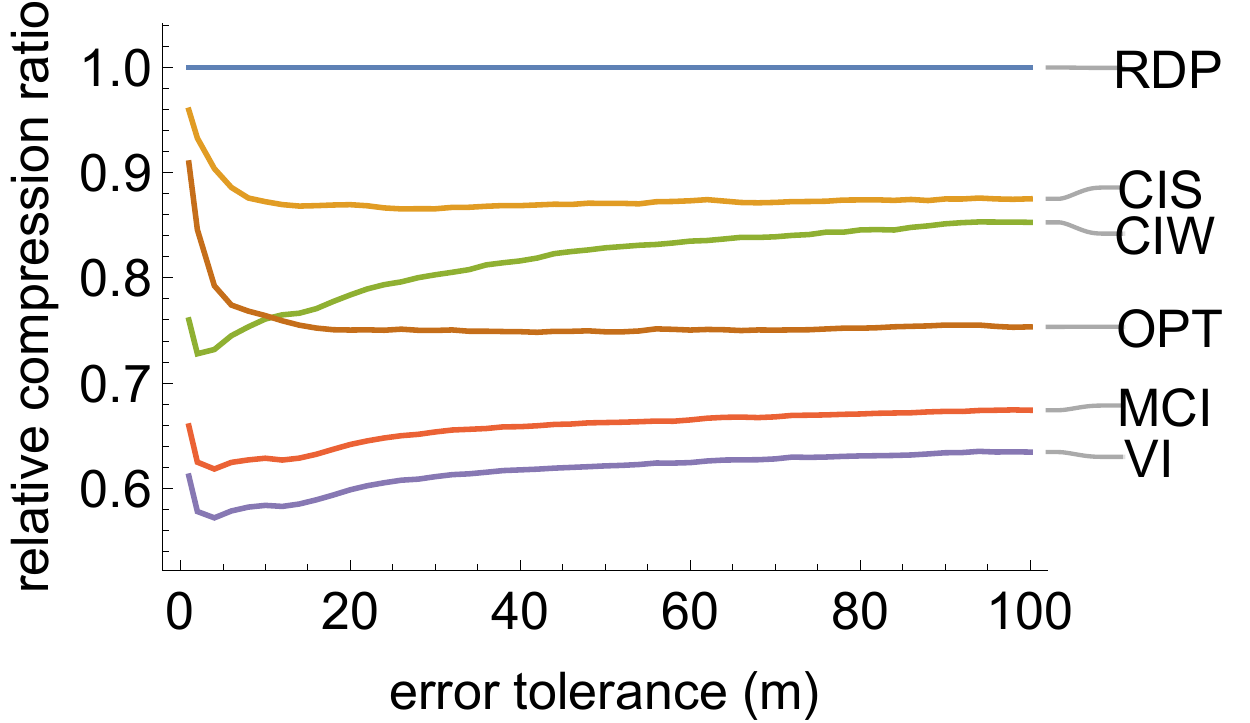}}
\subfigure[Bee 2D]{
\includegraphics[width=0.24\textwidth]{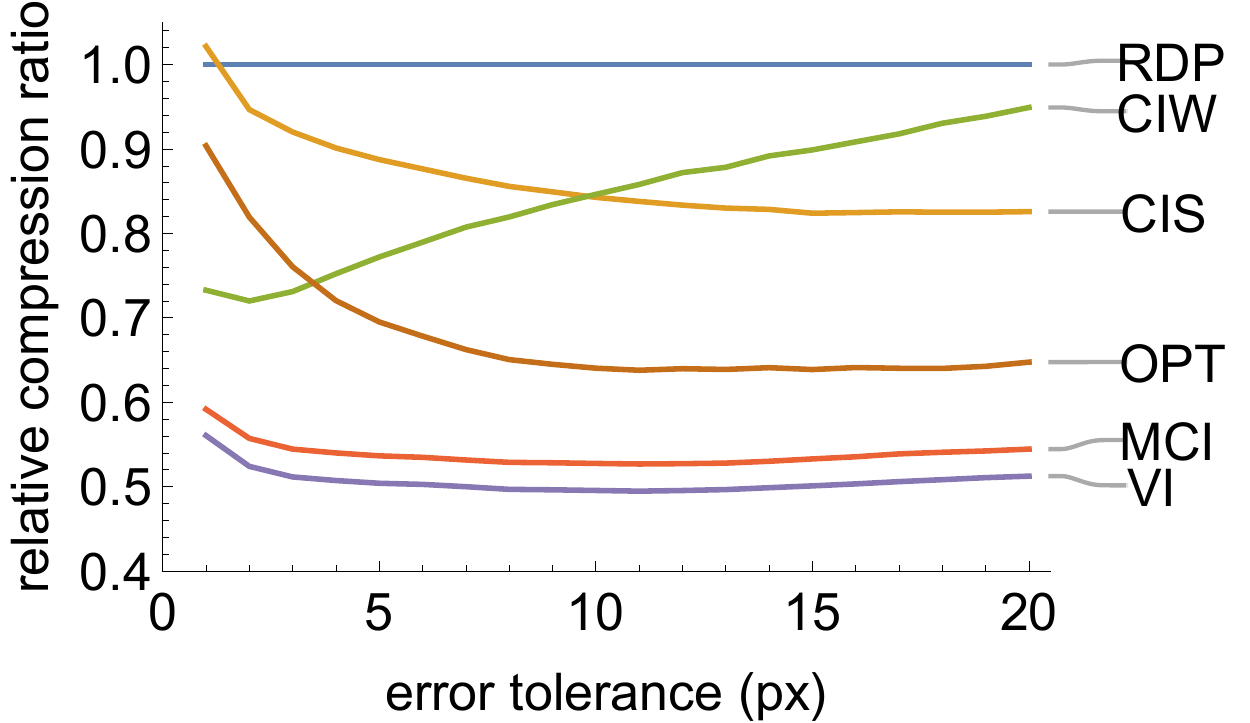}}
\subfigure[Danio 3D]{
\includegraphics[width=0.24\textwidth]{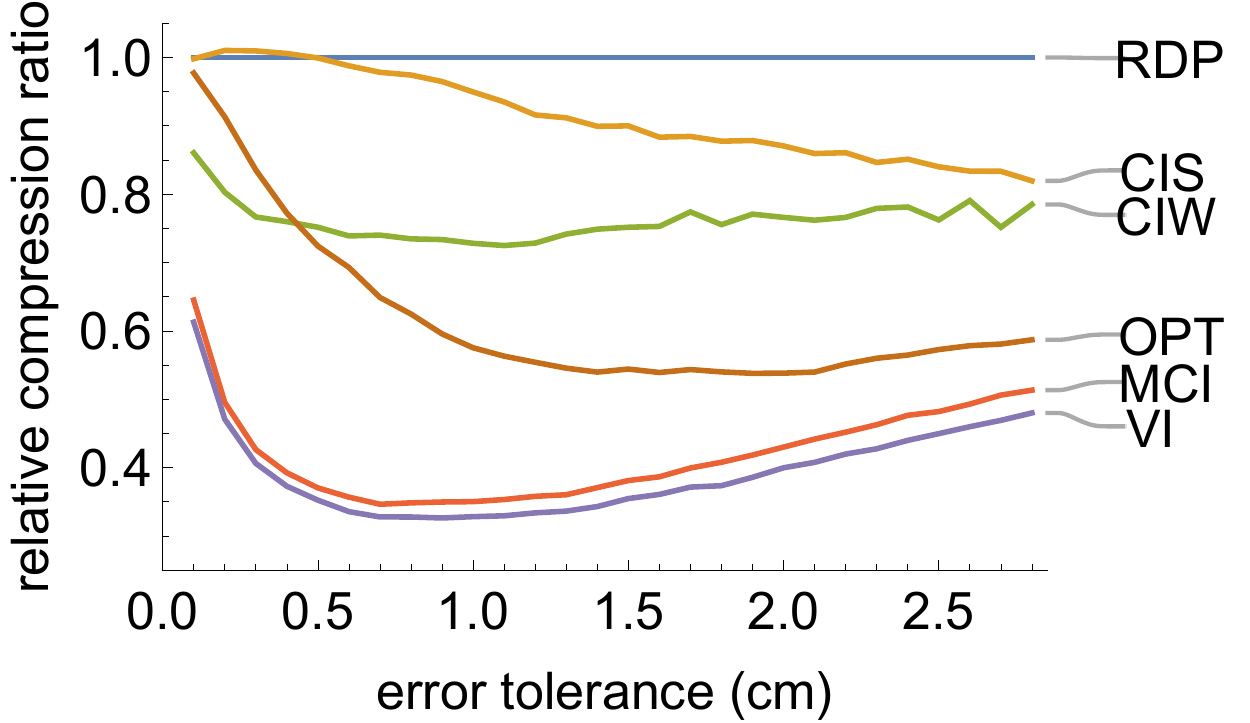}}
\subfigure[Wiener 2D]{
\includegraphics[width=0.24\textwidth]{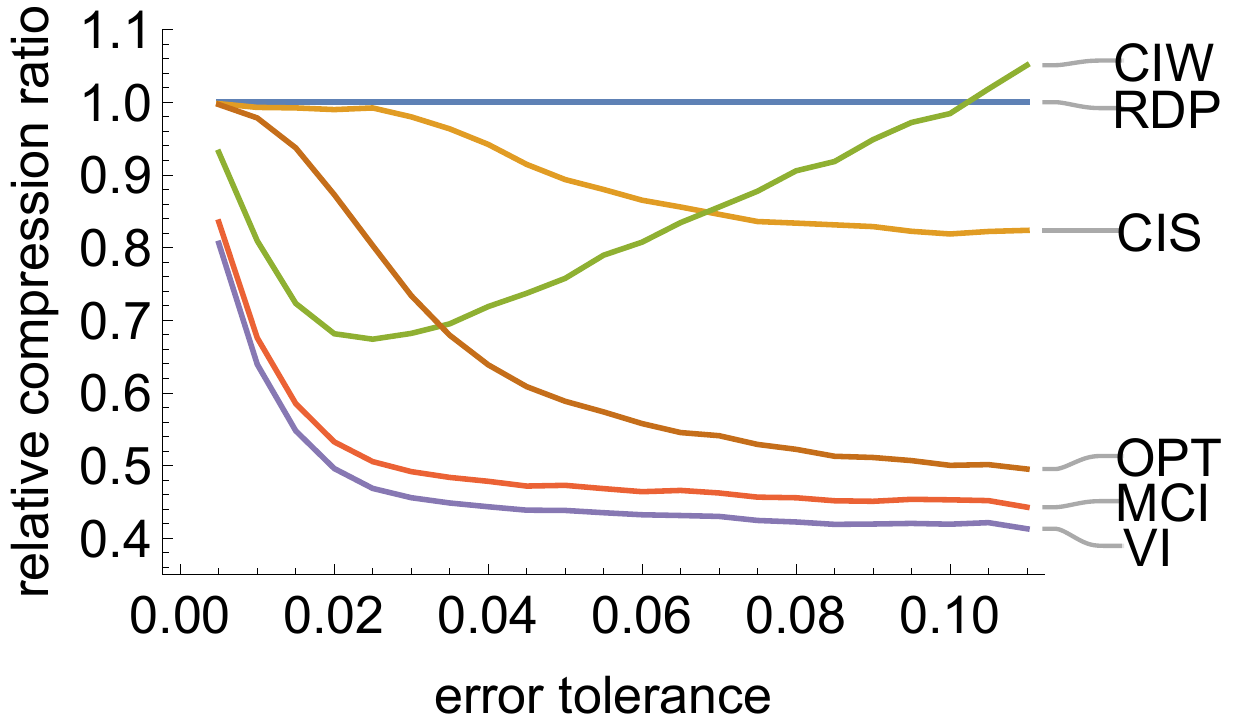}}
\subfigure[Wiener 3D]{
\includegraphics[width=0.24\textwidth]{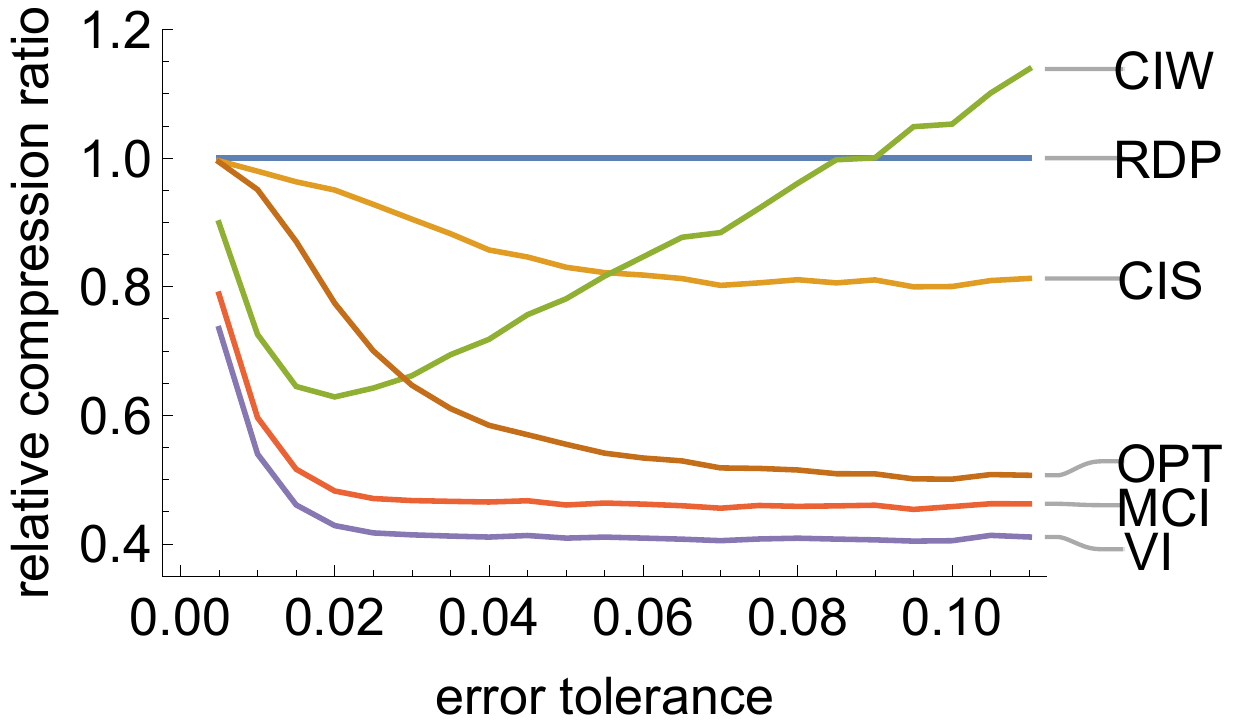}}
\captionsetup{font=bf}
\caption{Compact representation}
\label{fig:effiopt}
\end{figure*}
\begin{figure*}
\centering
\subfigure[Beijing 2D]{
\includegraphics[width=0.24\textwidth]{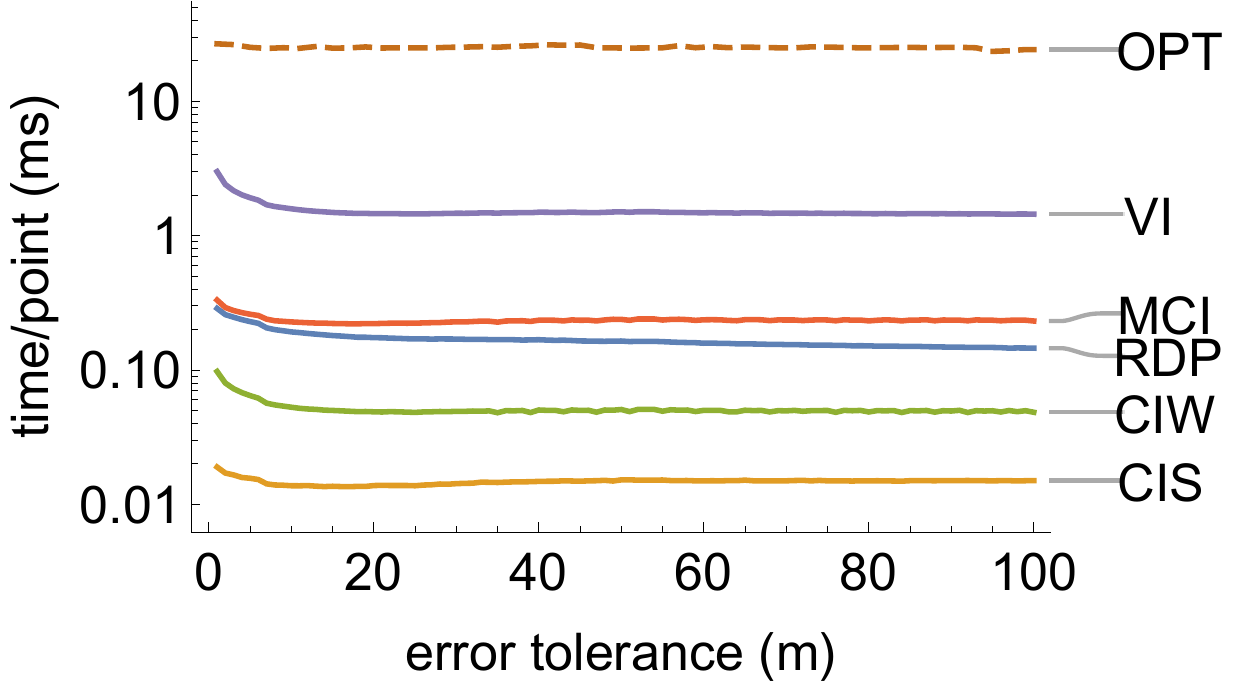}}
\subfigure[Beijing 3D]{
\includegraphics[width=0.24\textwidth]{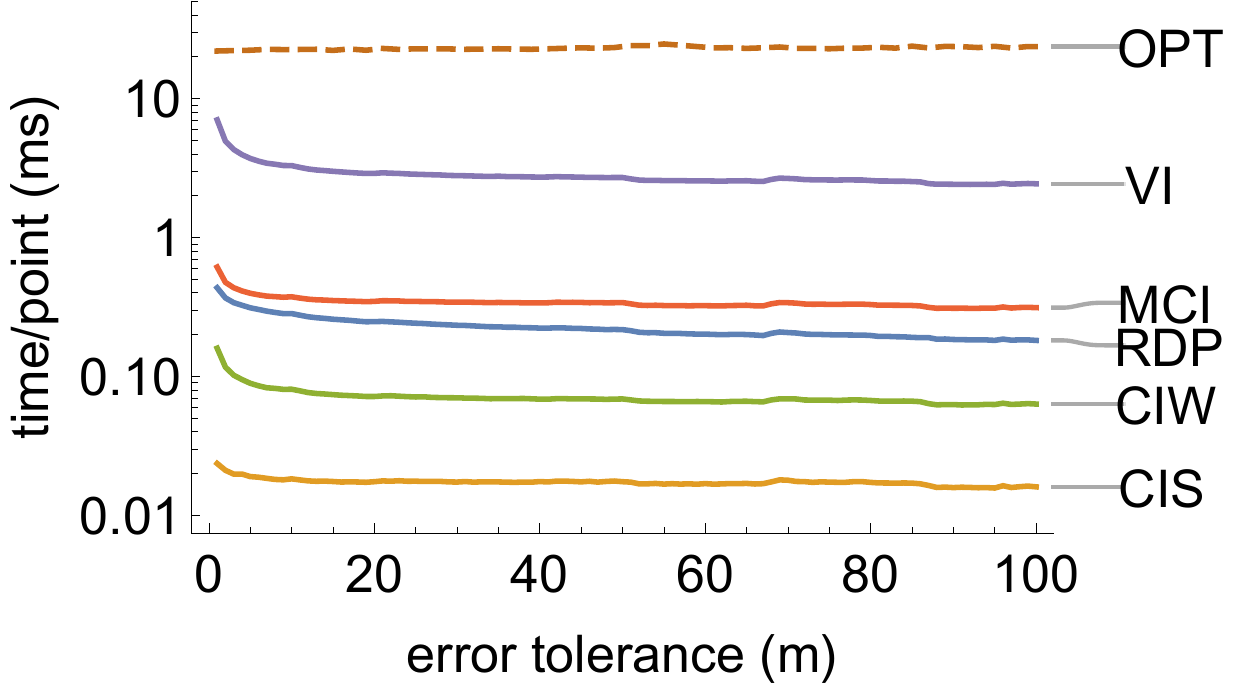}}
\subfigure[Mopsi 2D]{
\includegraphics[width=0.24\textwidth]{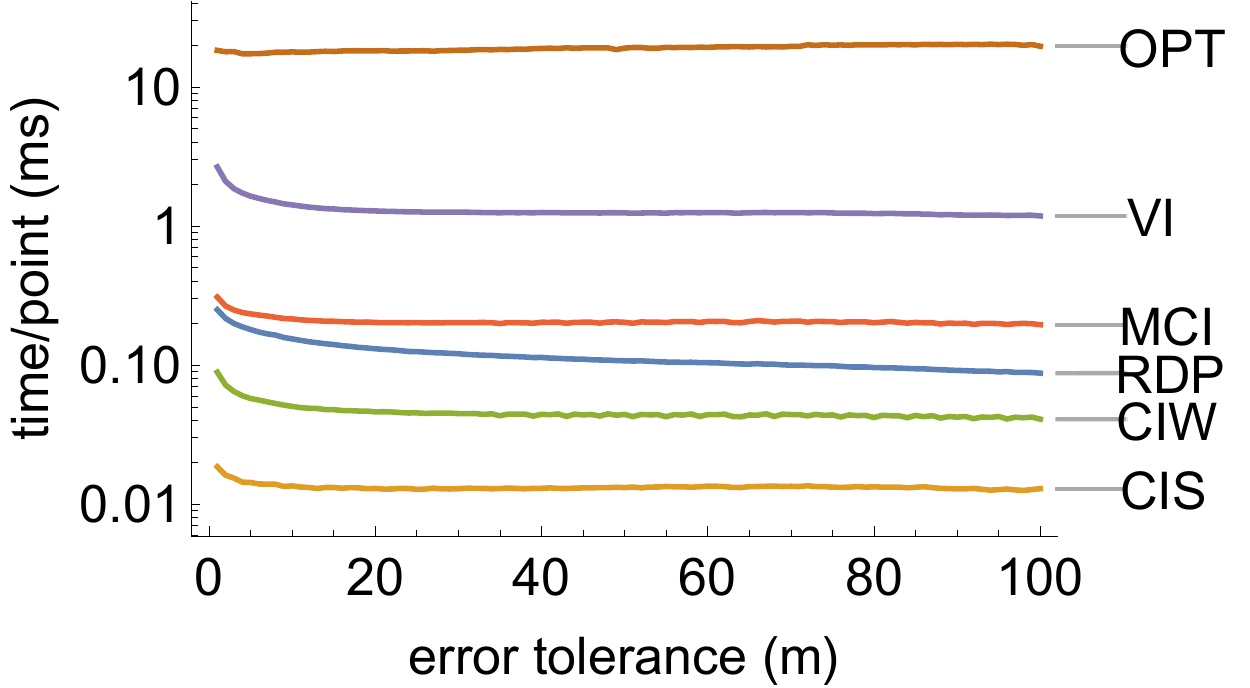}}
\subfigure[Mopsi 3D]{
\includegraphics[width=0.24\textwidth]{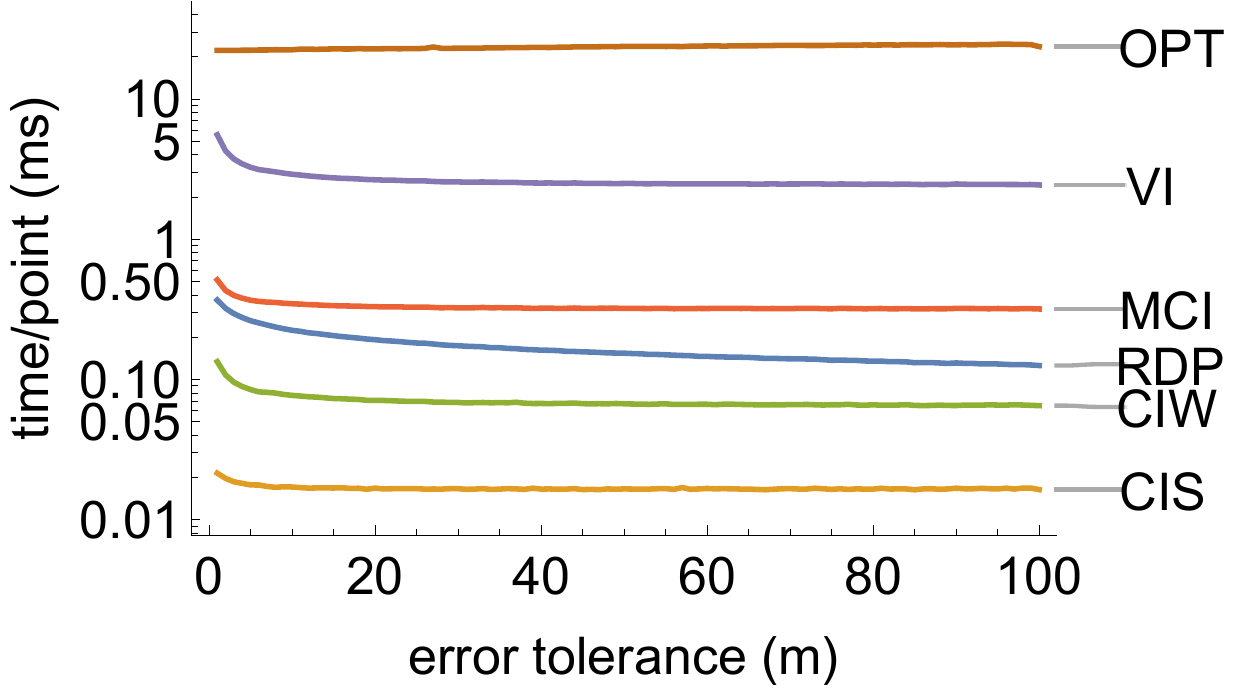}}
\subfigure[Bee 2D]{
\includegraphics[width=0.24\textwidth]{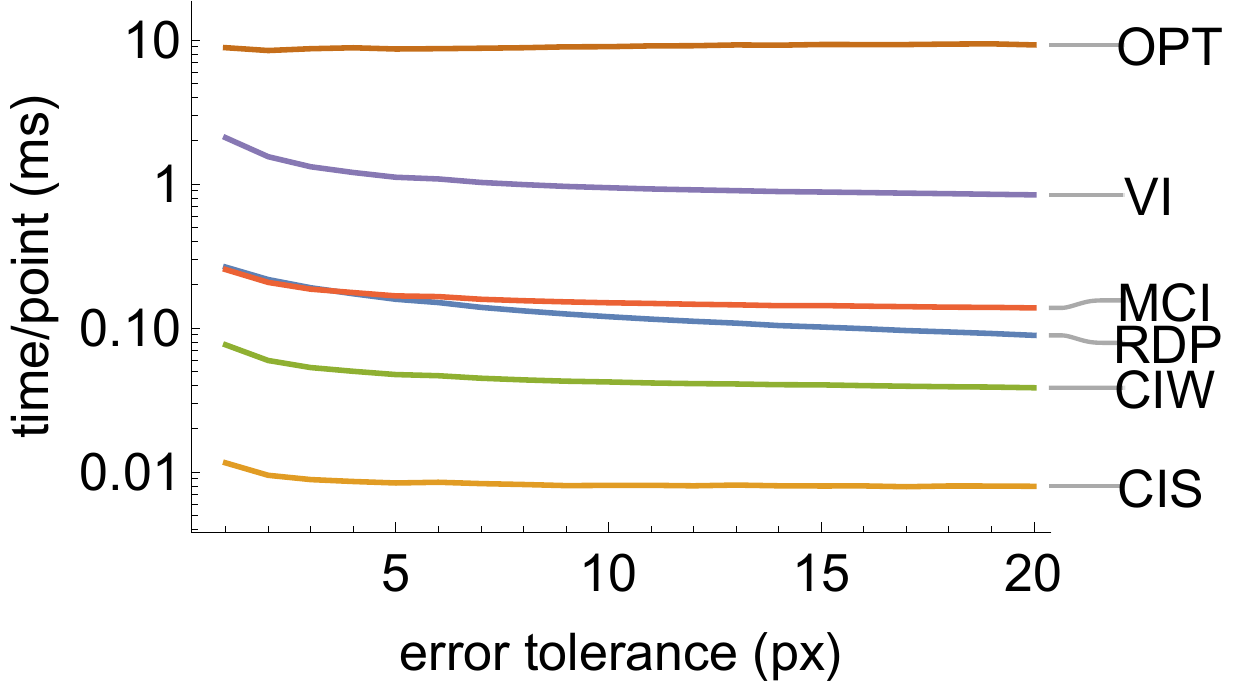}}
\subfigure[Danio 3D]{
\includegraphics[width=0.24\textwidth]{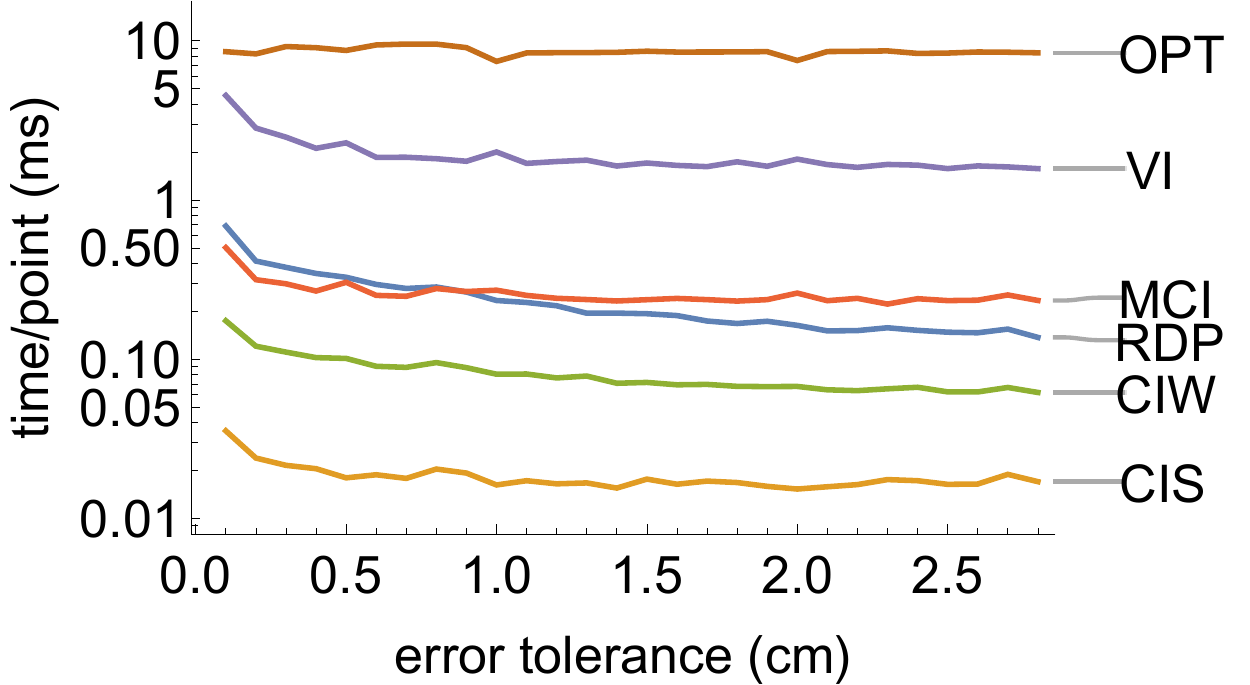}}
\subfigure[Wiener 2D]{
\includegraphics[width=0.24\textwidth]{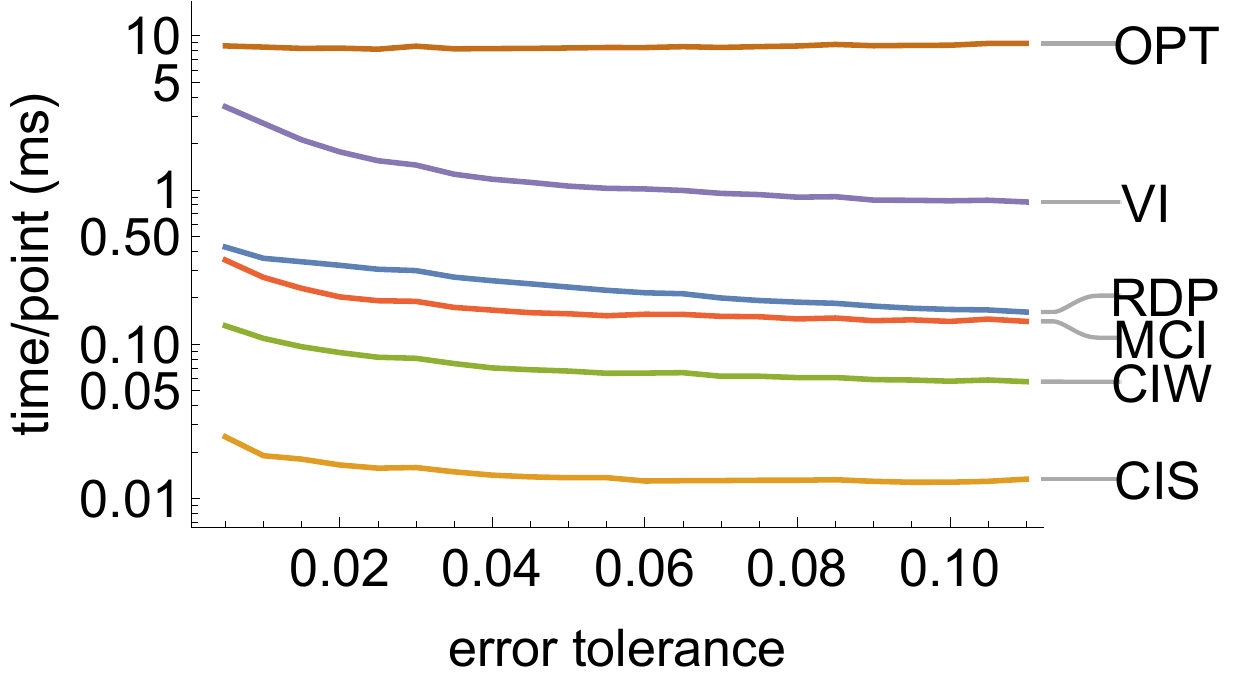}}
\subfigure[Wiener 3D]{
\includegraphics[width=0.24\textwidth]{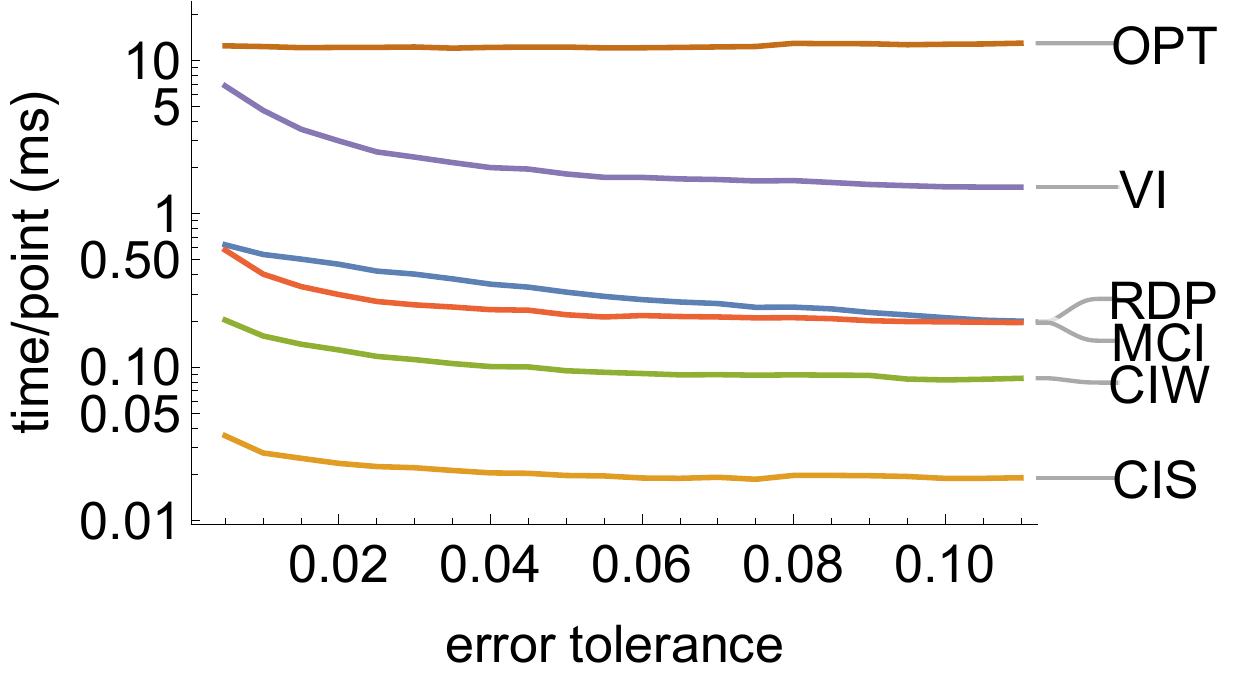}}
\captionsetup{font=bf}
\caption{Execution time}
\label{fig:effi}
\end{figure*}
\begin{figure*}
\centering
\subfigure[Mopsi 2D (ratio)]{
\includegraphics[width=0.24\textwidth]{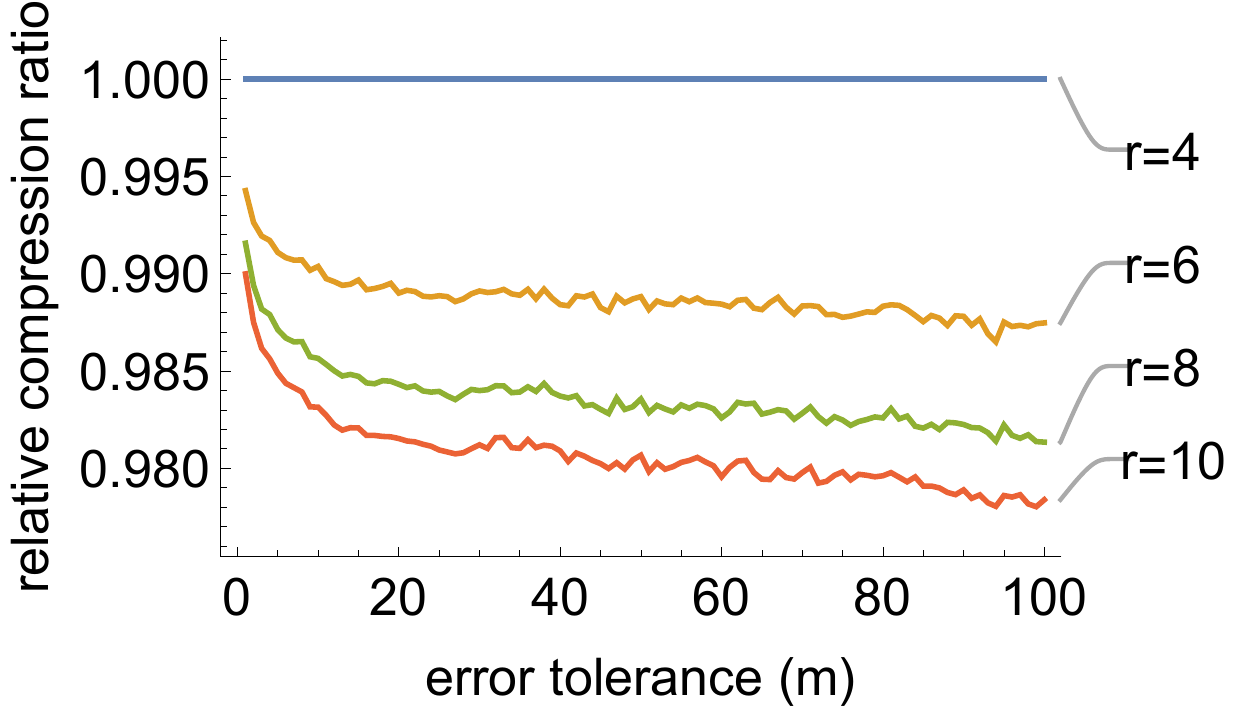}}
\subfigure[Mopsi 2D (time)]{
\includegraphics[width=0.24\textwidth]{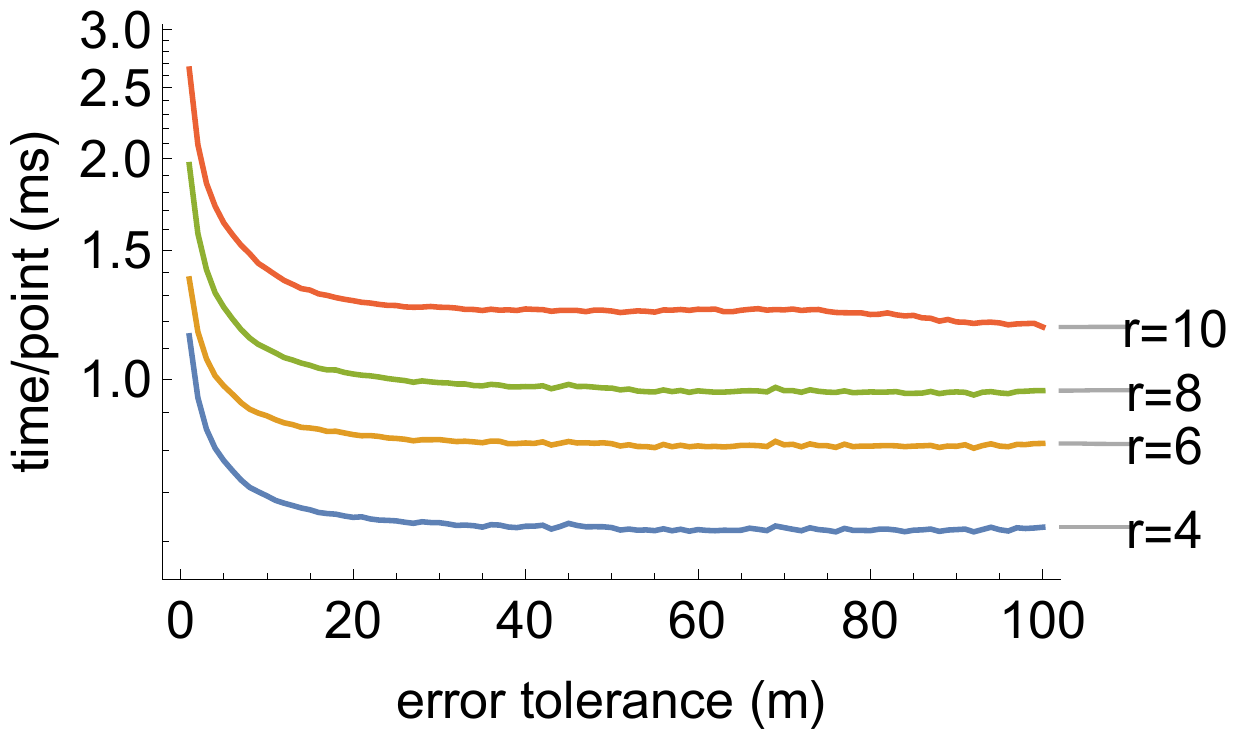}}
\subfigure[Mopsi 3D (ratio)]{
\includegraphics[width=0.24\textwidth]{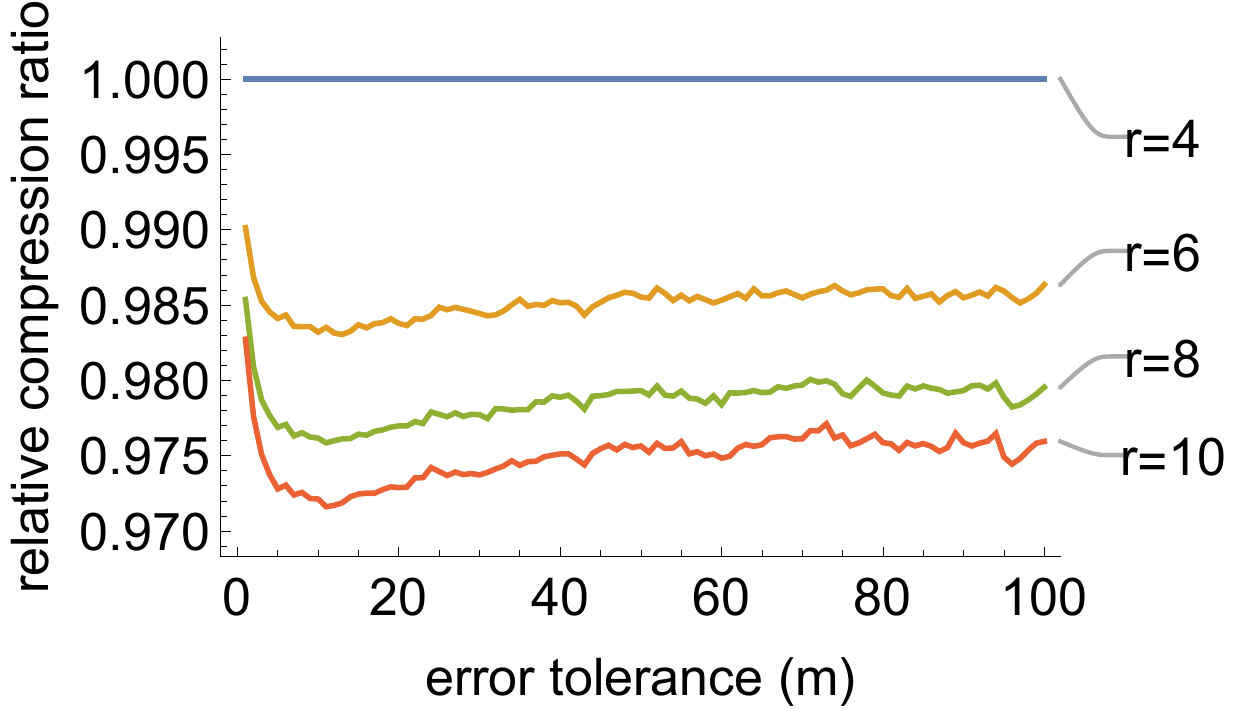}}
\subfigure[Mopsi 3D (time)]{
\includegraphics[width=0.24\textwidth]{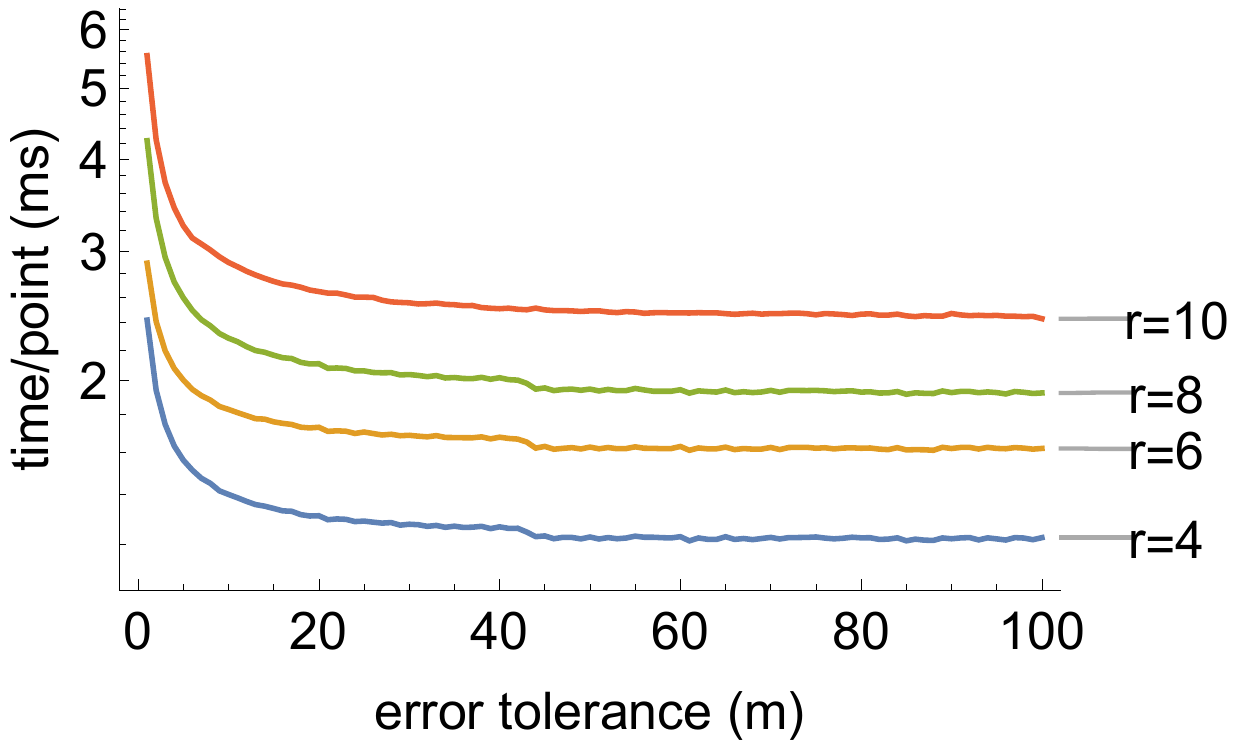}}
\captionsetup{font=bf}
\caption{Various interpolation}
\label{fig:r}
\end{figure*}

\subsection{Results}
We first evaluate the effectiveness of the algorithms in terms of compression ratios. 
We plot the \emph{relative compression ratios} with respect to the error tolerance $\epsilon$ in Figure~\ref{fig:effectr}. The relative compression ratio is defined as the ratio of the output size of an algorithm to that of the baseline algorithm RDP. For example, if RDP produces an output trajectory of 10 points while another algorithm produces only 7 points, the relative compression ratio of that algorithm is 70\%. We observe similar patterns in Figure~\ref{fig:effectr}. First, the relative performance of the algorithms for the different datasets is almost the same. Second, The performance of weak simplification methods is better than that of strong simplification methods, i.e., RDP and CIS are not as effective as MCI and VI. Third, we find the relative compression ratios of various algorithms (except CIW) to be relatively independent to the error tolerance when the error tolerance is not small. Moreover, the compression ratio of CIW becomes close to CIS as the error tolerance increases, and it is eventually even worse than RDP and/or CIS in the Bee and Wiener datasets.

Among the existing algorithms, OPT and CIW have the highest compression ratios. Moreover, CIW is more effective when the error tolerance is small while OPT is more effecive for great error tolerance. Regarding the implementations of LiMITS, we only plot MCI and VI because the DI is ineffective in practice. MCI usually outperforms other algorithms and VI is always the most effective. The compression ratio is further improved after we apply the compact format (see Figure~\ref{fig:cp}). We observe that both MCI and VI are significantly more effective than all other algorithms after using the compact representation. Note that OPT has a quadratic time complexity, and hence it is time-consuming to evaluate OPT on the largest dataset Beijing. Consequently, we sample a small part (10\%) and plot the results as dashed lines in Figures~\ref{fig:effectrbj2} and~\ref{fig:effectrbj3}.

Next, we report the efficiency of the algorithms. We plot the execution time per sample point with respect to the error tolerance in Figure~\ref{fig:effi}. Among all algorithms, CIS and CIW are the most efficient, because they are one-pass algorithms as they only access all sample points once. RDP has a time complexity of $O(n \log n)$ so it takes more time than CIS and CIW. The execution time of MCI is close to that of RDP though it is a linear algorithm. Because MCI involves complicated link distance computation, it has a greater constant factor. VI is $O(r^2mn)$ where $r$ is the sampling rate, $m$ is the dimension of the space, and $n$ is the number of vertices in the input trajectory. Consequently, VI takes even more time than MCI since it has a great constant factor. In fact, the dynamic programming in VI is very efficient. However, VI takes as $r+1$ times as much time as MCI, because the time is mainly spent on link distance computation and VI computes link distance $r+1$ times in each piece. Here the sampling rate of candidates in VI is 10, and VI takes approximately 11 times as much time as MCI on average. The optimal strong simplification method OPT outperforms RDP and CIS in terms of the compression ratio but is most time consuming. On the other hand, VI has a higher compression ratio than OPT, but only takes 10\% of its execution time on average.

We also demonstrate effects of the sampling rate of the candidates in VI. As Figure~\ref{fig:r} shows, the greater sampling rate leads to better compression ratios, but also results in greater execution time. Moreover, the execution time increases linearly with the growth of $r$. However, the difference in terms of compression ratios is not much, i.e., within 3\% in the Mopsi dataset. As a result, a moderate value of $r$ is sufficient to achieve high compression ratios.

To sum up, VI has the highest compression ratio among all approaches even without the compact representation, but takes more time to simplify a trajectory. MCI is less effective but more efficient than VI, and it completely outperforms the state-of-the-art algorithms (i.e., OPT and CIW) if the compact representation is used. Different from OPT and CIW, MCI and VI have high compression ratios no matter how small or large the error tolerance is.

\section{Related Work}
\label{related}
In this section, we review work closely related to the trajectory simplification problem, including data compression, error metrics, and existing algorithms exhibiting both strong
and weak simplification.

\subsection{Data compression}
LiMITS is a weak simplification approach for trajectory data, and thus belongs to data compression techniques. Data compression can be either lossless or lossy. The original data can be perfectly restored after lossless compression~\cite{huffman1952method, witten1987arithmetic, ziv1978compression}, but its compression ratio is limited by the information entropy~\cite{shannon1948mathematical}. Lossless algorithms for trajectory data~\cite{nibali2015trajic,cudre2010trajstore} are not effective when compared with lossy algorithms~\cite{zhang2018trajectory}, because trajectories are in the form of floating-point numbers. In other words, the alphabet size tends to be infinity when precision of data increases. Consequently, the statistical redundancy is very rare and thus lossless compression is ineffective, as shown in~\cite{han2017compress}. Lossy compression can achieve higher compression ratios by simply discarding unnecessary data, usually designed for images, audio and video~\cite{wallace1992jpeg, musmann2006genesis, rao2014video}. The quality of the data is preserved after lossy compression, i.e., users can hardly distinguish between the original and compressed data. Trajectories are similar to audio/video since they are all time series, and the majority of trajectory simplification algorithms are lossy, too. However, different error metrics should be used to measure the data quality.

\subsection{Error metrics}
There are a variety of error metrics in trajectory simplification. The first metric is the perpendicular Euclidean distance. It measures the perpendicular distance between output points and their associated line segments in the input, and then takes the maximum of all these values as the result. The perpendicular distance considers the spatial information of the trajectory data, so a small distance indicates that the shapes of two trajectories are similar. However, temporal information is as important, and thus two trajectories with similar shape may be very different in many other aspects like the starting/ending times, velocity, and acceleration. As a result, many other approaches bound errors through use of synchronized Euclidean distance, a simplified version of the \frechet distance, which is in the form of Equation~\ref{sd2} while using the $L_2$ metric. The synchronized distance takes into account both spatial and temporal information, and therefore the trajectories after simplification are similar in both space and time. Existing work uses the synchronized $L_2$ distance to measure the error. In this paper, we focus on using $L_\infty$, because the algorithms using this metric can be extended to higher dimensions. Besides the perpendicular and synchronized distances, other metrics based on direction~\cite{long2013direction} and velocity~\cite{lin2016velocity} are also useful to measure the similarity between trajectories.

\subsection{Simplification algorithms}
Existing trajectory simplification algorithms compatible with the synchronized $L_2$ distance include RDP~\cite{douglas1973algorithms}, SQUISH~\cite{Muckell2014}, OPW~\cite{meratnia2004spatiotemporal}, OPT~\cite{IMAI198631} and CI~\cite{Lin2019}, among which all except CI exhibit strong simplification. RDP is a well-known algorithm for trajectory simplification. The worst-case time complexity of RDP is $O(n^2)$, but can be reduced to $O(n\log n)$ and it is very efficient in practice. SQUISH maintains a priority queue of points, in which the priority of a point is given by the error introduced upon removing the point from the trajectories. Taking advantage of the priority queue, it can remove the point with the lowest priority efficiently, and takes $O(n\log n)$ time to simplify a trajectory with $n$ points. OPW is a greedy algorithm that always tries to simplify as many points as possible with one line segment at each iteration. The time complexity of OPW is $O(n^2)$ but it can be reduced to $O(n)$ through sector intersection~\cite{song2014press,Lin2019}. OPT is an optimal algorithm in terms of the compression ratio for strong simplification~\cite{IMAI198631}, which has an $O(n^3)$ time complexity but is able to produce minimum subsets. Similar to OPW, the time cost of OPT can be reduced to $O(n^2)$ through sector intersection as well~\cite{han2017compress}. All these algorithms are strong simplification algorithms, and therefore the compression ratio is limited. The state-of-the-art approach is CI, which improves OPW via sector intersection. Moreover, it introduces both strong and weak simplification algorithms, namely CIS and CIW, respectively. Although CIW is a weak simplification, it still constricts the interpolation points in the temporal dimension, whereas there are no such constraints in LiMITS. 

Besides strong and weak simplifications, we also decompose all algorithms into two groups according to their methodology. Trajectory data, in the form of a series of sample points, are discrete geometric objects, so all algorithms involve a varying degrees of knowledge from computational geometry. More precisely, the only geometric operation used in RDP, SQUISH, and naive OPW/OPT is the evaluation of distances between points, while improved OPW/OPT and CI implement the more complicated sector intersection operation in order to achieve better performance. Consequently, it is simple to generalize algorithms like RDP to other metrics in high dimensions by changing the distance evaluation function, while other algorithms using sector intersection are difficult to generalize, because the intersection of hyper-balls with respect to $L_p$ metrics in $m$-space takes more than linear time when $m>1$ and $1<p<\infty$. In 2-space, CI tackles $L_2$ by approximating circles with regular polygons of $n$ vertices so that the computation of sector intersections is simplified, which is equivalent to $L_\infty$ when $n = 4$. In this paper, we use $L_\infty$ directly without any approximation.

In~\cite{zhang2018trajectory}, the authors conducted a comprehensive experimental study on simplification algorithms invented in the last several decades, including other algorithms that do not use the synchronized distance and also those that do not use error tolerances. On the basis of the experimental study, the authors point out several directions worthy of further exploration. First, it is suggested to trade execution time and memory space for effective simplification. Second, it is useful to study new error metrics which are suitable for real applications. Finally, inter-trajectory redundancy has not been exploited yet, which could lead to even higher performance. In our work, we followed the first two directions. We first explored the metric $L_\infty$, making simplification in high dimensions feasible. Next, we proposed LiMITS, a linear but more complicated algorithm, to simplify trajectories more effectively.

\section{Conclusion}
\label{conclusion}
In this paper, we explored the problem of trajectory simplification under the $L_\infty$ metric, which makes it possible to generalize existing algorithms to higher dimensions. Next, we proposed three linear weak simplification algorithms based on multidimensional interpolation, namely LiMITS. It also allows a compact representation leading to an even higher compression ratio. Finally, we conducted extensive experiments on real datasets showing that LiMITS outperforms existing algorithms in terms of the compression ratio.

Notwithstanding, LiMITS is not an optimal algorithm, i.e., the number of points in the output is not minimized. Although the optimal weak simplification is reducible to the link distance problem, link distance computation in high dimensions is still an open problem~\cite{minlinkpath2016}. Nonetheless, the polytopes in the trajectory simplification problem are not arbitrary but tube-shaped. As a result, it is interesting to investigate whether there is an efficient algorithm to compute link distance inside the tube-shaped polytopes, and then simplify trajectories with the minimum link paths. Another direction for future work is exploring the relationship between trajectories that we obtain with those generated as a result of performing spatial network queries (e.g.,~\cite{sankaranarayanan2009path, sankaranarayanan2006distance, sankaranarayanan2010query}) and trajectory compression in networks~\cite{kellaris2013map,song2014press, han2017compress,li2020compression,cao2005nonmaterialized} as a great proportion of trajectories is from spatial networks.

\balance
\bibliographystyle{ACM-Reference-Format}
\bibliography{sample}

\end{document}